\documentclass{aa} 
\usepackage{graphicx}
\usepackage{txfonts}
\usepackage{soul}
\usepackage{ulem}
\usepackage{color}

\begin{document} 

\title{Effect of Coronal Loop Structure on Wave Heating by Phase Mixing}

\author{P. Pagano\inst{\ref{inst1}} \and
          I. De Moortel\inst{\ref{inst1},\ref{inst2}} \and
          R. J. Morton\inst{\ref{inst3}}
          }

\authorrunning{Pagano et al.}
\titlerunning{Wave heating}

\institute{School of Mathematics and Statistics, University of St Andrews, North Haugh, St Andrews, Fife, Scotland KY16 9SS, UK \label{inst1}
            \and
            Rosseland Centre for Solar Physics, University of Oslo, PO Box 1029  Blindern, NO-0315 Oslo, Norway \label{inst2}
            \and
            Department of Mathematics, Physics and Electrical Engineering, Northumbria University, Newcastle upon Tyne, NE1 8ST UK \label{inst3}
\\
      \email{pp25@st-andrews.ac.uk}
      }

   \date{ }

  \abstract
    {The mechanism(s) behind coronal heating still elude(s) direct observation and modelling of viable theoretical processes and the subsequent effect on coronal structures is one of the key tools available to assess possible heating mechanisms.
    Wave-heating via phase-mixing of Magnetohydrodynamics (MHD) transverse waves has been proposed as a possible way to convert magnetic energy into thermal energy but increasingly, MHD models suggest this is not a sufficiently efficient mechanism.}
   {We model heating by phase-mixing of transverse MHD waves in various configurations, to investigate whether certain circumstances can enhance the heating sufficiently to sustain the million degree solar corona and to assess the impact of the propagation and phase-mixing of transverse MHD waves on the structure of the boundary shell of coronal loops.}
   {We use 3D MHD simulations of a pre-existing density enhancement in magnetised medium and a boundary driver to trigger the propagation of transverse waves with the same power spectrum as measured by the Coronal Multi-Channel Polarimeter (COmP). We consider different density structures, boundary conditions at the non-drive footpoint, characteristics of the driver, and different forms of magnetic resistivity.}
   {We find that different initial density structures significantly affect the evolution of the boundary shell and that some driver configurations can enhance the heating generated from the dissipation of the MHD waves. In particular, drivers coherent on a larger spatial scale and higher dissipation coefficients can generate significant heating, although it is still insufficient to balance the radiative losses in this setup.}
   {We conclude that while phase-mixing of transverse MHD waves is unlikely to sustain the thermal structure of the corona, there are configurations that allow for an enhanced efficiency of this mechanism. We provide possible signatures to identify the presence of such configurations, such as the location of where the heating is deposited along the coronal loop.}


   \maketitle
%

\section{Introduction}
\label{introduction}

Whether the energy carried by magnetohydrodynamics (MHD) waves in the solar corona can contribute to maintaining the million degree plasma against its radiative losses remains a puzzle \citep[e.g. ][]{ParnellDeMoortel2012, Arregui2015}.
MHD waves have been detected in the solar corona for more than a decade \citep[e.g.][]{Tomczyk2007, McIntosh2011, 2013SSRv..175....1M} and
are reported to carry a significant amount of energy, of the order of 50-200 W/m$^2$  \citep[e.g.][]{McIntosh2011, 2014ApJ...790L...2T, 2015NatCo...6.7813M} in comparison with the energy requirements to balance the continuous radiative losses of the coronal plasma.
However, we still have no observational confirmation, nor full modelling corroboration that this wave energy can be converted into thermal energy efficiently and effectively.
These investigations are primarily carried out observing and modelling coronal loops that are arch-like dense and hot magnetic structures in the solar corona \citep{Reale2010}.
Although coronal loops do not fill up the entire coronal volume, as they are denser than the surrounding corona, they are responsible for most of the coronal emission and thus most of its radiative losses.

Moreover, recent observations of the solar corona have further constrained the properties of propagating transverse waves. In a series of work the power spectrum of these oscillations has been derived and confirmed to be only marginally dependent on time and locations, confirming that waves are an inherent and ubiquitous property of the solar corona \citep{Morton2016,2019NatAs...3..223M}.
Although several mechanisms have been proposed to convert wave energy into thermal energy, realistic models of these mechanisms in coronal loops have so far not shown that they can indeed support the thermal structure of the corona.
An example of such models is the mode coupling and subsequent phase-mixing of Alfv\'en waves \citep{HeyvaertsPriest1983}.
\citet{Pascoe2010} have shown that the mode-coupling of kink and Alfv\'en azimuthal modes, and their following phase-mixing, can lead to the concentration of energy into thin boundary shells across which the Alfv\'en speed varies. In this model the boundary shell is assumed to be pre-existent to the wave propagation.
Subsequent works \citep{Pascoe2011, Pascoe2013} have confirmed the robustness of these modelling results showing
that such an energy concentration naturally occurs when a boundary shell is present.
Although this mode-coupling is effective in concentrating the wave energy, when the dissipation of waves is included in the model, the heating power derived from the conversion of the wave energy into thermal energy is largely insufficient to balance the radiative losses. Indeed, \citet{PaganoDeMoortel2017} and \citet{Pagano2018} have shown that there is insufficient heating of the coronal plasma to counteract the plasma radiative losses in a simplified model.
The same conclusion is reached in \citet{Pagano2019} when propagting MHD waves are excited in the domain through the buffeting of the coronal loop footpoint with a driver derived from the multi-frequency spectrum observed by \citet{Morton2016}.
Other modelling efforts have reached similar conclusions \citep[e.g.][]{Karampelas2017,Karampelas2019},
also addressing how the mix of cold and hot plasma can show apparent heating.
In contrast, other studies claim  that MHD waves can efficiently heat the coronal plasma \citep{Srivastava2017,2018A&A...614A.145L}. These models either rely on very high frequency oscillations or they only focus on the energy stored in the waves, without addressing the conversion mechanism. However, analysing SDO/AIA observations, \cite{2019NatAs...3..223M} found that although high-frequency oscillations have larger amplitudes, they do not appear to occur regularly in the corona. Hence, their time-averaged power is small, implying they probably make little contribution to the energy supply.

At the same time, another line of research based on both modelling and observations has focused on the effect of propagating or standing transverse waves on the pre-existing coronal loop structure. Several papers have shown that standing waves cause Kelvin-Helmholtz instabilities (KHIs) at the boundary shell
when a density contrast is present between the interior and the exterior region of a waveguide, leading to fragmentation of the boundary shell \citep[e.g.][]{BrowningPriest1984, Terradas2008, Antolin2015, Okamoto2015}.
These results have been shown to be quite robust as KHIs develops even when resistivity and viscosity are included in the model \citep{Howson2017}, when standing modes are setup by the reflection of propagating waves \citep{Karampelas2017},
when gravity stratification is included \citep{2018ApJ...856...44A, Karampelas2019}
and that KHI can occur in conjunction with parametric instabilities \citep{2019MNRAS.482.1143H}.
Similar structures have been found in the simulations of \citet{PaganoDeMoortel2017} and \citet{Pagano2019} with purely propagating waves and \citet{2017NatSR...714820M} provide a first step to describe the dynamics in this setup as 'uniturbulence'. Either way, the presence of MHD waves in dense loops seems to lead to the development of small scale structures and the erosion of the density enhancement of the loop. Finally, these small scale structures across which the density and magnetic field intensity vary rapidly could be favourable places to enhance phase-mixing of waves or where the magnetic field is sufficiently entangled \citep[e.g.][]{Reale2016,2017ApJ...837..108P} such that nanoflares \citep{1988ApJ...330..474P} can occur.

In summary, so far 3D numerical models seem to conclude that the direct dissipation of MHD waves is not able to supply enough energy to counteract the losses through radiation and thermal conduction in active region coronal loops. However, waves could play a key role in shaping coronal loops and developing the small-scale structures required for their own dissipation or other heating mechanisms to become more efficient. The present paper builds on the work of \citet{Pagano2019}. We expand our investigation of the heating induced by propagating transverse waves by considering a number of configurations for the loop density structuring, open or closed footpoints, and the extent of the driver.
To do so, we drive an observed spectrum of Alfv\'en waves in the corona into a loop structure and we investigate the subsequent heating and boundary shell evolution. We investigate how various parameters can affect i) the heating deposited from the dissipation of waves and ii) how the boundary shell evolves and whether it becomes more effective. In particular, we focus on loop structures which are either homogeneous or change along the field-aligned direction, to model pre-existing loops that have been completely filled by plasma or loops where the filling through evaporation is still ongoing \citep{2019ApJ...882....7R}, or which are short-lived structures such as spicules. We note here though that the process of evaporation or the motion of spicules are not included self-consistently. Conditions at the top boundary are modified to model both open and closed structures and the horizontal extent of the boundary driver is altered to examine how the spatial extent of the footpoint motions compared to the extent of the flux tube affects the energy budget in the loop. Finally, we consider different power spectra in addition to the observed spectrum to show how model results are sensitive to the power spectrum used.

 The paper is structured as follows. In Sec.\ref{modeldriver}, we illustrate the loop models and the driver setup and in Sec.\ref{MHDsimulation} we show the result for one representative simulation. In Sec.\ref{loopstructure}, we address the problem of the boundary shell evolution, followed by Sec.\ref{haetingthecorona} where we explain how different configurations impact on the heating deriving from waves. Finally, in Sec.\ref{spectrum} we show how the wave heating depends on the input power spectrum. Conclusions are presented in Sec.\ref{conclusions}.

\section{Model and driver}
\label{modeldriver}

In order to study how the loop density structure affects the heating by transverse MHD waves in the solar corona, we run a number of 3D numerical simulations where we solve the non-ideal MHD equations using the AMRVAC code \citep{Porth2014}. In these experiments, we vary (i) the prescribed initial condition for the loop structure, (ii) the boundary conditions that govern how transverse MHD waves are introduced or retained in the computational domain, (iii) the dissipation coefficients, and (iv) the driver we use to generate these waves. In this section, we describe in detail how these different experiments are constructed.

\subsection{Loop structure}
\label{sectionloop}
We treat the coronal loop as a magnetised cylinder which consists of a dense interior region embedded in a less dense environment. The loop interior is surrounded by a boundary shell across which the density decreases gradually until it matches the exterior density. The Alfv\'en speed is therefore uniform in the interior region and the exterior of the loop, but varies through the boundary shell. We neglect the effects of gravity and curvature,
and focus on the coronal segment of the loop, without modelling the chromosphere and transition region.

We use a Cartesian reference frame, where $z$ is the direction along the cylinder axis.
The cylinder configuration we use is defined between $z_{min}=-20$ $Mm$ and $z_{max}=180$ $Mm$. The centre of the loop footpoint is placed at the origin of the axes. The cylinder has a radius $a$ and the radius of the dense interior region is given by $b$. Hence, the boundary shell is the interface region between the interior and the exterior, i.e. between radii $b$ and $a$.

\begin{figure*}
\centering
\includegraphics[scale=0.23]{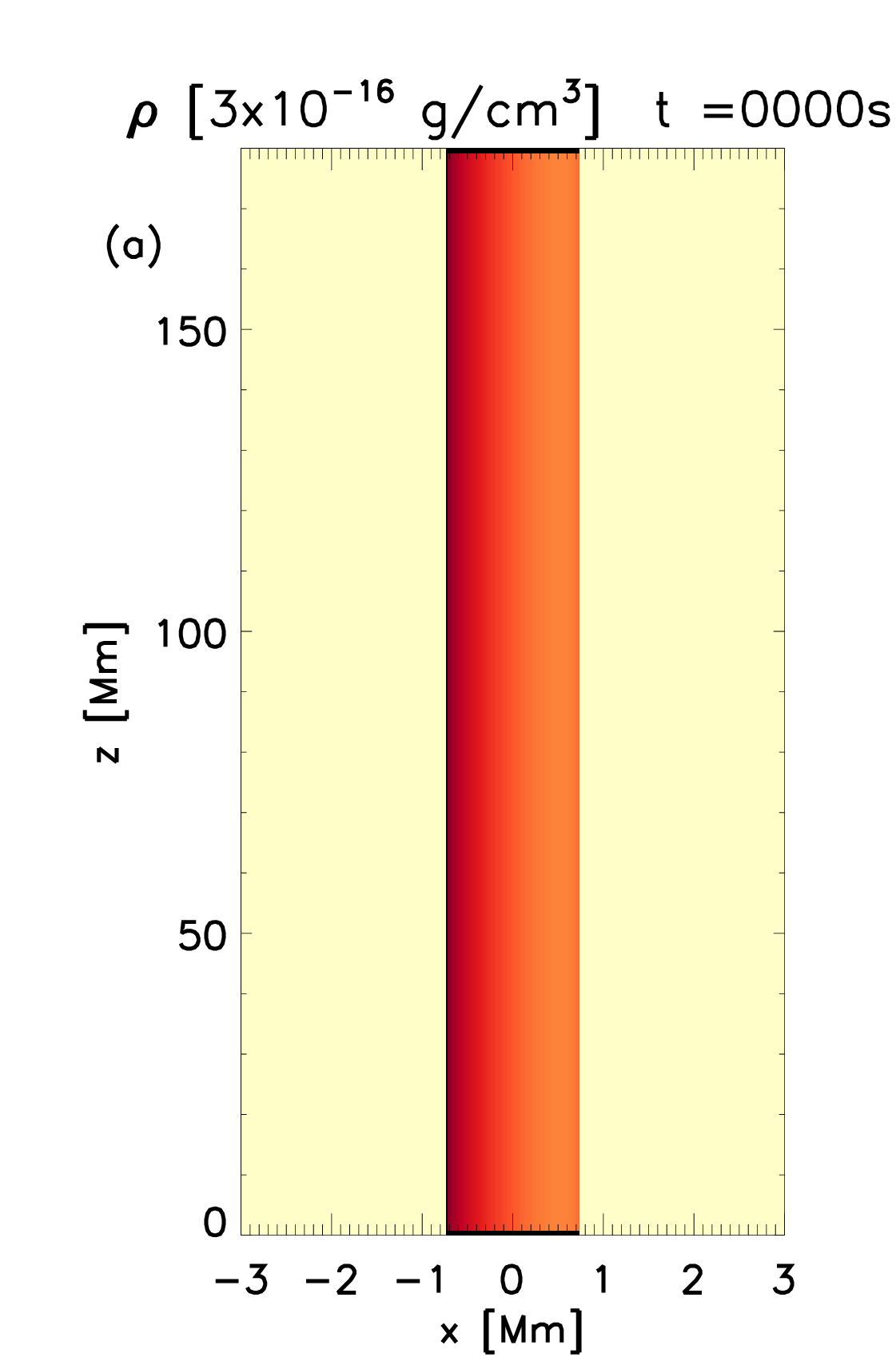}
\includegraphics[scale=0.23]{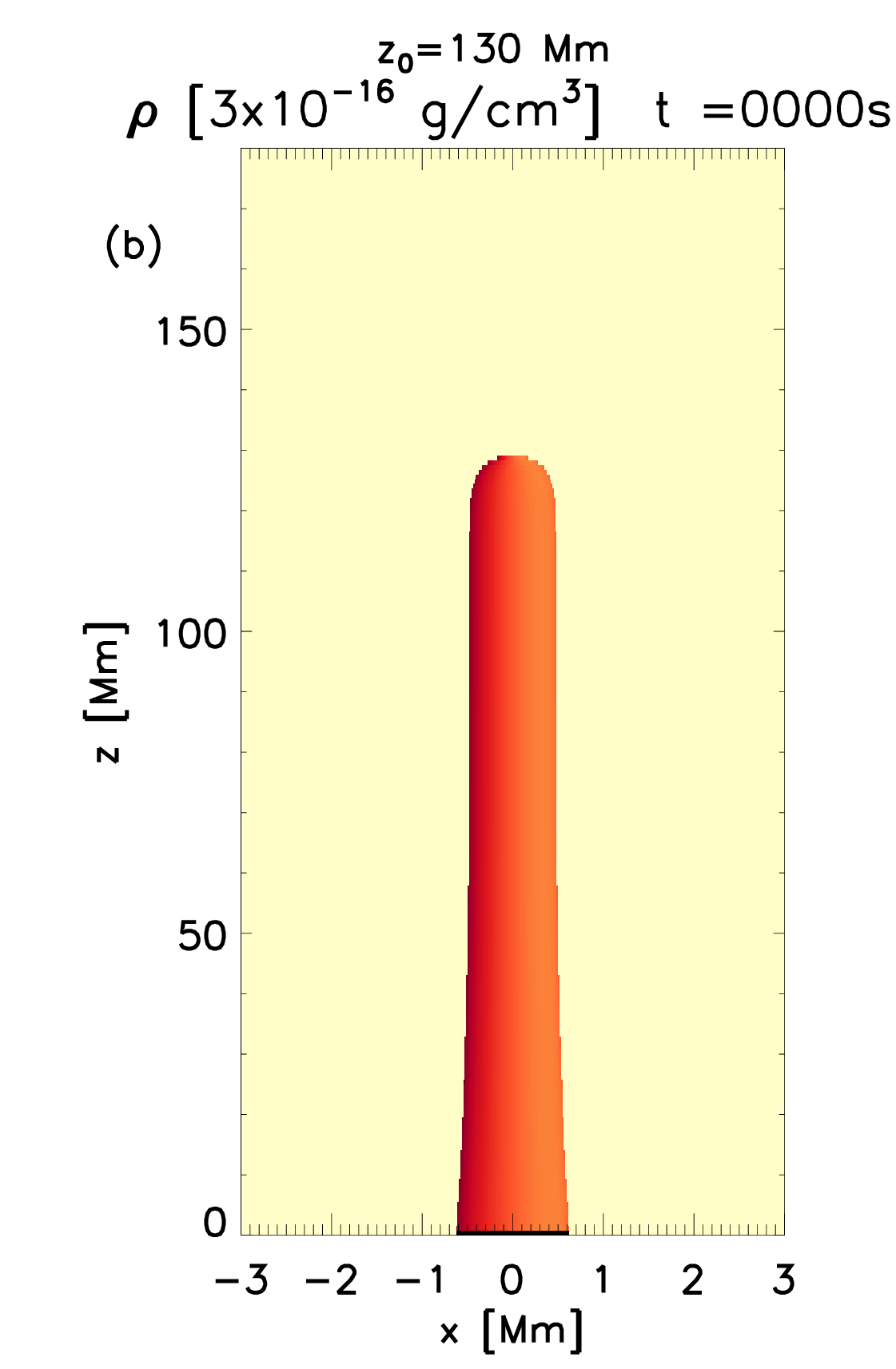}
\includegraphics[scale=0.23]{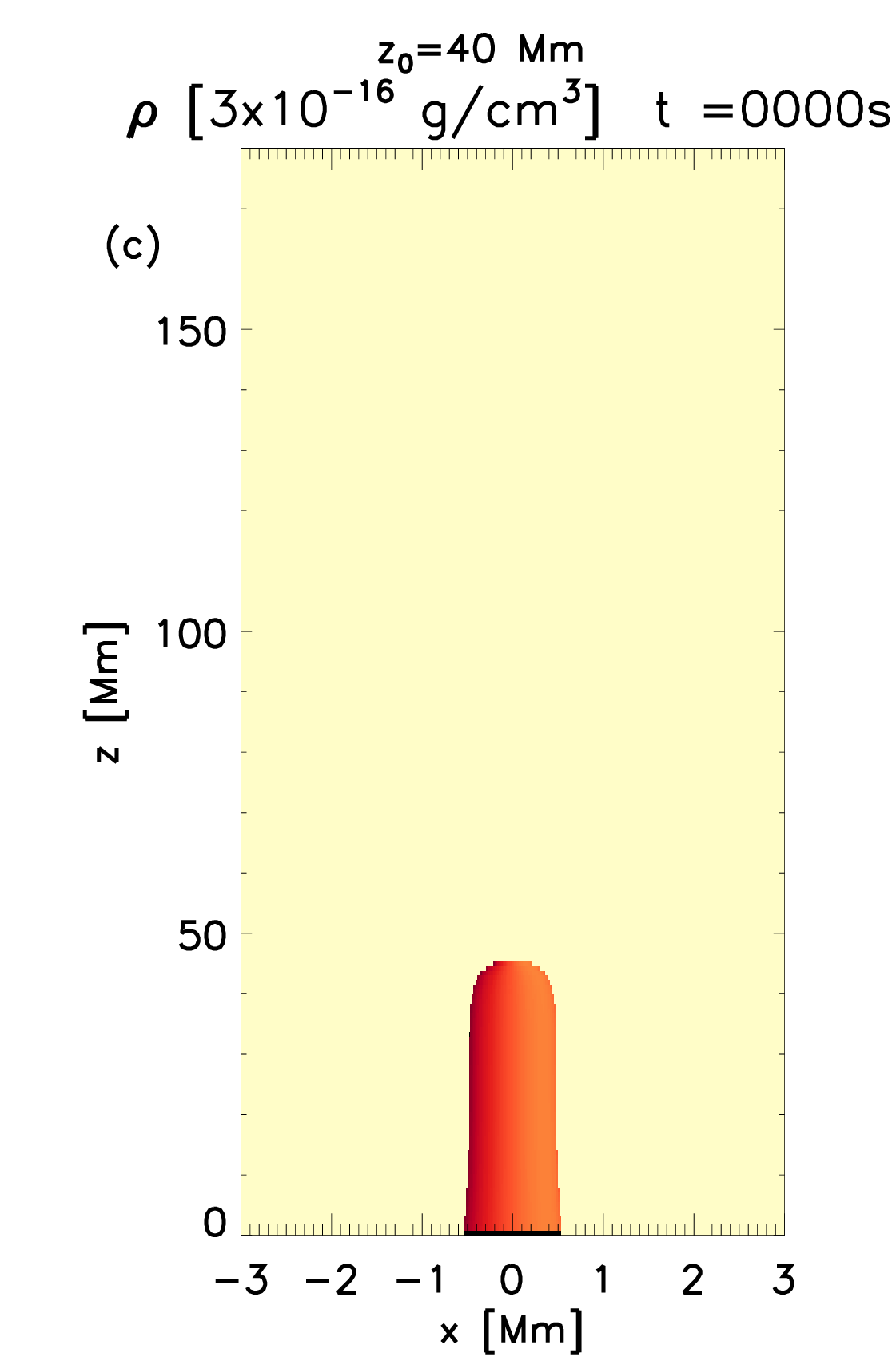}

\includegraphics[scale=0.23]{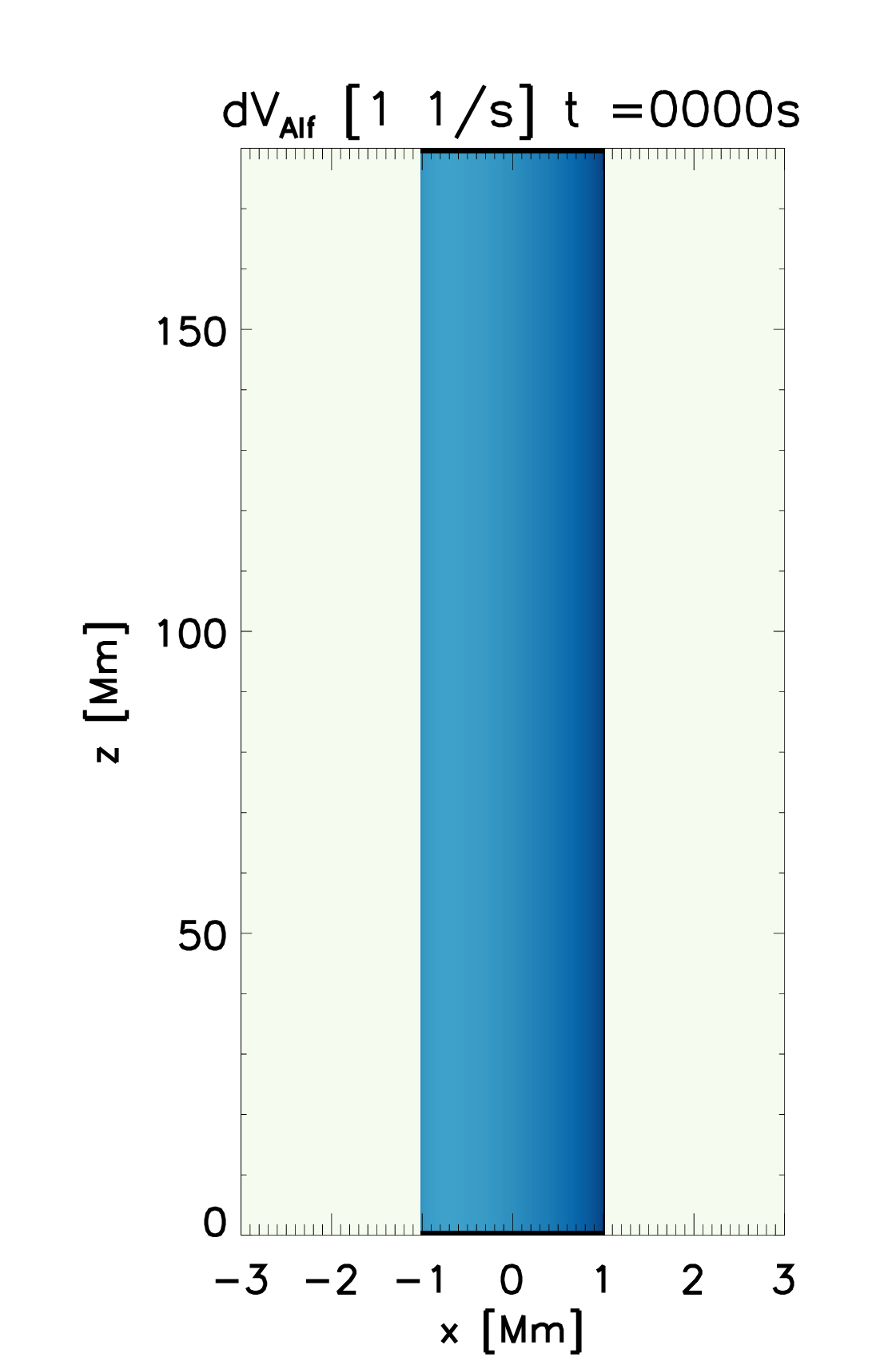}
\includegraphics[scale=0.23]{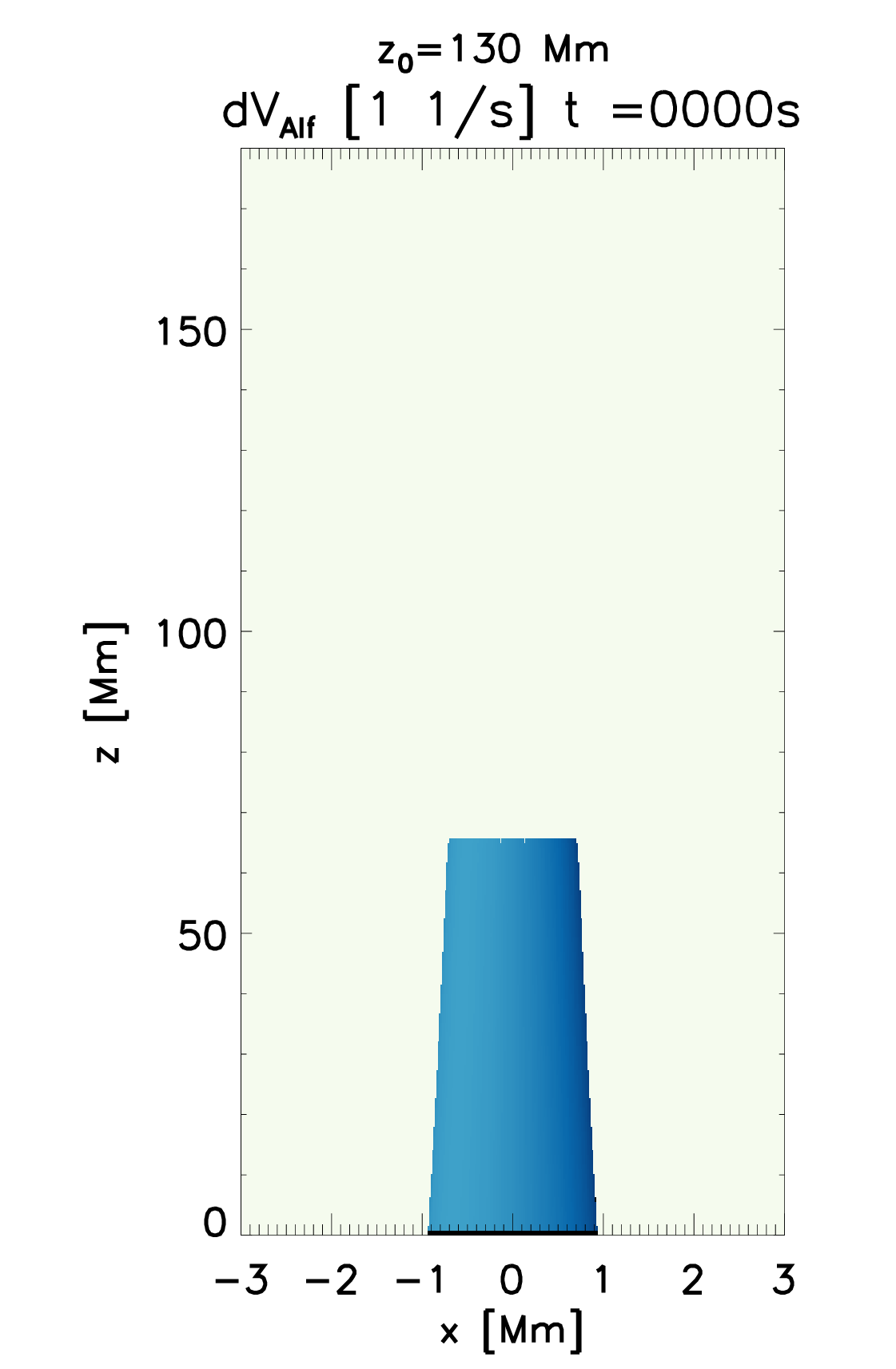}
\includegraphics[scale=0.23]{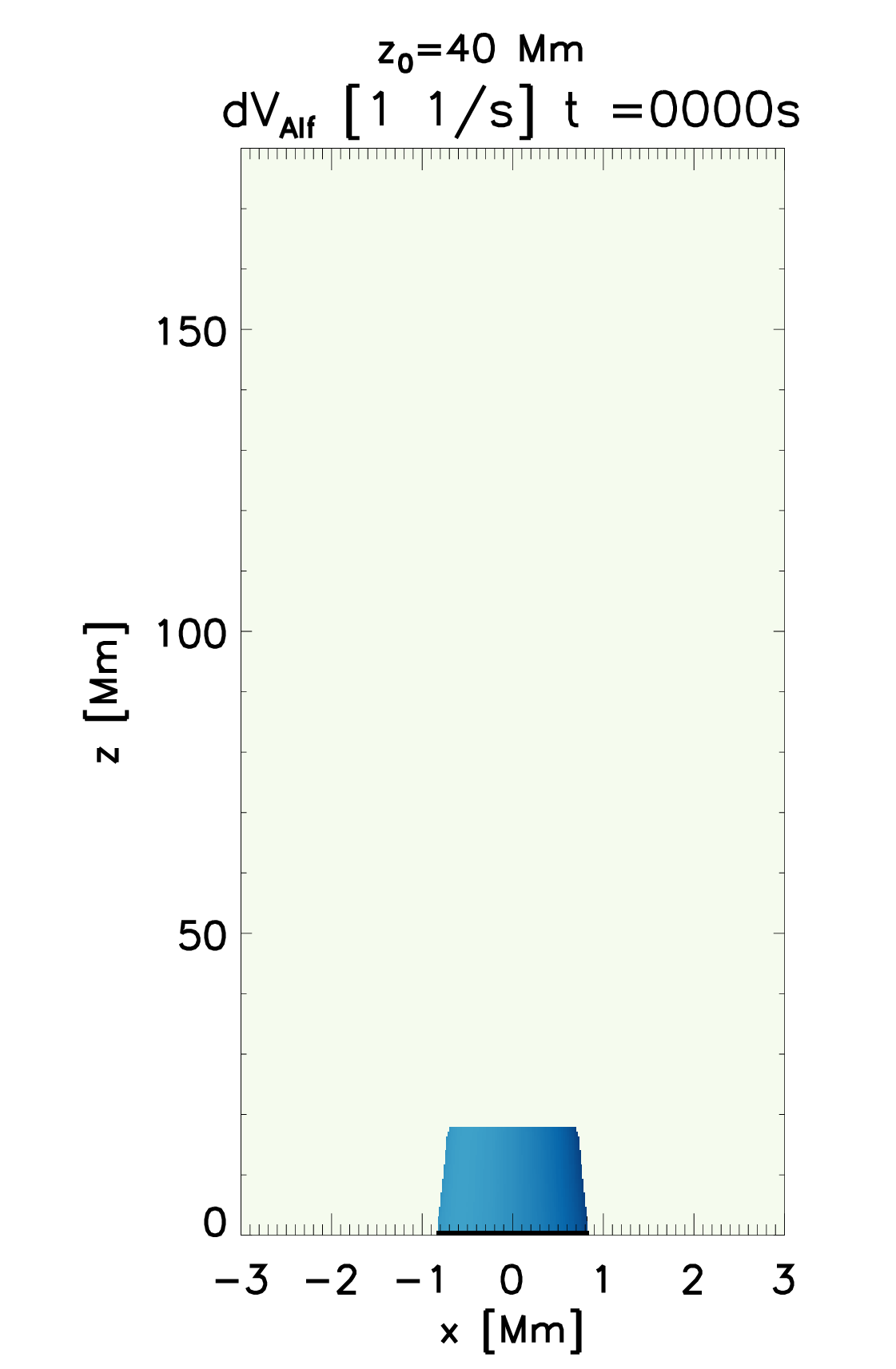}
\caption{Isosurfaces of density (top row) and Alfv\'en speed gradient (bottom row) for three different loop configurations: uniform (left hand side), $z_0=130$ $Mm$ (centre), and $z_0=46$ $Mm$ (right hand side).}
\label{initialloop}
\end{figure*}

In order to study the effect of the initial coronal loop structure, we consider two different density distributions.
In the uniform one, the interior region radius $b$ is uniform along the length of the cylinder and we take $a=1$ Mm and $b=0.5$ Mm.
The density, $\rho$, decreases across the boundary shell and is 
defined as the following function of $\rho_e$, $\rho_c$, $a$, and $b$:
\begin{equation}
\label{densitylayer}
\displaystyle{\rho(\rho_e,\rho_c,a,b)=\rho_e\left[1+\left(\frac{\rho_c-1}{2}\right)\left[1-\tanh\left(\frac{e}{a-b}\left[r-\frac{b+a}{2}\right]\right)\right]\right]},
\end{equation}
where $r=\sqrt{x^2+y^2}$ is the radial distance from the centre of the cylinder,
$\rho_e=1.16\times10^{-16}$ g cm$^{-3}$ is the external density and $\rho_c$ is the density enhancement between the exterior and interior region which we set to be $\rho_c=4$.
Fig.\ref{initialloop}a shows a density isosurface for this uniform configuration.

\begin{figure}
\centering
\includegraphics[scale=0.23]{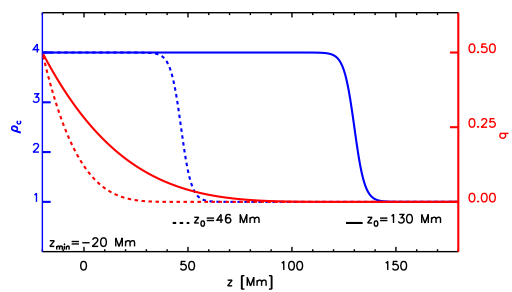}
\caption{Profile of the density contrast $\rho_c$ (blue curves) and interior region radius $b$ (red curves),
as a function of z for two loop density profiles with $z_0=46$ $Mm$ (dashed curves)
and $z_0=130$ $Mm$ (continuous curves).}
\label{brhocfig}
\end{figure}

In addition, we also consider a non-uniform cylinder where the interior region is only present near the lower boundary (for $z \le z_0$), defining its radius as 
\begin{equation}
b\left(z\le z_0\right)=0.5\left(1-\frac{z-z_{min}}{z_0-z_{min}}\right)^4
\label{bequation}
\end{equation}
and $b=0$ for $z>z_0$.
Using this profile of $b\left(z\le z_0\right)$ the interior region radius rapidly shrinks above $z=z_{min}$ and it disappears at $z=z_0$ (red curves in Fig.\ref{brhocfig}).
Additionally, at $z=z_0$ the density quickly drops as
\begin{equation}
\rho_c=1+\frac{3}{2}\left[\tanh\left(\frac{z_0-z}{\Delta z}\right)+1\right]\,.
\label{rhocequation}
\end{equation}
This profile for the density contrast is close to $\rho_c=4$ for $z<z_0$ and then decreases to $\rho_c=1$ over a length $\Delta z$ near $z=z_0$ (blue curves in Fig.\ref{brhocfig}).
Fig.\ref{initialloop}b shows the density contour for this non-uniform configuration, where we use $z_0=130\,Mm$ and $\Delta z=5\,Mm$ and Fig.\ref{initialloop}c for $z_0=46\,Mm$.
The density structure narrows above $z_{min}$ because the interior region gets thinner and the loop structure disappears above $z=z_0$, where the density contrast becomes $\rho_c=1$.
It should be noted that the density distribution at the cross section $z=-20\,Mm$ is the same for both the uniform and non-uniform setup.

Such non-uniform configurations resemble the scenario where the coronal loop density enhancement occurs as consequence of evaporation, which lifts plasma from the dense chromosphere into the corona. We assume that this local density enhancement is present on a time scale comparable to the wave propagation along the loop (see e.g. \citet{2018ApJ...856...44A} for a similar model of a spicule).

The magnetic field, $\vec{B}$, in the cylinder is initially uniform and aligned with the $z$-direction. With an initial magnetic field strength of $B_0=5.68$ G, we have a plasma $\beta=0.02$. The initial plasma temperature $T$ is set by the equation of state:
\begin{equation}
\label{eos}
\displaystyle{p=\frac{\rho}{0.5 m_p} k_b T}
\end{equation}
where $m_p$ is the proton mass and $k_b$ is the Boltzmann constant. The initial temperature ranges between $0.34$ MK (interior) and $1.35$ MK (exterior).

\subsection{Driver}
\label{sectiondriver}
To reproduce the observed power distribution across different frequencies, we use a boundary driver to trigger propagating, transverse MHD waves along our magnetised cylinder, implemented as a 2D velocity field (along $x$ and $y$ directions) at the lower boundary of the domain, $z=-20$ $Mm$.
In our reference setup, the velocity field is applied within a radius $a$ (the outer radius of the cylinder) of the centre of the lower boundary.

Our 2D velocity driver is based on the average observed spectrum of transverse oscillations in the solar corona 
as measured by \citet{2019NatAs...3..223M} which is obtained as a spatial and time average of the observed transverse oscillations
in the low corona observe by the
Coronal Multi-channel Polarimeter \citep[COMP, ][]{Tomczyk2008}.
This observed spectrum consists of a power law with a power index of $\alpha_p$
and an additional Gaussian power enhancement distributed around $\nu_0=4$ $mHz$
with maximum $\Delta p=0.002$ in normalised units
and standard deviation $\sigma_p=0.2$ $Hz$.
Our power spectrum is constructed such that the time integral of the velocity is zero, i.e.~so the net displacement over the duration of the simulation is zero.
In this study, we use the average observed spectrum as reference for our investigation but also consider two modified spectra in order to analyse the role of the key features of the observed power spectrum (see Table \ref{spectrumpar} for the parameters used for the spectra). In the first modified spectrum, the power enhancement at $\nu_0=4$ $mHz$ is removed and uniformly redistributed across the spectrum to understand the role of waves generated by the localised power enhancement 
on the loop structure and heating. For the second modified spectrum, we use a smaller power law index to prescribe a different distribution, with relatively more power at high frequencies than in the observed spectrum, in order to verify the efficiency of the heating if more high-frequency waves propagate into the corona from lower layers of the solar atmosphere.

\begin{table}[!h]
\centering                          
\begin{tabular}{c c c}        
\hline\hline                 
Spectrum & Power Law Index     & Power enhancement  \\    
         & $\alpha_p$          & $\Delta p$ \\    
\hline                        
   Observed          & -1.36022, &  0.00213022 \\      
   No Enhancement    & -1.36022, &  0 \\      
   Powered High - $\nu$ & -1,       &  0.0003 \\
\hline                                   
\end{tabular}
\caption{Parameters of the observed spectrum.}             
\label{spectrumpar}      
\end{table}

\begin{figure}
\centering
\includegraphics[scale=0.32]{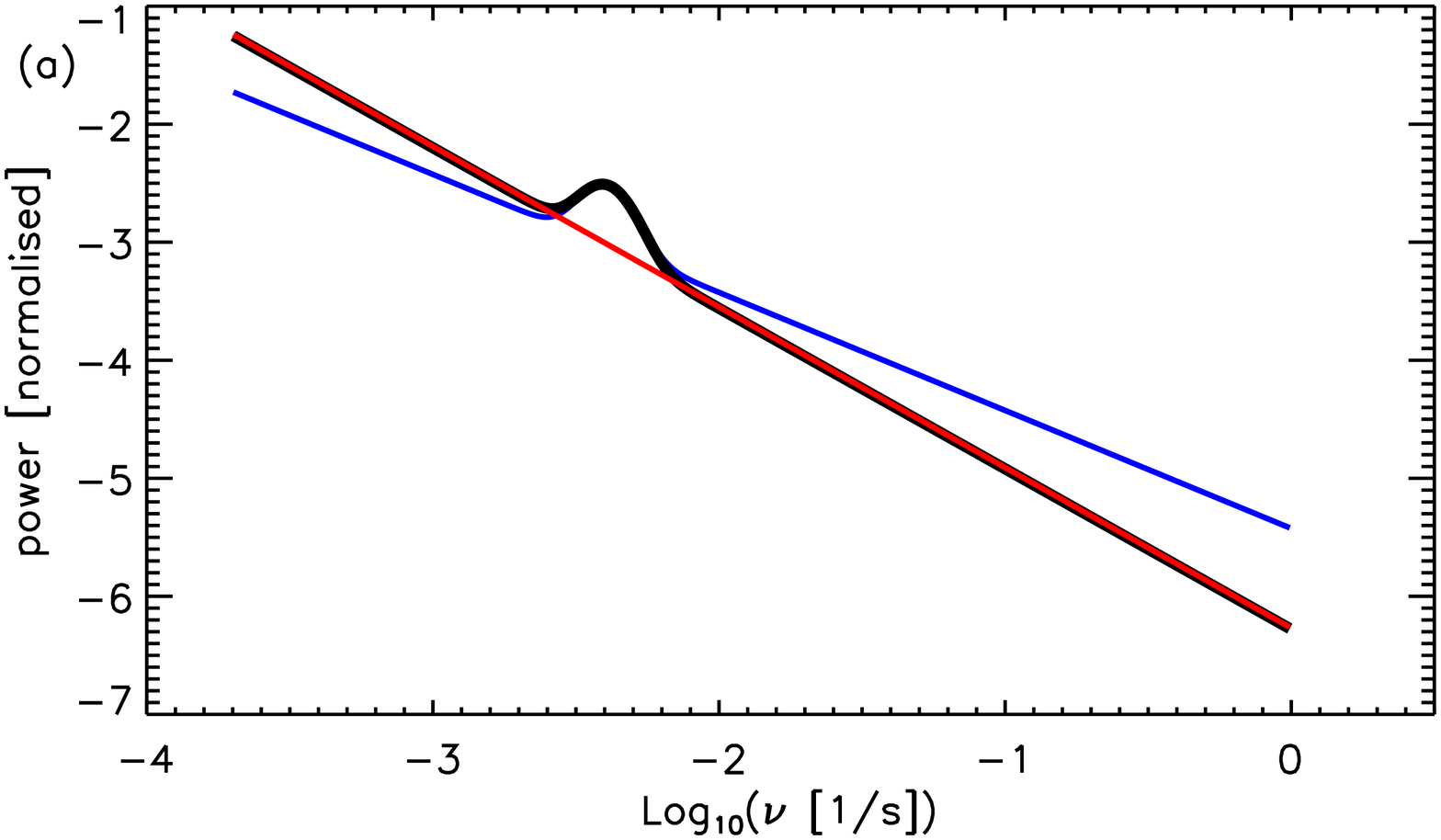}
\includegraphics[scale=0.32]{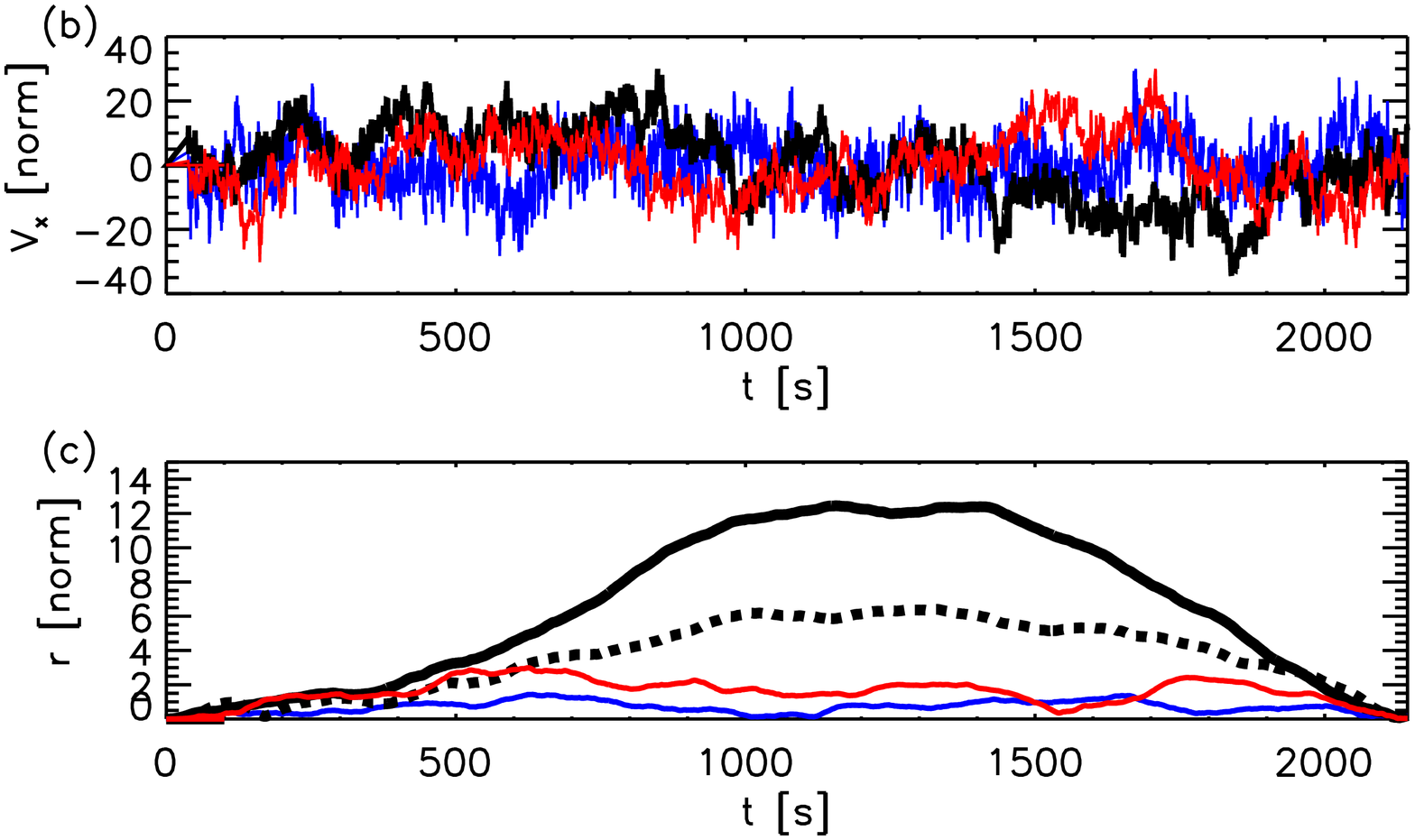}
\caption{(a) Spectra for the transverse waves we use in this study. The black curve represent the averaged observed spectrum, the red curve represent a spectrum where the power enhancement at $\nu_0=4$ $mHz$ is removed, and the blue curve represent a spectrum where we use a shallower power law index.
(b) Velocity $V_x$ time series for the three spectra in normalised units with same colour legend.
(c) The displacement associated with the velocity time series in normalised units and with the same colour legend. The black dashed curve represent the displacement derived from an additional velocity profile obtained from the same observed power spectrum.}
\label{spectrumtimeseries}
\end{figure}

Fig.\ref{spectrumtimeseries}a shows the power distribution for our reference study based on the average observed spectrum (black), the spectrum without the local power enhancement (red) and the redistributed power spectrum (blue).

To construct the 2D velocity driver, we derive two random time series from the given power spectra,
one for the x-component of the velocity $v_x\left(t\right)$ and 
one for the y-component $v_y\left(t\right)$.
Fig.\ref{spectrumtimeseries}b
shows the x-component $v_x\left(t\right)$ for the 3 cases in Tab.\ref{spectrumpar}
and Fig.\ref{spectrumtimeseries}c shows the corresponding radial displacement of the centre of the cylinder obtained from the time integration of 
$v_x\left(t\right)$ and $v_y\left(t\right)$.
We notice that the observed spectrum leads to significantly higher displacements than other spectra and we have tested this results through a number of randomised time series.
For example, the black dashed line represents a different displacement profile derived from the same observed power spectrum and its displacement is still larger than the ones from the two modified spectra. In order to be able to draw conclusions on the relation between the driver spectrum and the MHD evolution, we run several simulations with the same spectra but where we vary the derived time series.

In \citep{Pagano2019}, we modelled a velocity time series based the same observed power spectrum using a large number of pulses of different period and amplitude.  In addition, \citet{Pagano2019}, assumed a uniform distribution of the period of the pulses which, combined with the power spectrum, meant that low frequency pulses had a substantially larger amplitude than high frequency ones,  which contradict more recent observational results by \citet{2019NatAs...3..223M}. The approach presented in this current study is more general as it is not based on deconstructing the velocity driver into single pulses and hence, no assumptions are made about any potential correlation between period and amplitude.

\subsection{MHD simulation}

To study the evolution of this system, we solve the MHD equations numerically, where thermal conduction, magnetic diffusion, and joule heating are included as source terms as follows:
\begin{equation}
\label{mass}
\displaystyle{\frac{\partial\rho}{\partial t}+\vec{\nabla}\cdot(\rho\vec{v})=0},
\end{equation}
\begin{equation}
\label{momentum}
\displaystyle{\frac{\partial\rho\vec{v}}{\partial t}+\vec{\nabla}\cdot(\rho\vec{v}\vec{v})
   +\nabla p-\frac{\vec{j}\times\vec{B}}{c}=0},
\end{equation}
\begin{equation}
\label{induction}
\displaystyle{\frac{\partial\vec{B}}{\partial t}-\vec{\nabla}\times(\vec{v}\times\vec{B})=\eta\frac{c^2}{4\pi}\nabla^2\vec{B}},
\end{equation}
\begin{equation}
\label{energy}
\displaystyle{\frac{\partial e}{\partial t}+\vec{\nabla}\cdot[(e+p)\vec{v}]=-\eta j^2-\nabla\cdot\vec{F_c}}, 
\end{equation}
where $t$ is time, $\vec{v}$ velocity,
$\eta$ the magnetic resistivity, $c$ the speed of light, $j=\frac{c}{4\pi}\nabla\times\vec{B}$ the current density, and
$F_c$ the conductive flux \citep{Spitzer1962}.
The total energy density $e$ is given by
\begin{equation}
\label{enercouple}
\displaystyle{e=\frac{p}{\gamma-1}+\frac{1}{2}\rho\vec{v}^2+\frac{\vec{B}^2}{8\pi}},
\end{equation}
where $\gamma=5/3$ denotes the ratio of specific heats.
In \citet{Pagano2019}, we have described in detail why using values for the magnetic resistivity derived from plasma theory is currently unfeasible in state-of-the-art MHD simulations. Therefore, in this set of numerical experiments we use an anomalous resistivity, parametrised to meet the following two competing requirements. On the one hand, the dissipation of electric currents must be sufficiently efficient to convert a noticeable amount of energy into heating. On the other hand, the dissipation of currents must not damp the wave propagation too rapidly, otherwise they would not be observed. An anomalous magnetic resistivity is designed to meet these requirements
and thus
\begin{equation}
\eta=\eta_0
\hspace{1 cm}
\left(|\vec{j}| > j_0\right)
\label{resistivity}
\end{equation}
where we use $j_0=2.5$ G/s as the threshold current and
$\eta_0$ is $10^8 \eta_S$, where $\eta_S$ is the magnetic resistivity according to \citet{Spitzer1962} at $T=2$ MK.
Additionally, this description of the dissipation mechanism allows us to run the simulations on a coarser grid.
Generally, a rather small grid size is required to fully resolve the phase-mixing or resonant absorption of MHD waves \citep[e.g.][]{2014ApJ...787L..22A,PaganoDeMoortel2017,Karampelas2017,2019A&A...631A.105H}. Such a high resolution allows for the description of the development of small scale structures and the conversion of the wave energy into heating.  
However, in this work, we do not focus on the detailed description of the phase-mixing \citep[already addressed in][]{Pascoe2010,PaganoDeMoortel2017}, but instead are interested in the amount of wave energy that this can convert. Here, the threshold $j_0$ is used to activate the dissipation mechanism only at the locations where the phase-mixing happens and the high value of the magnetic resistivity allows for a significant and rapid conversion of wave energy into heating.
Moreover, even if all the numerical experiments presented here
are inevitably affected by numerical diffusion, with the chosen values of $j_0$ and $\eta_0$ the effects of the magnetic resistivity dominate on the effect of the numerical diffusion.

The computational grid has a uniform resolution of $\Delta x=\Delta y=0.15$ Mm and $\Delta z=0.78$ Mm.  The simulation domain extends from $z=-20$ Mm to $z=180$ Mm in the direction of the initial magnetic field and horizontally from $x=-3$ Mm to $x=3$ Mm and from $y=-3$ Mm to $y=3$ Mm, in order to contain the loop structure presented in Sect.\ref{sectionloop}.
The boundary conditions are treated with a system of ghost cells, where we have implemented periodic boundary conditions at both the $x$ and $y$ boundaries. For the upper $z$ boundary, we use outflow
boundary conditions, except when we want to focus on the dynamics
occurring in closed loops, when we use reflective boundary conditions, i.e. where the magnetic field components, density, and energy are symmetrically copied in the ghost cells
and the velocity components are symmetrically copied and changed sign). The driver is set as a boundary condition at the lower $z$ boundary. In addition, we add a damping layer between $z=-20$ $Mm$ and $z=0$ $Mm$ which only allows the propagation of transverse oscillations, by multiplying the field-aligned component of the velocity $v_z$ by a damping factor $0.75$ after each time integration iteration in this part of the domain. We have applied this improvement in order to prevent the propagation of slow waves as this can significantly alter the energy budget. For this reason, we will only show the numerical domain for $z\ge 0$, as we regard the solution that we find for $z<0$ as non-physical. Finally, we scale the amplitude of the driver to obtain transverse velocities of the order of $v_x\sim v_y \sim 15$ $km/s$
beyond the damping layer (i.e.~above $z \ge 0$), which is the order of magnitude of  transverse displacements observed in the solar corona \citep{Threlfall2013}.
We run all our numerical experiments for $2224$~$s$ of
physical time, which is sufficient for the driving to finish plus an additional $80$ seconds which allows the last oscillations driven at the lower boundary to travel into the domain.

\section{MHD evolution}
\label{MHDsimulation}

We start by analysing the simulation with a uniform loop structure, the driver is applied within the radius $a$
at the lower boundary and where we set open boundary conditions at the upper $z$-boundary.
This is not necessarily the setup that best describes the conditions we find in the solar corona, but it is useful reference simulation to compare with other simulations to measure the effect of the density structure and the driver on the wave heating and the evolution of the boundary shell.

\begin{figure*}
\centering
\includegraphics[scale=0.32]{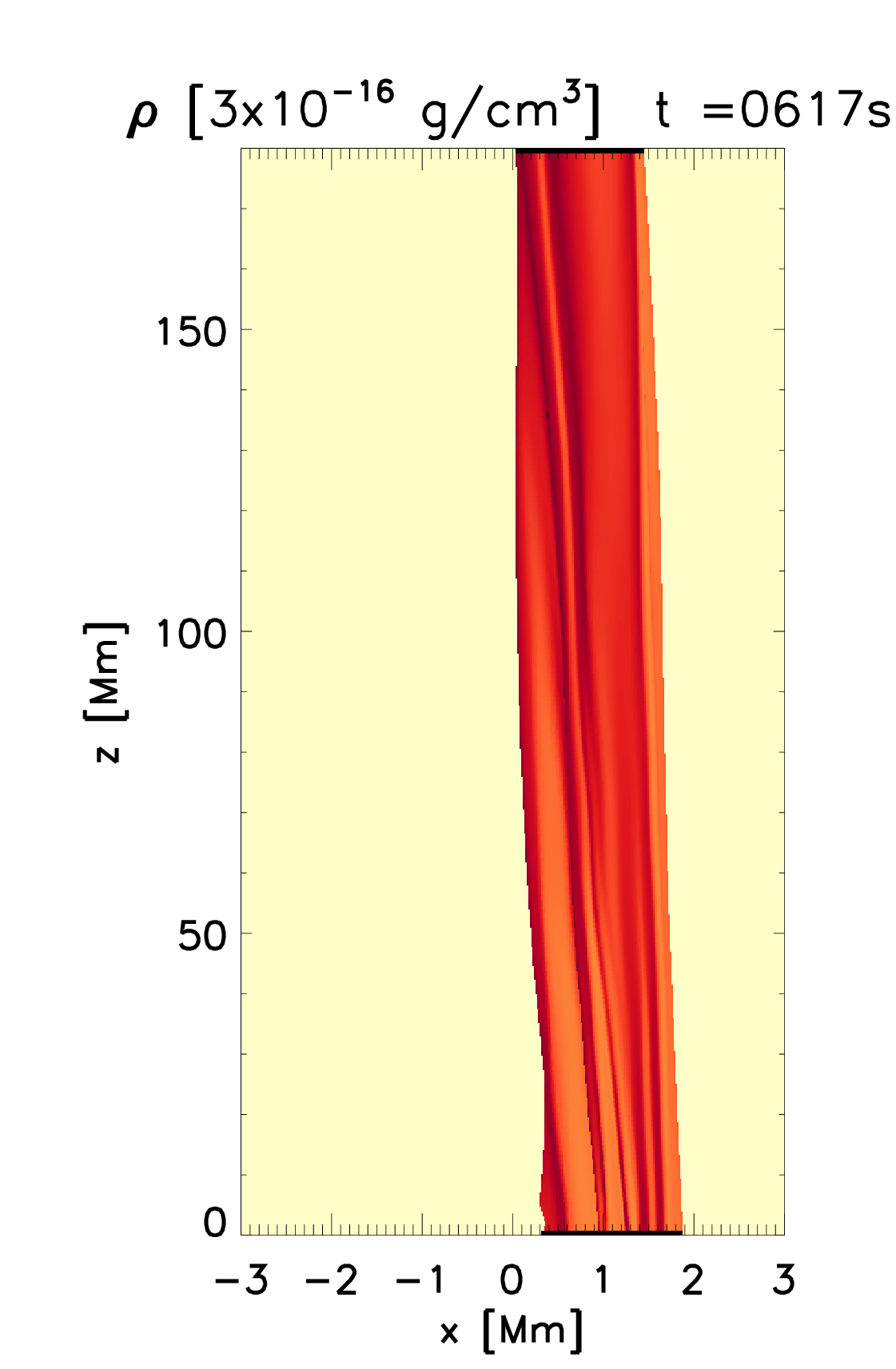}
\includegraphics[scale=0.32]{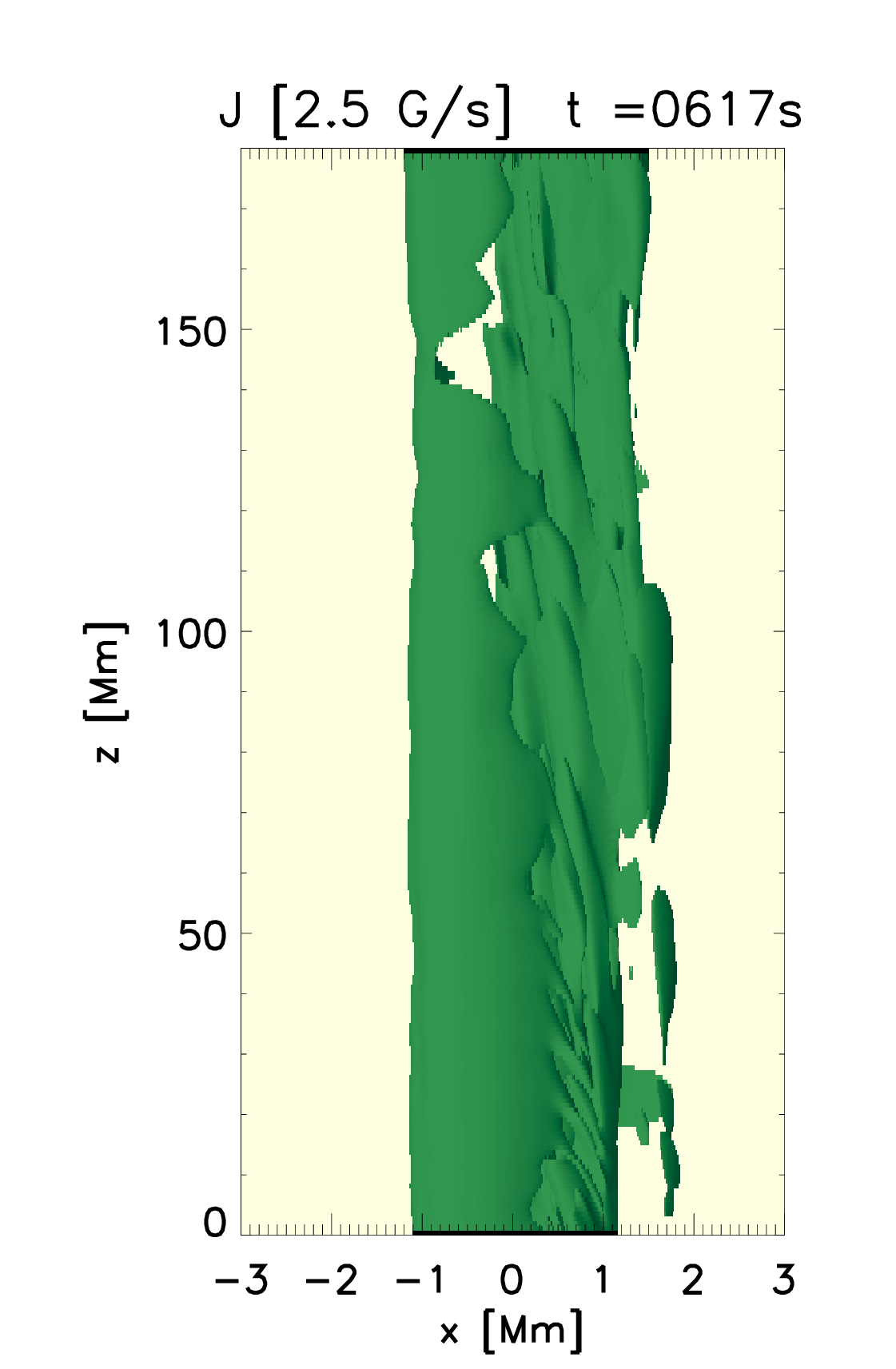}

\includegraphics[scale=0.32]{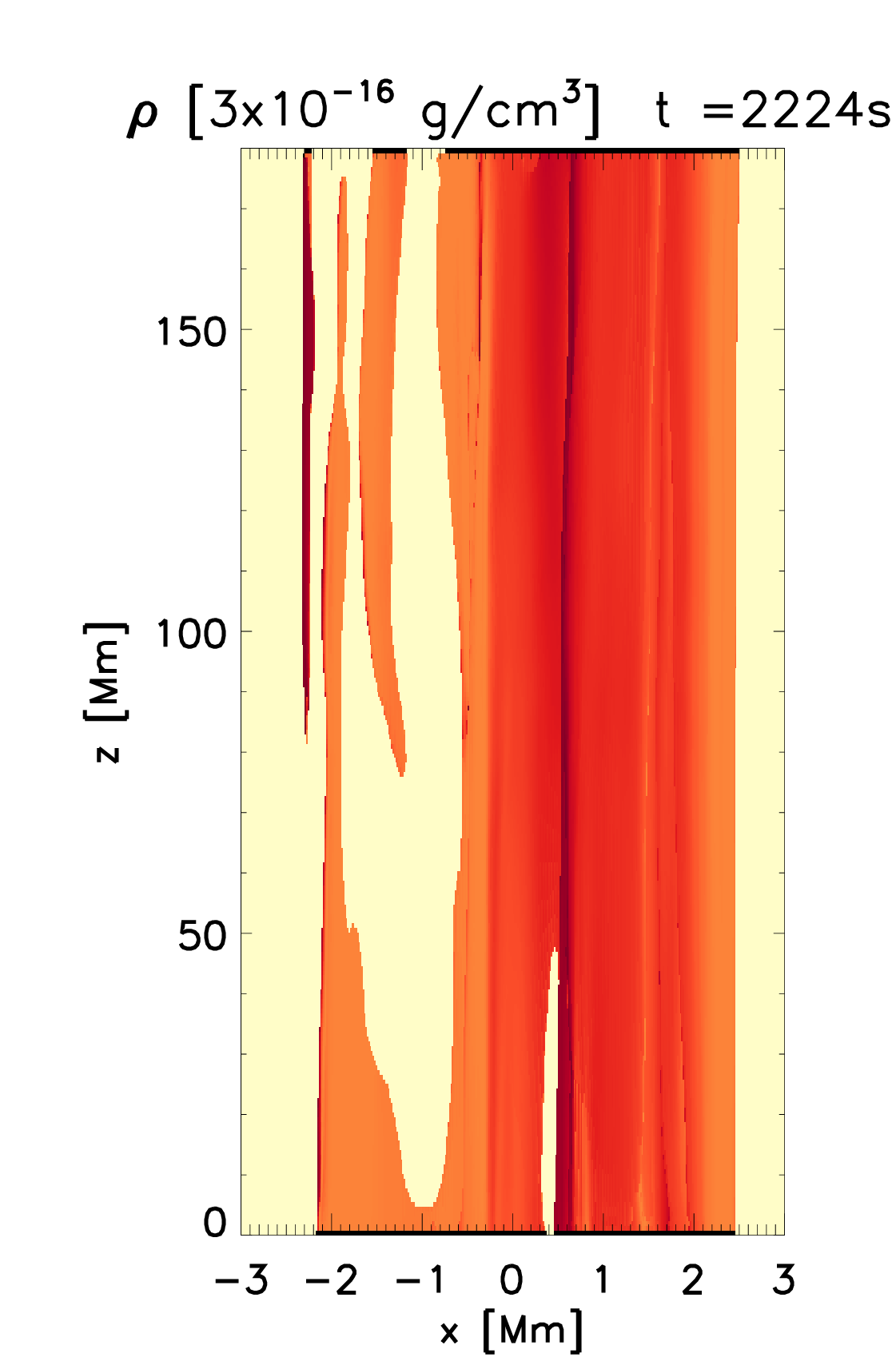}
\includegraphics[scale=0.32]{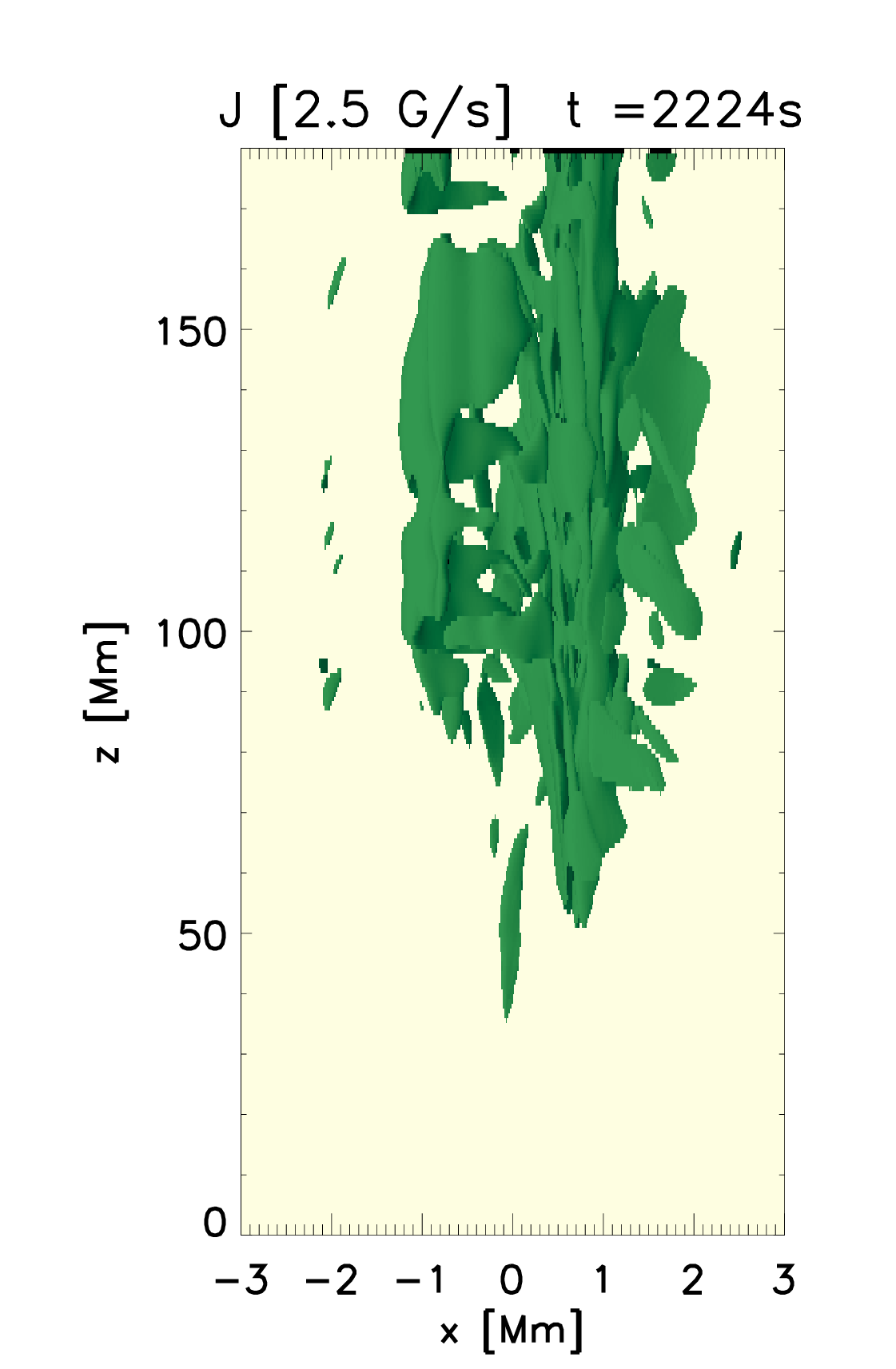}

\caption{Density and electric current density isosurfaces at two different times ($t=617\,s$ and $t=2224\,s$) for the simulation with a uniform loop density structure.
A movie of the density evolution is available online.}
\label{opensimevol}
\end{figure*}

As the driver is applied at the lower boundary, the cylinder is displaced from its initial position and transverse waves propagate along the loop, with velocities of the order of $10$ $km/s$. In addition, some remnant longitudinal perturbations of the order of $1$ $km/s$ still propagate past the damping layer as well. The transverse waves propagate at different speeds across the boundary shell and the standard phase-mixing features develop. At the same time, the displacement of the magnetised cylinder leads to the interaction between the denser interior and boundary shell regions with the background generating compression and rarefaction just outside the boundary shell.
Fig.\ref{opensimevol} shows the evolution of density and electric currents at an earlier stage ($t=617 \,s$) and at the end of the simulation ($t=2224 \,s$). We find that the density structure evolves significantly throughout the simulation 
and the initial circular cylindrical structure is replaced by a more fragmented one that shows signatures of eddies at the boundary shell. Hence, our loop structure is strongly disrupted as a result of the transverse waves and filamentary structures are created. Standing transverse are known that be unstable to the KHI \citep{Terradas2008}, and propagating transverse waves show uniturbulence \citep{2017NatSR...714820M}. Since we are considering propagating waves, the occurrence of uniturbulence is more likely here, but the connection or difference between KHI and uniturbulence is currently not well understood.

Electric currents are generated across the boundary shell and dissipated by the effect of magnetic resistivity. Fig.\ref{opensimevol} shows the contour of the modulus of the electric current at the level of the threshold current at which the anomalous resistivity is triggered.
Electric currents are generated by both the phase-mixing of Alfv\'en waves and the shearing between different regions. 
While near the beginning of the simulation the electric currents are more uniformly distributed around the magnetised cylinder, their distribution becomes significantly more fragmented later on.
\citet{Pagano2019} provide a discussion on how different frequencies of the boundary driven oscillations trigger different kinds of electric currents.
At the end of the simulation, the magnetised structure has been altered substantially and the density structure is very fragmented, having lost any resemblance of a cylinder. For the purpose of this study, we will focus on the evolution of the boundary shell and the distribution of the plasma heating in time and space.

\subsection{Boundary shell}

As the density structure is altered by the propagating transverse waves, the boundary shell evolves in time.
In order to follow this evolution, we need an operative definition
of this region. For the purposes of this study, the key property of the boundary shell is the gradient of the Alfv\'en speed. Here, we define the boundary shell as the region where $\nabla V_A>0.5\,s^{-1}$, a threshold chosen to match the boundary shell at $t=0\,s$.

\begin{figure}
\centering

\includegraphics[scale=0.41]{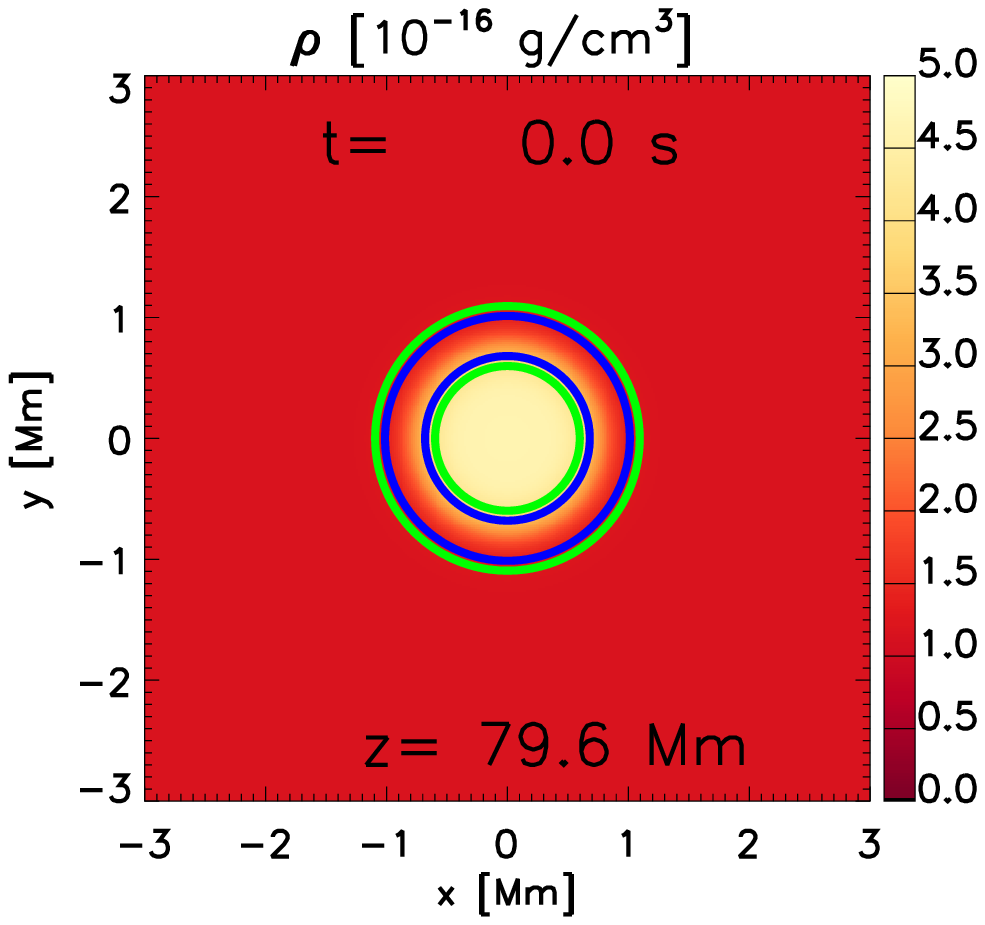}
\includegraphics[scale=0.41]{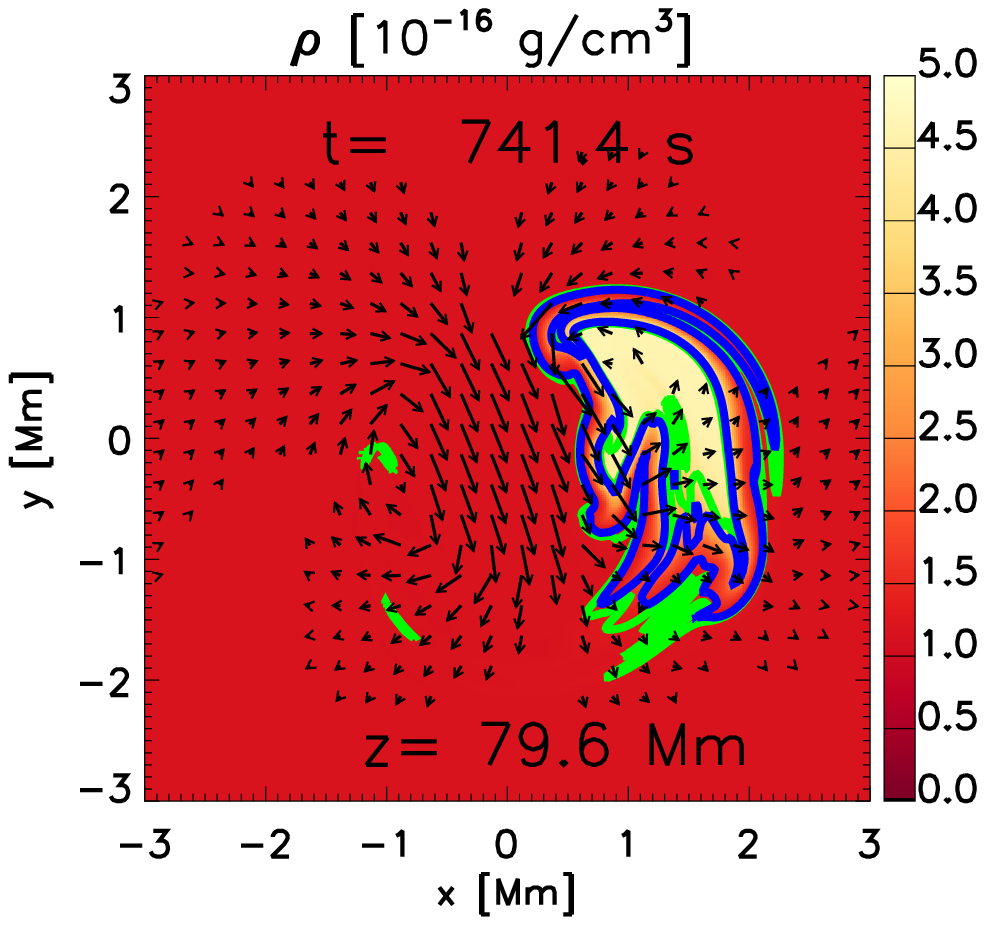}
\includegraphics[scale=0.41]{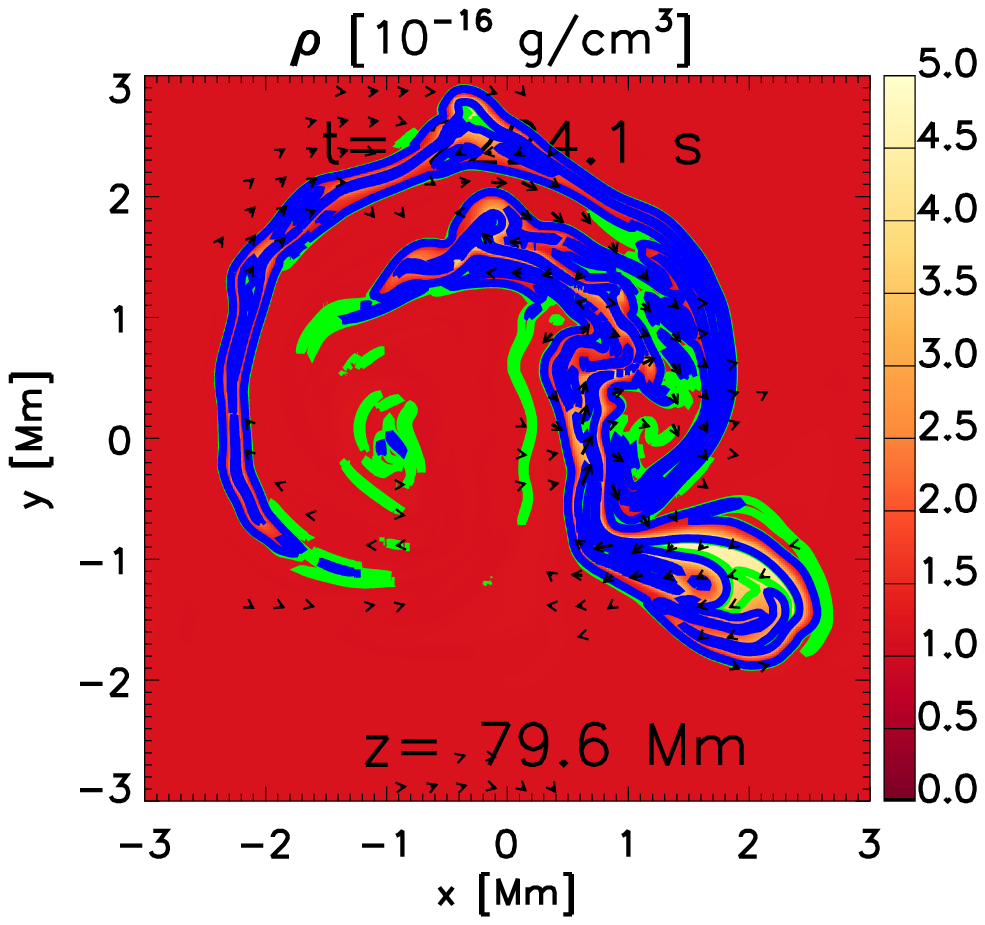}

\caption{Density cross sections at $z=79.6$ $Mm$ at three different times (t=0, $t=741.4$ $s$, and $t=2241$ $s$). Overplotted are velocity vectors where  $|\vec{v}|>5$ $km/s$ and contours of the gradient of the Alfv\'en speed $\nabla V_A=0.5\,s^{-1}$ (green contour) and $\nabla V_A=1\,s^{-1}$ (blue contour).}
\label{dalf15oe}
\end{figure}

Fig.\ref{dalf15oe} shows the density distribution at the middle of the loop, $z=79.6$, at $t=0\,s$ and at two times during the evolution. Contours for $\nabla V_A=0.5\,s^{-1}$ (green contour)
and $\nabla V_A=1\,s^{-1}$ (blue contour) are overplotted, as well as velocity vectors where $|\vec{v}|>5\,km/s$.
At $t=0\,s$, the boundary shell is a ring around the loop interior (region between the green contours), but it has significantly transformed at $t=741.4$ $s$, when it is both displaced and distorted, although it still surrounds a denser, i.e. interior, region. Later, the boundary shell becomes a filamentary structure ($t=2241$ $s$), indicating the presence of uniturbulence eddies around the loop structure. At this stage, it is no longer possible to identify a proper loop interior. Similar dynamics are present along the entire loop structure.

In order to follow the distortion of the boundary shell, we continue to measure where $\nabla V_A>0.5\,s^{-1}$ and consequently this is an evolving structure that can also appear or disappear where the gradients of the Alfv\'en speed vary significantly \citep[thus a dynamic boundary shell similar to][]{Antolin2015,Howson2017,Karampelas2017,2017NatSR...714820M}.
It is not immediately clear whether such evolution increases the efficiency of the boundary shell in terms of providing favourable conditions for the dissipation of transverse waves.
To gain more insight, we will investigate three characteristics of the boundary shell, i) its spatial extent, ii) its efficiency, i.e. the Alfv\'en speed jump across the boundary shell and iii) its consistency, i.e. the coherence of the boundary shell in the field-aligned direction.

\begin{figure}
\centering
\includegraphics[scale=0.28]{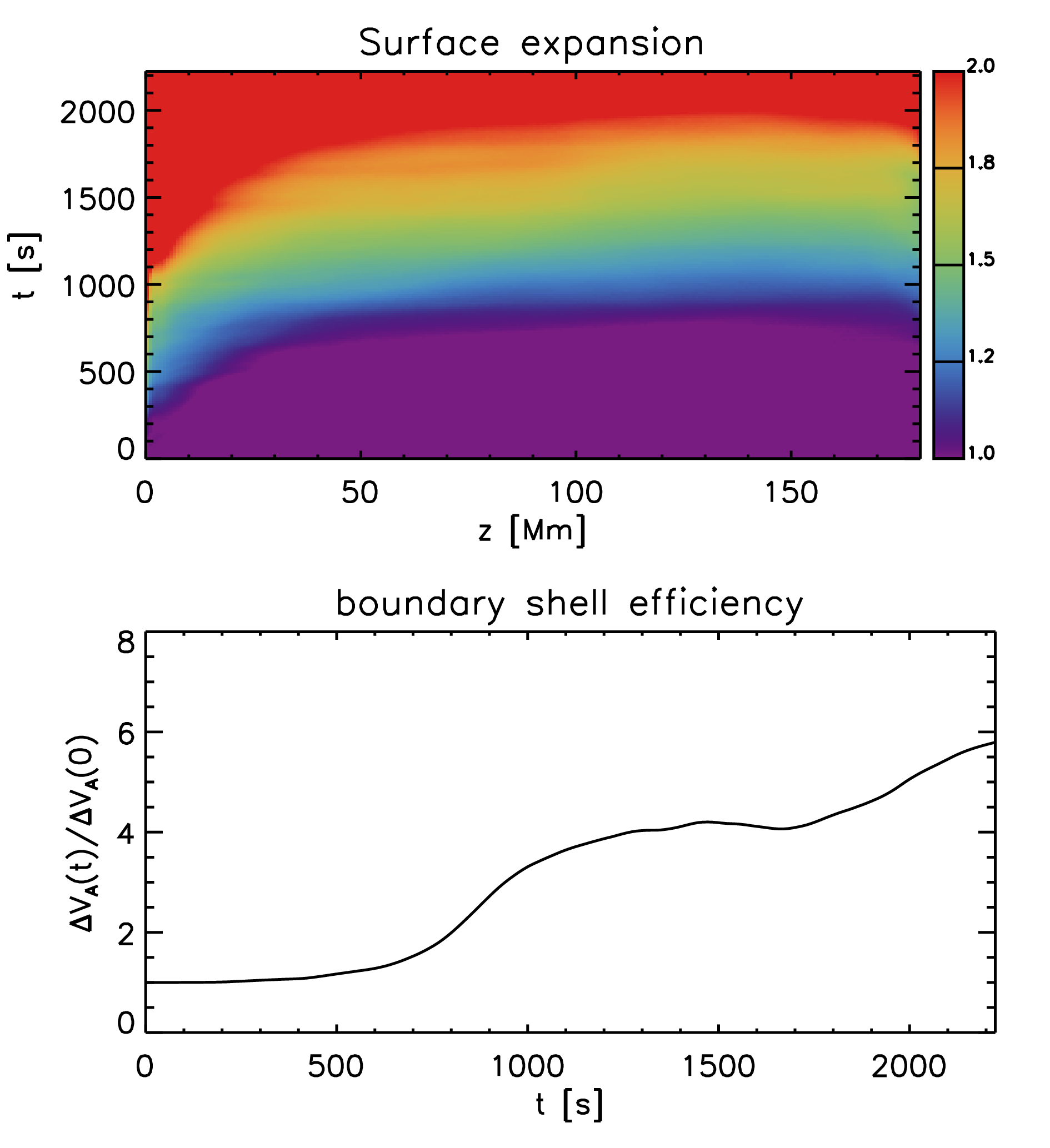}
\caption{Map of the expansion of the boundary shell as a function of $z$ and $t$
for the simulation with a uniform loop (upper panel).
Evolution of the boundary shell efficiency in the same simulation (lower panel).}
\label{shellexp}
\end{figure}

First we focus on the area that is covered by the boundary shell (i.e.~its spatial extent) at different z-coordinates along the loop. The upper panel of Fig.\ref{shellexp} shows the area expansion of the boundary shell as a function of the z-coordinate and time, normalised to the initial surface area. We find that the area covered by the boundary shell remains roughly unchanged for the first $600$ seconds throughout the cylinder, apart from near the lower (driven) boundary, where it starts expanding earlier. After $t=600\,s$, the boundary shell starts expanding everywhere along the magnetised cylinder, doubling its surface near the end of the simulation. 
This is the time when the uniturbulence is fully developed (see Fig.\ref{opensimevol}).

The volume of the boundary shell follows a similar evolution, remaining roughly constant for the first $600$ $s$ and then linearly expanding as soon as the uniturbulence sets in, reaching double its initial volume by the end of the simulation.

To examine the boundary shell efficiency, we consider the total Alfv\'en speed variation within the boundary shell, i.e. $\int \nabla V_A\left(t\right) dV$, normalised to the value of the same integral at $t=0\,s$ (bottom panel Fig.~\ref{shellexp}). The larger the Alfve\'n speed variation, the faster the phase-mixing of Alfv\'en waves can occur \citep[see e.g.][]{HeyvaertsPriest1983}. We find that this value becomes up to 6 times larger by the end of the simulation, after a more efficient boundary shell starts to develop from around $t=600$ $s$.
Hence, after the first transitory phase of the simulation, the boundary shell expands and becomes more efficient, in principle creating more favourable conditions for wave heating to occur.

\begin{figure*}
\centering
\includegraphics[scale=0.40]{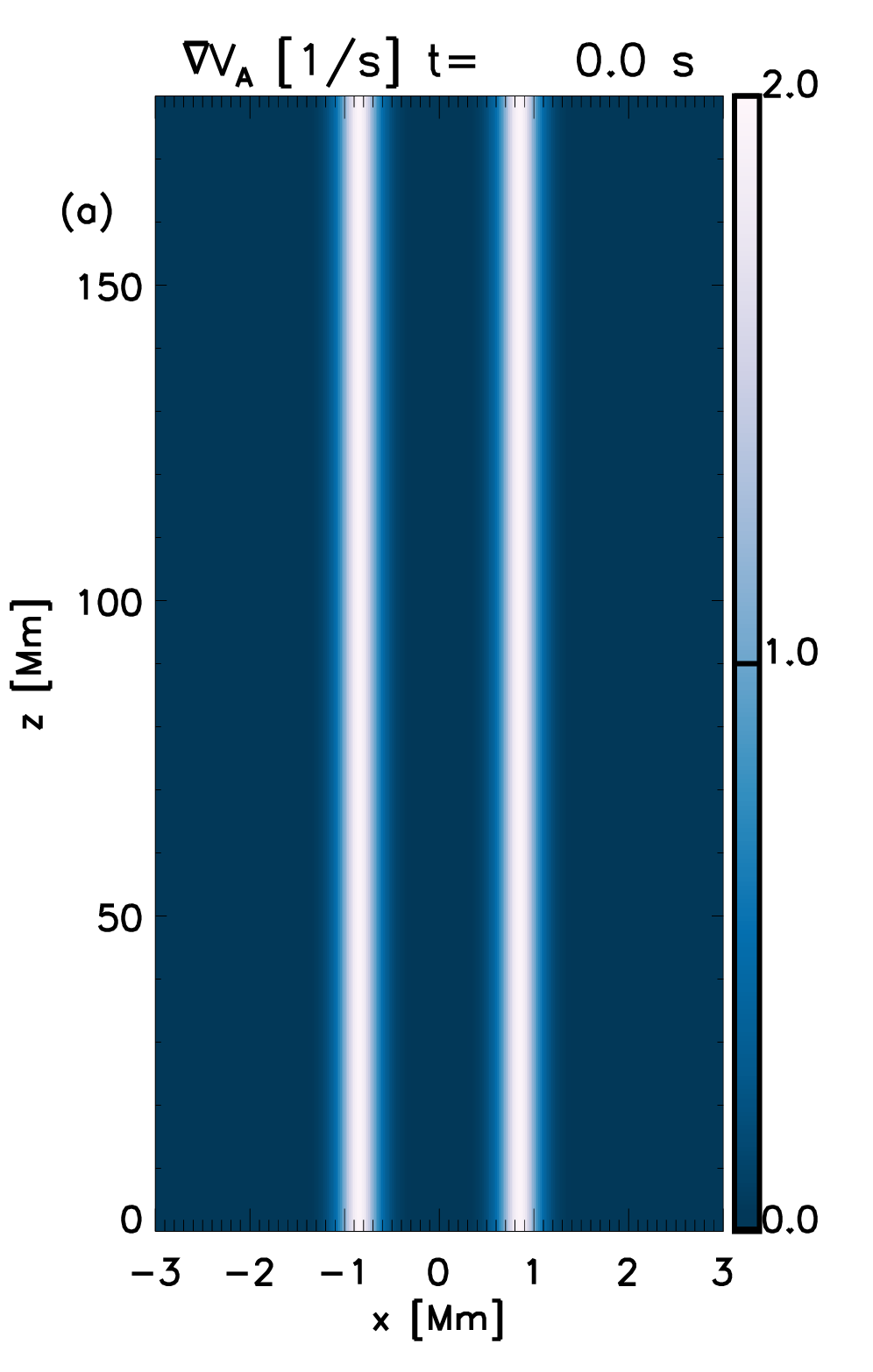}
\includegraphics[scale=0.40]{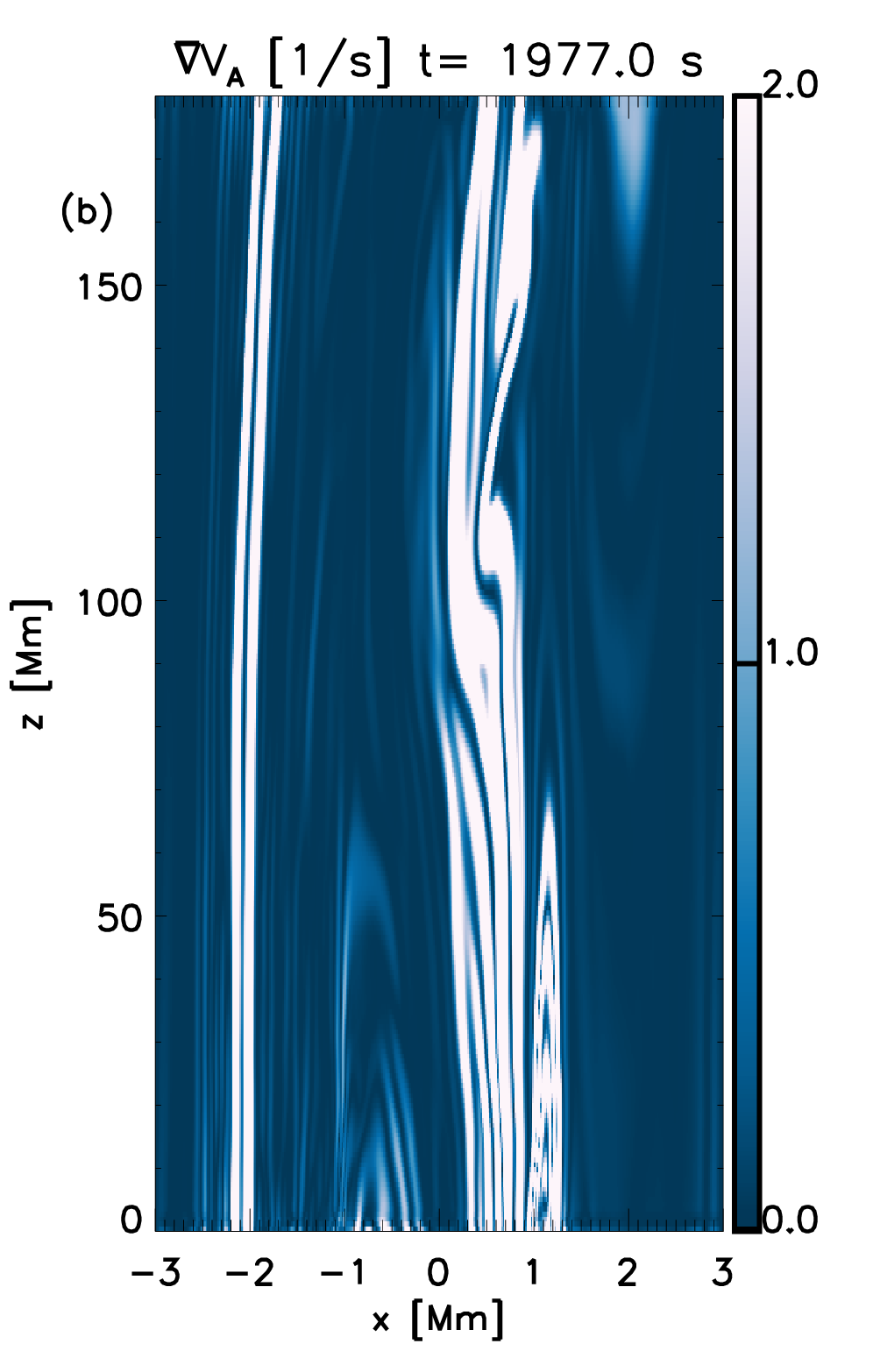}
\includegraphics[scale=0.30]{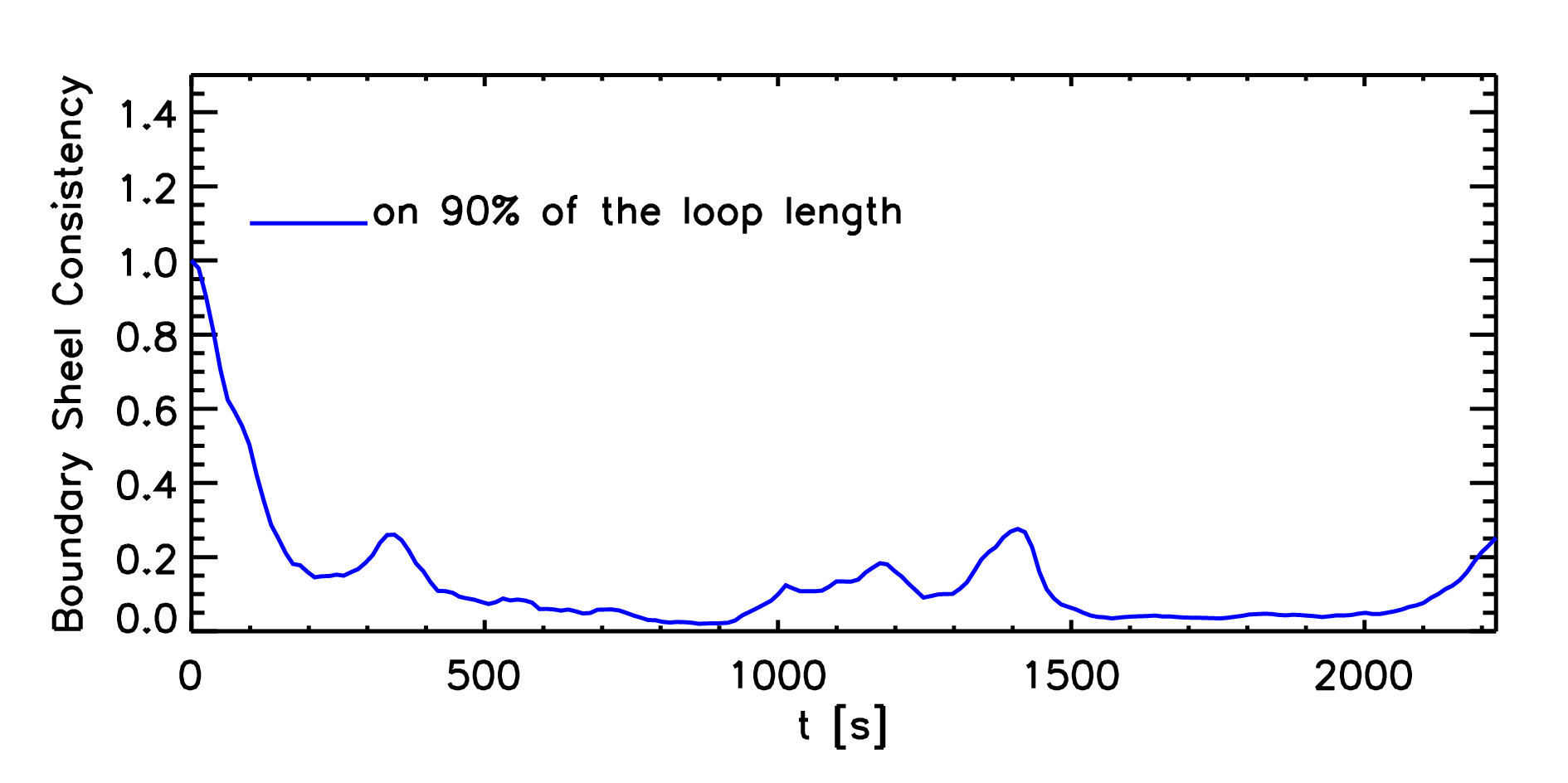}
\caption{Cross section of the gradient of the Alfv\'en speed at $y=0$
at the initial condition $t=0\,s$ (a) and at $t=1977\,s$ (b).
Evolution of the boundary shell consistency for the simulation with uniform loop.}
\label{shellcons}
\end{figure*}

Finally, we investigate the consistency of the boundary shell
$\left(\int \nabla V_A dz\right)$. In order for phase-mixing of the waves to develop, neighbouring propagating waves need to travel over some distance whilst a gradient in the Alfv\'en speed remains present. 
Fig.\ref{shellcons} (top left) show a cut of the gradient of the Alfv\'en speed along a vertical plan at $y=0$ at $t=0$. The gradient of the Alfv\'en speed is uniform in the z-direction and waves travelling from the lower boundary will phase-mix over the full longitudinal extension of the loop.
However, this configuration significantly changes over the course of the simulation, as shown in Fig.\ref{shellcons} (top right), at $t=1755\,s$.
As the propagation of the waves alters the loop structure, the gradient of the Alfv\'en speed changes significantly in the z-direction
and transverse waves have a smaller distance over which to consistently develop phase-mixing. This dynamic is specific to this footpoint driver, which continuously changes, hence breaking any initial invariance along the z-direction. In such a setup, the boundary shell constantly evolves in different ways, at different heights along the loop, leading to a reduced consistency. We now define the projected area onto the $xy$-plane of the region where the boundary shell covers at least 90\% of its initial z-extension as a measure of the boundary shell consistency. The lower panel of Fig.\ref{shellcons} shows that the boundary shell consistency drops to as low as 10\% of its initial value during the simulation, increasing again to about 20\% near the end, in the time interval when the simulation runs after the driver is switched off.

In conclusion, the propagation of the transverse waves along the loop alters the initial boundary shell significantly throughout the simulation.
The boundary shell expands and the gradient of the Alfv\'en speed increases over time, making the boundary shell larger and more efficient, in principle increasing the efficiency of phase-mixing. However, at the same time, the boundary shell becomes less consistent, reducing the opportunity for waves to phase-mix. In the next section, we focus on the plasma heating in the boundary shell and compare it with the expected radiative losses.

\subsection{Heating}

During the simulation, most of the electric currents form in the boundary shell.
As we have stated in Eq.\ref{energy} and Eq.\ref{resistivity}, 
when electric currents grow above a designated threshold, they are dissipated and Ohmic heating occurs, where magnetic energy is transformed into internal energy of the plasma.

\begin{figure*}
\centering
\includegraphics[scale=0.32]{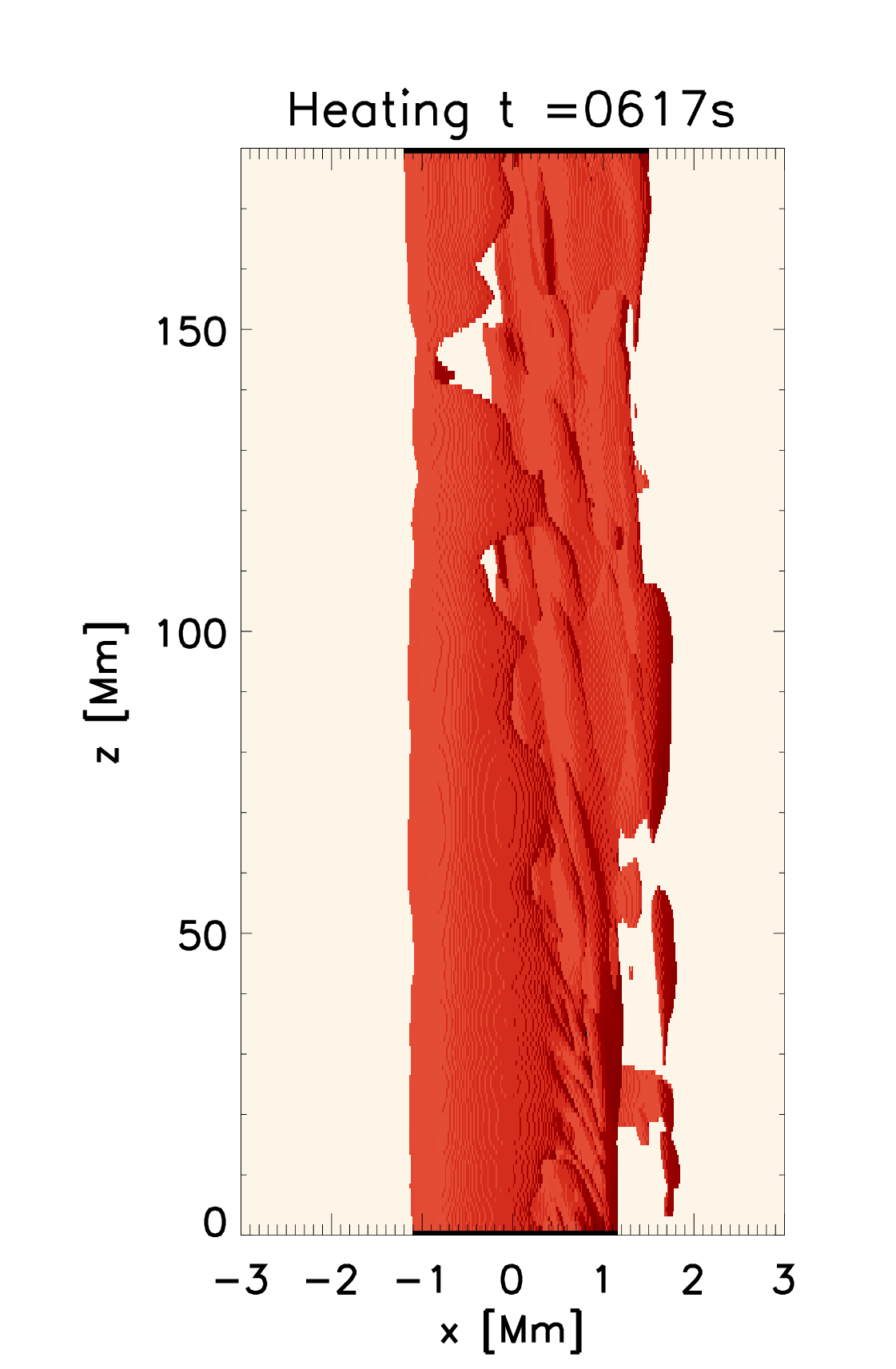}
\includegraphics[scale=0.32]{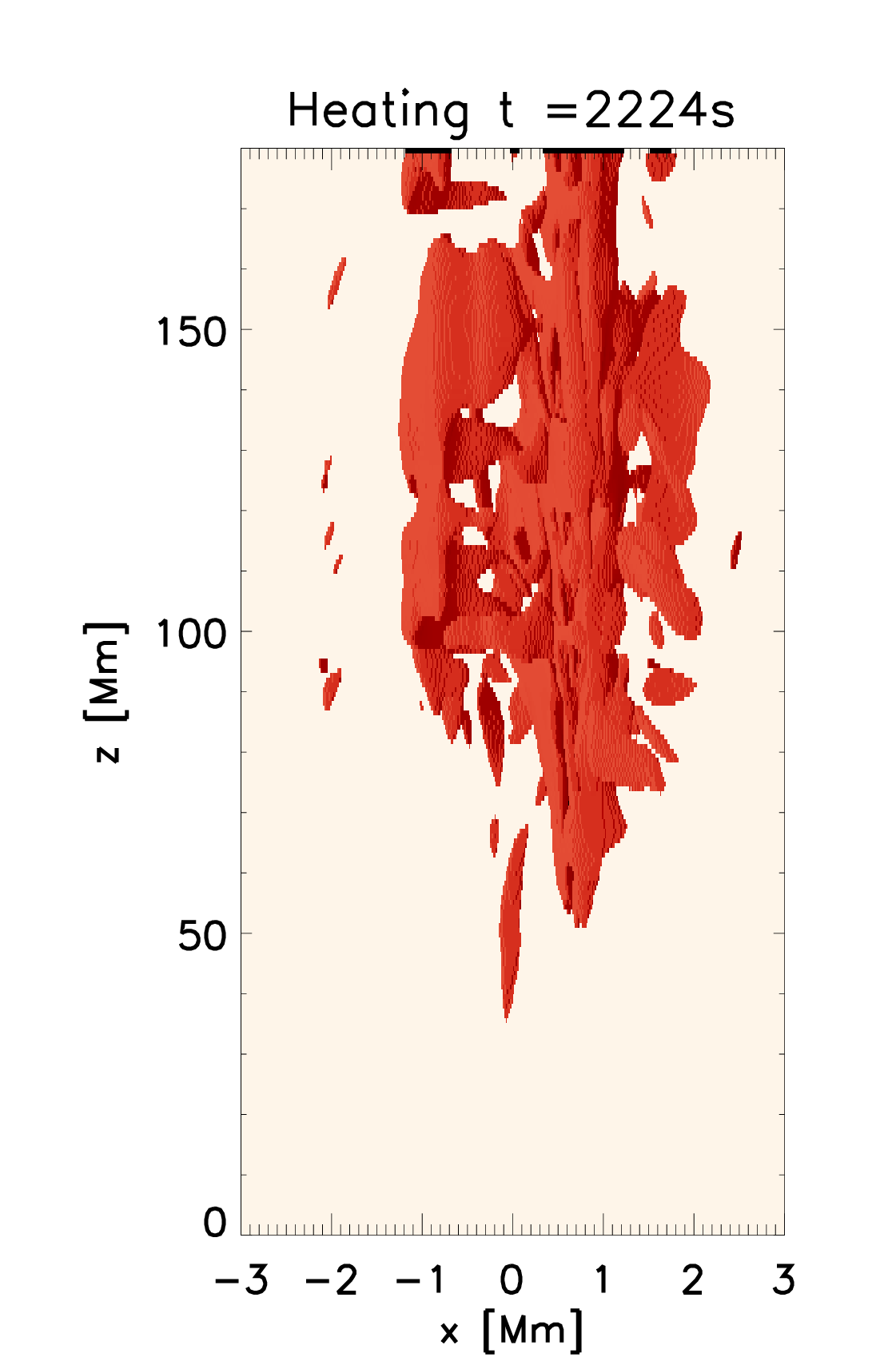}
\caption{Isosurface of the ohmic heating at $H=10^8$ $erg/cm^2/s$
at two different times ($t=617\,s$ and $t=2224\,s$) in the simulation with a uniform loop. A movie of the heating evolution is available online.}
\label{heat15oe}
\end{figure*}

Fig.\ref{heat15oe} shows the distribution of this Ohmic heating at two representative times in the simulation, namely at $t=617\,s$, just after the uniturbulence has developed, and at the end of the simulation ($t=2224\,s$), when the lower boundary driver has already stopped.
We find that at earlier times, the heating is mostly uniformly distributed
around the boundary shell, although the random footpoint motions have already created some fragmentation.
In contrast, at later stages, the heating is distributed 
in a more heterogeneous way and the heating region is now very fragmented.

\begin{figure}
\centering
\includegraphics[scale=0.32]{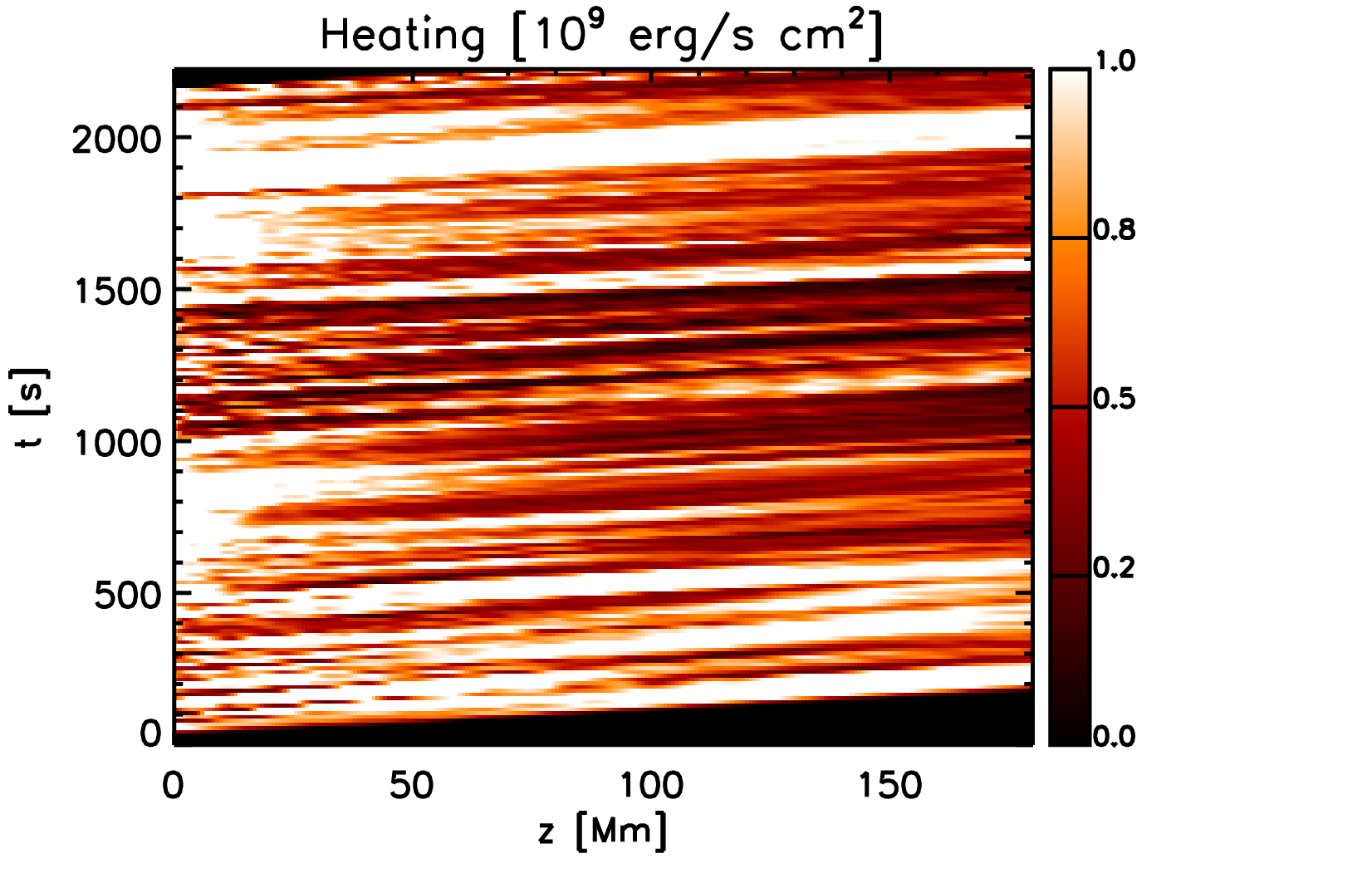}
\caption{Map of the average heating in the boundary shell along the loop as a function of time for the simulation with a uniform loop. Although the heating is generally within the range shown here, the maximum heating is $6\times10^9$~$erg/s cm^2$}
\label{heatingmap}
\end{figure}

In this experiment, the magnetic resistivity is sufficient to allow for the dissipation of the waves, but it is not strong enough to damp the wave propagation fully. Fig.\ref{heatingmap} shows the heating averaged over the boundary shell cross section
as a function of the $z$-coordinate along the loop and time.
By comparing the evolution of the wave energy and thermal energy of this simulation with a corresponding ideal simulation, we confirmed that the effect of the anomalous resistivity is significantly larger than the effect of the numerical dissipation.

We find that several distinct heating events can be identified, even if waves are continuously travelling along the loop. Each event leaves a distinct signature in the time-distance map, as the heating at different z-coordinates occurs at different times.
Similarly, it is clear that a large portion of the heating is deposited closer to the footpoint, as this is where stronger currents are generated.
While the majority of the heating events start from the footpoint and are continued along the loop structure as the waves propagate, there are some that only form at higher z-coordinate.

\begin{figure}
\centering
\includegraphics[scale=0.28]{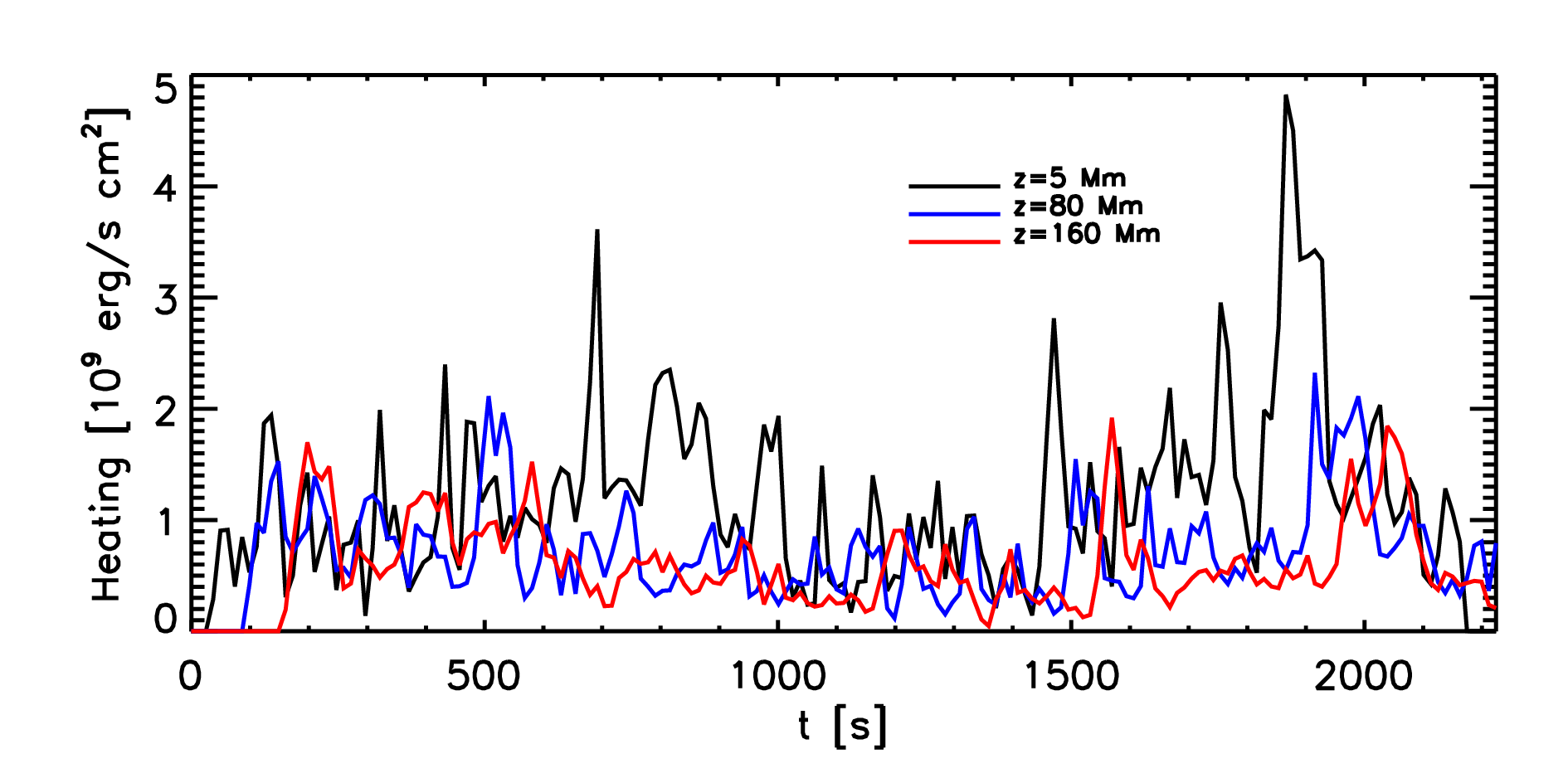}
\caption{Average heating in the boundary shell as a function of time for the simulation with a uniform loop, at three different coordinates along the loop.}
\label{heatinginterm}
\end{figure}

The discrete nature of the heating is also evident from the time evolution of the average heating in the boundary shell at different $z$-coordinates, as shown in Fig.\ref{heatinginterm}.

We find that the time evolution of the heating can be described as a sequence of distinct events, where different amounts of energy are released for each event. While Fig.\ref{heatingmap}
clearly illustrates that these single peaks are part of more elongated heating structures, it should be noted that this pattern is not evident when focusing on fixed z-coordinates. Fig.\ref{heatinginterm} illustrates again that the heating near the driven footpoint (black curve) is 
larger than the heating occurring further along the loop.

Finally, we compare this heating deposition
with an estimate of the radiative losses, in order to understand
to what extent this heating can maintain the million degrees corona. 
\begin{figure}
\centering
\includegraphics[scale=0.32]{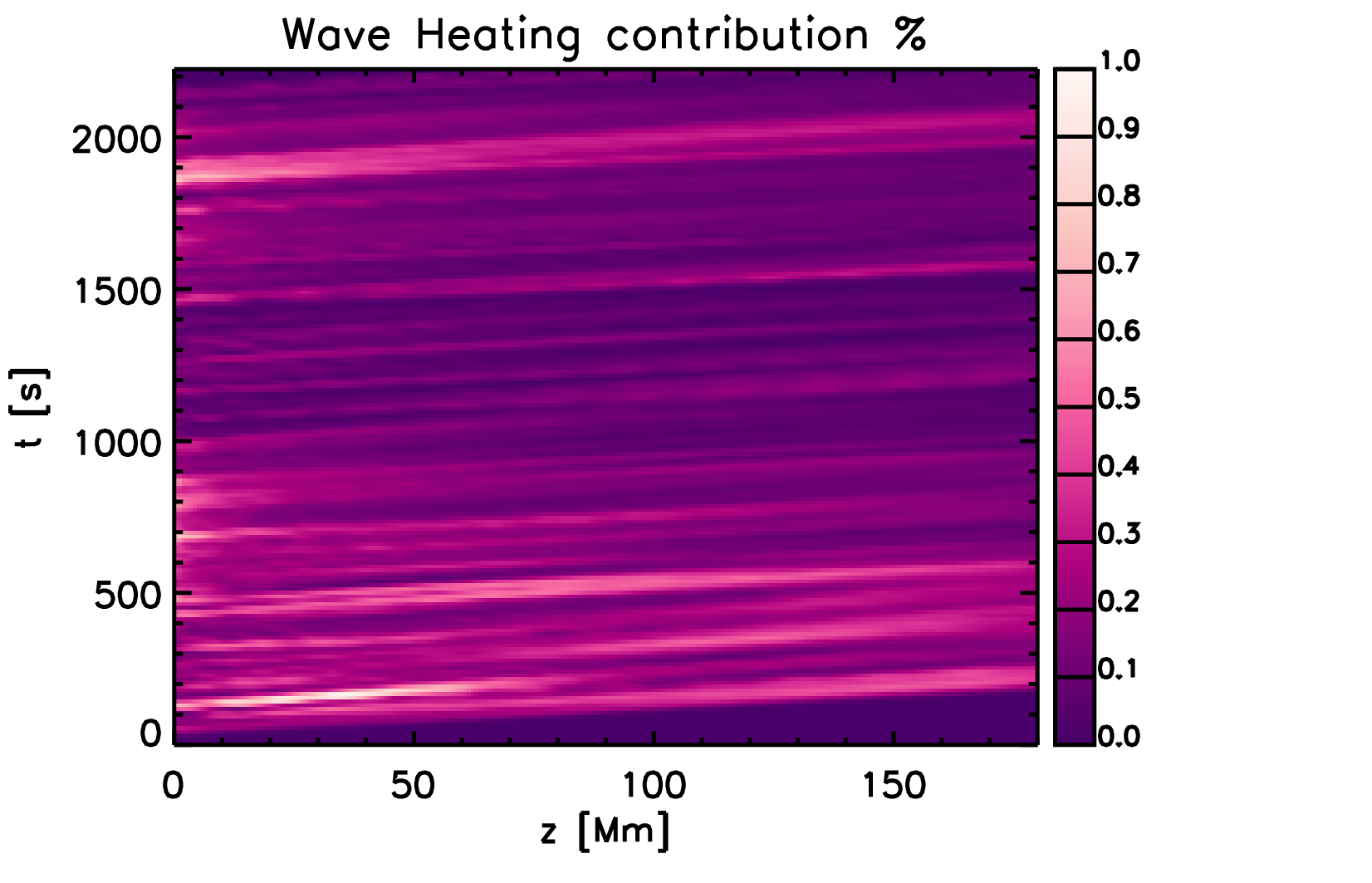}
\caption{Map of the ratio of the averaged heating and the average radiative losses in the boundary shell as a function of loop position and time, for the simulation with a uniform loop.}
\label{radloss}
\end{figure}
In Fig.~\ref{radloss}, we compare the radiative losses with the heating as a function of position along the loop and time. Both quantities are integrated over the boundary shell at each $z$-coordinate.
In particular, we compare the time integral of the average heating and average radiative losses in a cross section ($z$) of the boundary shell between two simulations snapshots around the time $t$. For the purpose of the time integral, the heating is approximated as a step function between the simulation snapshots. The time integral of the radiative losses is computed for a plasma whose temperature and density linearly evolve from the average density and temperature in the boundary shell of one simulation snapshot to the other. For this purpose the radiative losses from a plasma are computed using the
piecewise continuous function by \citet{2008ApJ...682.1351K}. It should also be noted that the temperature variations are rather small in this simulation and the variation in the plasma radiative losses are mostly due to density variations.
The radiative losses decrease in time because the boundary shell becomes less dense as it evolves. In contrast, the heating evolves in a less monotonic way but in this simulation, the heating power from the transverse waves always remains only a small percentage of the power required to balance the radiative losses (of the order of 1\% of the radiative losses, with one short-lived peaks of 10\%).

Driving the lower boundary with a different velocity time series derived from the same power spectrum but with a smaller displacement (dashed line in Fig.\ref{spectrumtimeseries}b), we find very similar results. Althgouh the maximum contribution of the wave heating goes up to $13\%$ of the expected radiative losses, overall it remains a small percentage of the radiative losses, indicating that our conclusions do not significantly depend on the chosen time series.

When we compare the rate of change of the thermal energy contained in the computational domain in this simulation with a corresponding ideal simulation ($\eta=0$), we find that, as expected, the increase of thermal energy in the non-ideal simulation initially exceeds the thermal energy increase in the ideal simulation because of the ongoing heating.
However, after some time, the situation reverses, because the thermal pressure excess in the non-ideal simulation leads to a gradient of the thermal pressure in the $z$-direction and consequently, mass flows across the upper boundary of the domain. These flows lead to a mass loss and, hence, a thermal energy loss through this boundary.
Such dynamics are compatible with the wave heating of open field lines in the corona and the triggering of plasma outflows. In our simulations, we find a mass loss of $6\times10^7$ $g$ over a time of $2200$ seconds, corresponding to an average mass loss rate of $2.5\times10^4$ g/s across the $36$\,Mm$^2$surface of the outer boundary. If we consider a characteristic density of $10^{-16}$ $g/cm^3$, we need a steady outflow of $\sim0.01$ $km/s$ to account for this mass loss.

\section{Boundary shell evolution}
\label{loopstructure}

In this section, we investigate how different initial loop structures affect the evolution of the boundary shell. In particular, we consider the non-uniform density distributions (with $z_0=130$\,Mm and $z_0=46$\,Mm) presented in Fig.\ref{initialloop}, where both the density along the loop and the extent of the interior region decrease with height.

\begin{figure*}
\centering
\includegraphics[scale=0.22]{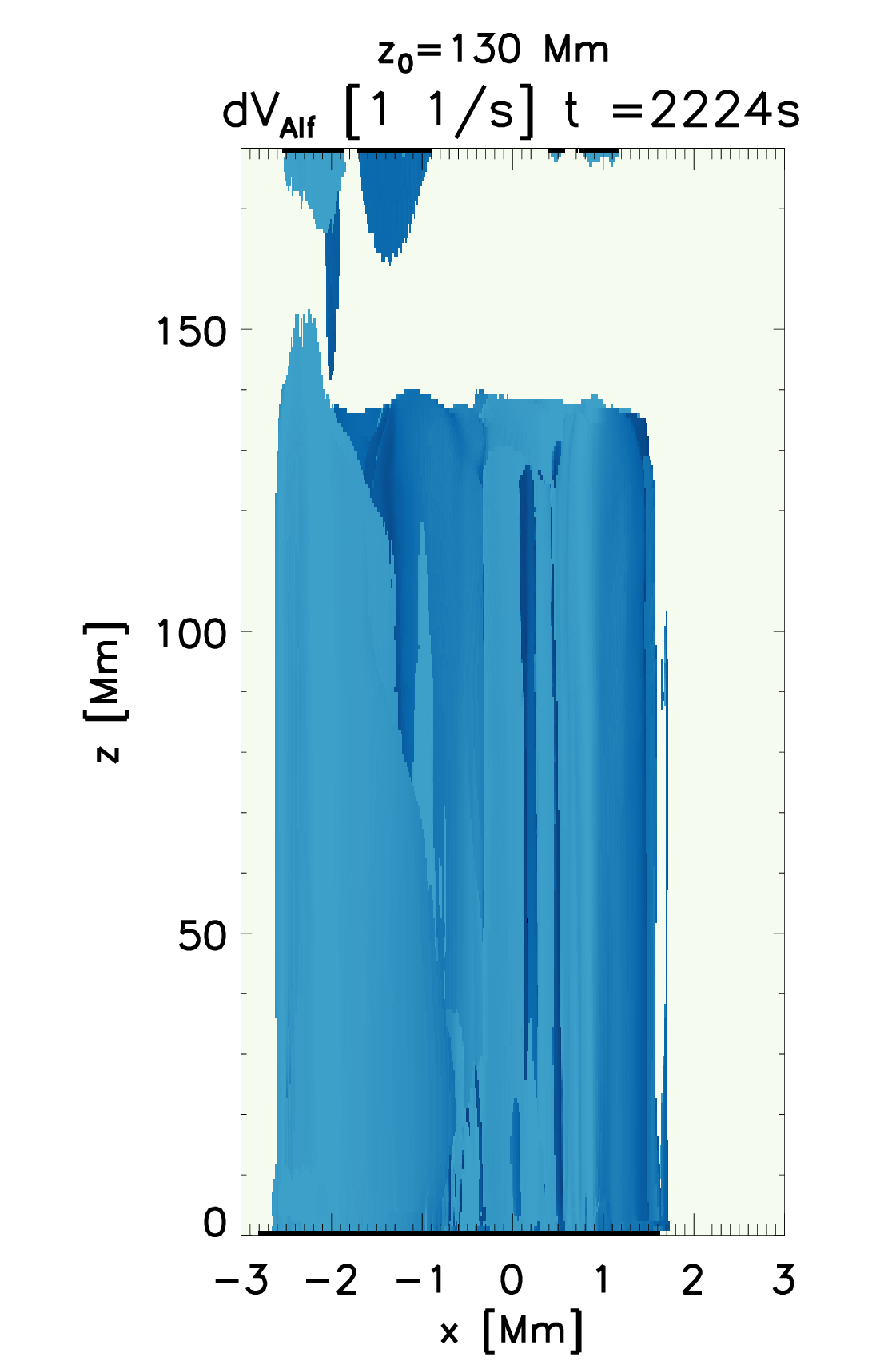}
\includegraphics[scale=0.22]{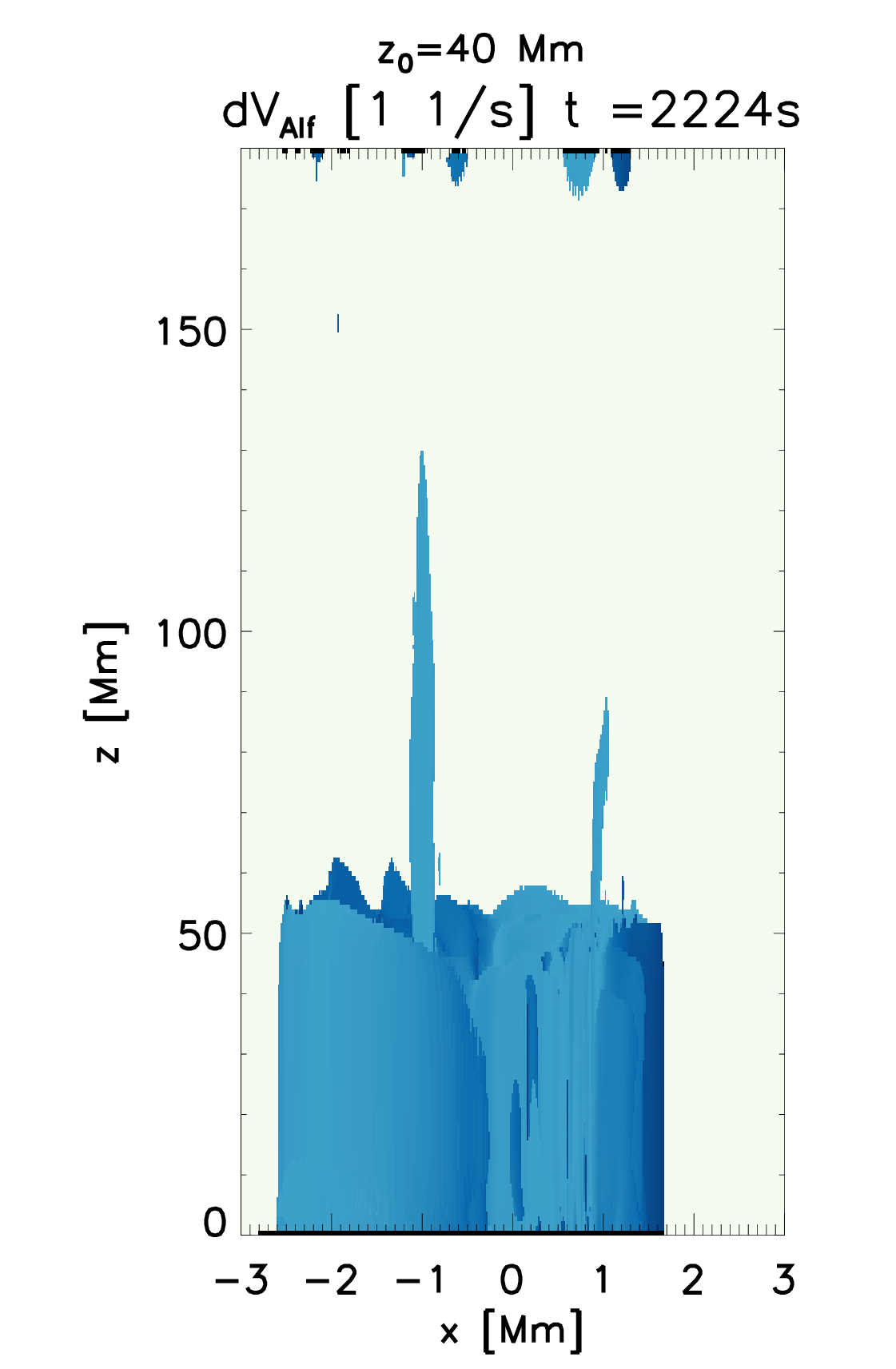}

\caption{Isosurfaces of the gradient of the Alfv\'en speed at $t=2224$ $s$
for the simulations with $z_0=130$ $Mm$ (left hand side) and $z_0=46$ $Mm$ (left hand side).}
\label{tiploops_dalf}
\end{figure*}

Fig.\ref{tiploops_dalf} show the final contour of the boundary shell for these two simulations. Compared to their initial structure in Fig.\ref{initialloop}, we find that in both cases, the boundary shell has expanded in the $x$ and $y$ directions and extended further upwards in the $z$-direction.
Although the expansion is already present at the lower boundary, 
it is most significant at higher z coordinates where the relatively shallow
boundary shell was previously not present. Additionally, the contours in Fig.\ref{tiploops_dalf} extend further in the $z$-direction compared to the initial configuration as the propagation of transverse waves along the loop structure leads to (i) steeper gradients of the Alfv\'en speed
and (ii) the generation of a boundary shell where this was initially not present.
The first mechanisms occurs quickly, as the gradient of the Alfv\'en 
speed steepens as soon as the waves propagate
($\sim120$ $s$ for both simulations) and the entire loop structure up until $z_0$ is surrounded by a cylindrical boundary shell.
In contrast, the second process is slower and becomes visible 
only after about $1000$ $s$, by which time, uniturbulence has already started affecting the boundary shell.

\begin{figure*}
\centering
\includegraphics[scale=0.20]{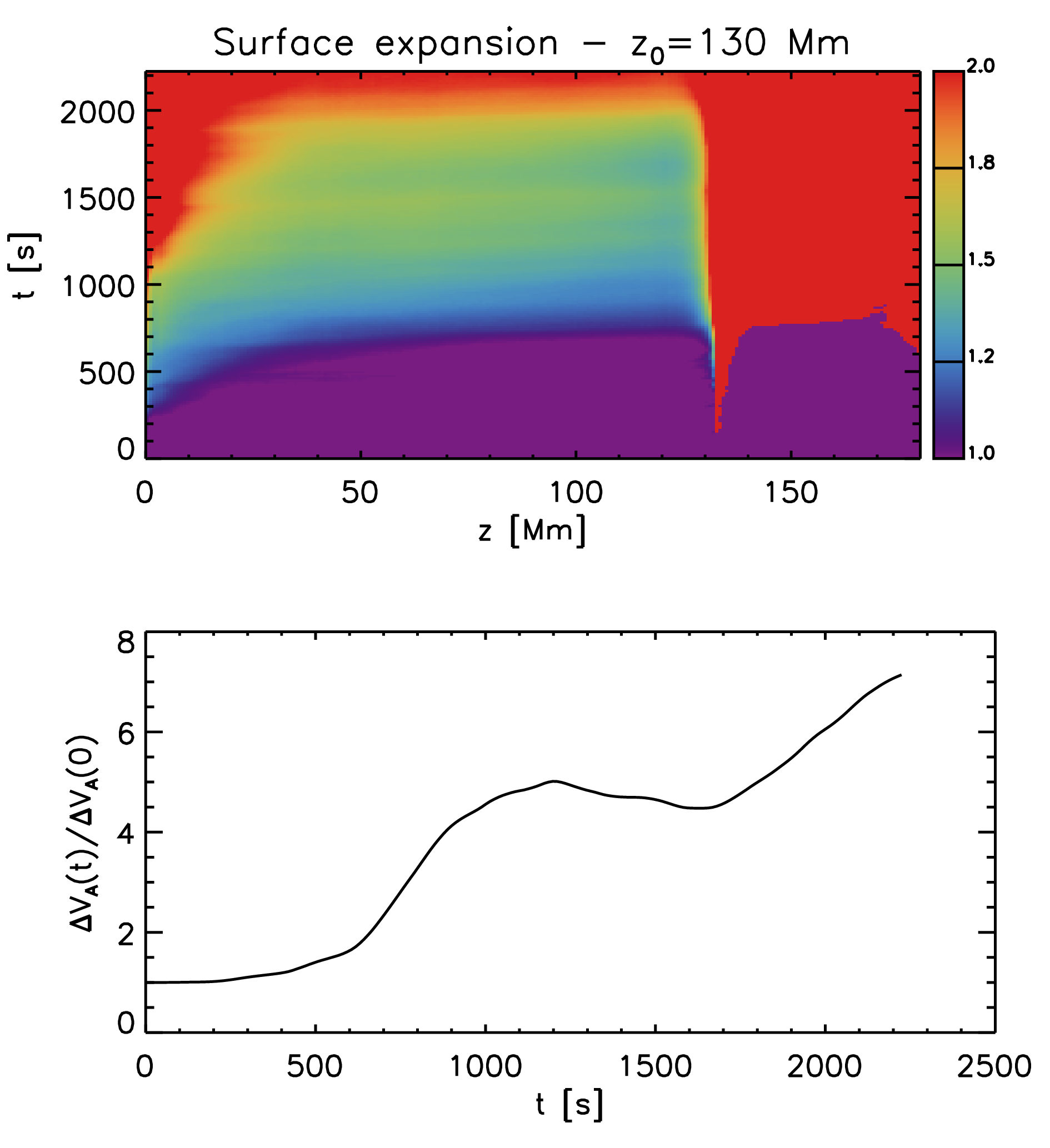}
\includegraphics[scale=0.20]{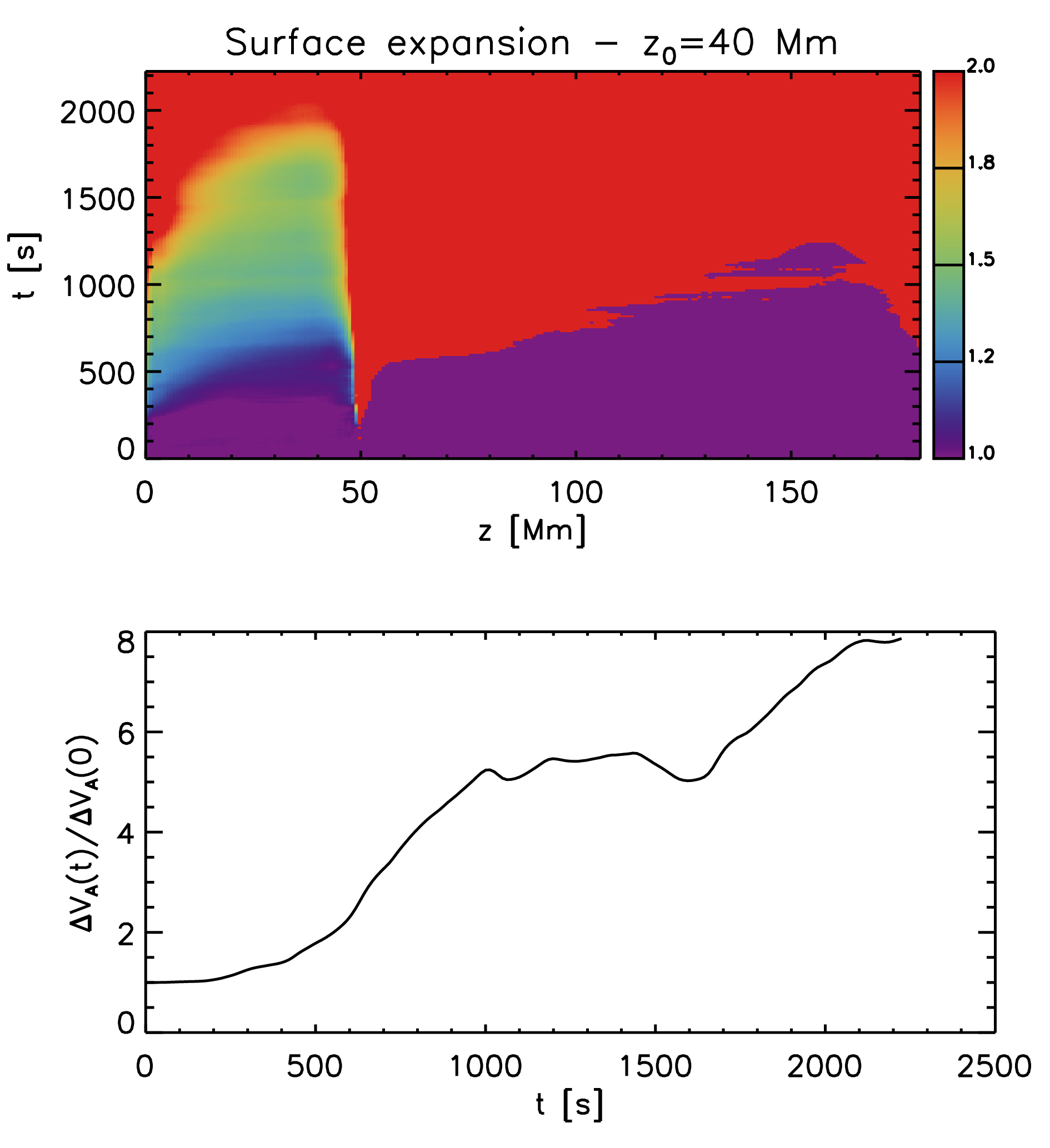}
\caption{Map of the expansion of the boundary shell as a function of $z$ and $t$ (upper row) and evolution of the boundary shell efficiency (lower row).
for the simulations with $z_0=130$ $Mm$ (left hand side) and $z_0=46$ $Mm$ (left hand side).}
\label{tiploops_expa}
\end{figure*}

\begin{figure*}
\centering
\includegraphics[scale=0.30]{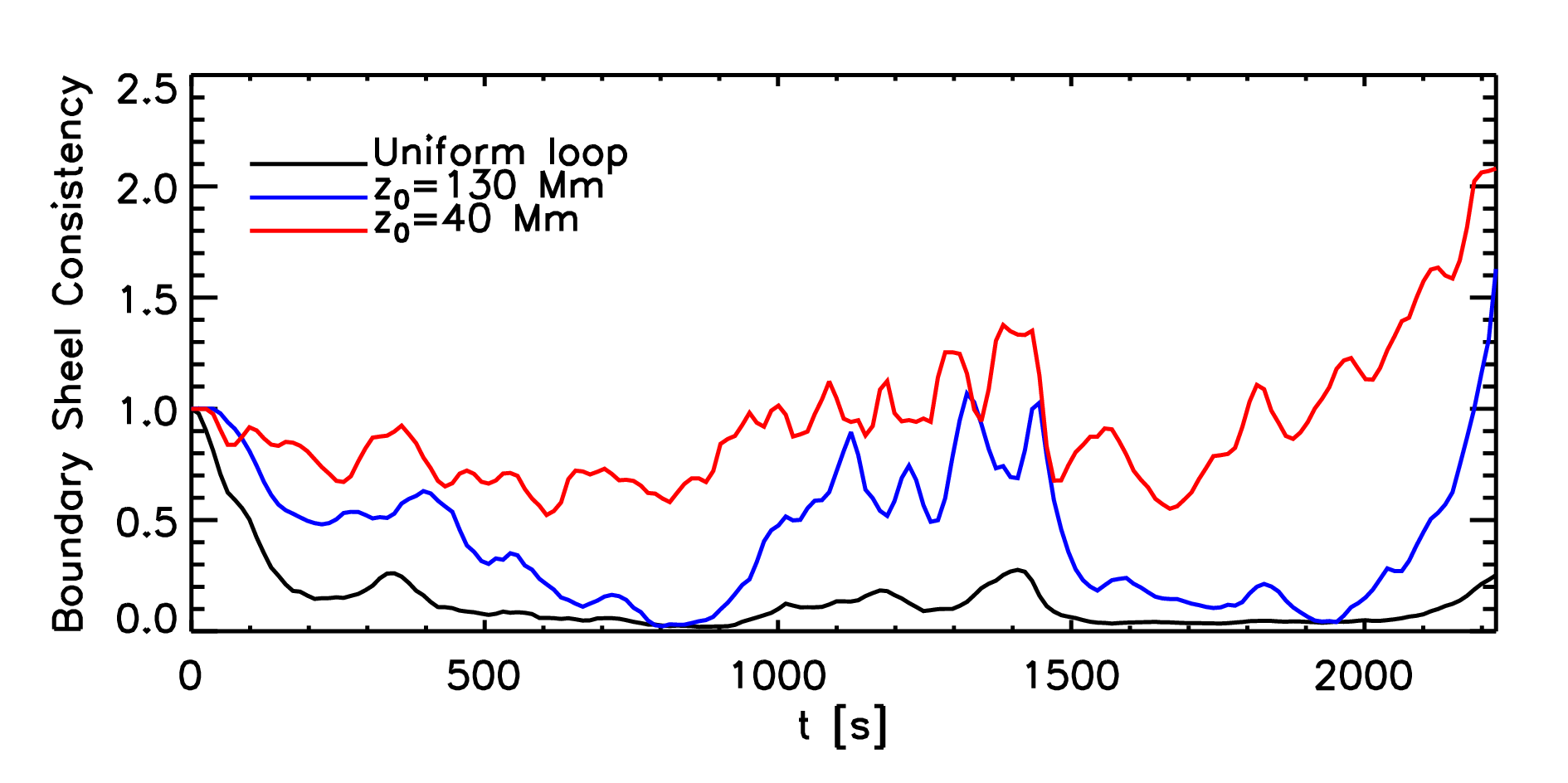}
\caption{Evolution of the boundary shell consistency for the simulation with uniform loop, with $z_0=130$ $Mm$, and $z_0=46$ $Mm$.}
\label{tiploops_cons}
\end{figure*}

The expansion of the boundary shell in these simulations (Fig.\ref{tiploops_expa}) follows a similar pattern as in Fig.\ref{shellexp} (where the loop structure is present from the beginning)
as it slowly expands to cover a surface twice as large at the initial one.
However, the expansion appears much larger where no initial boundary shell was present and this is illustrated by the red region (colour saturation) above $z_0$. This effect is uniform from $z_0$
to the end of the domain for the simulation with $z_0=130$ $Mm$ after $t=500$ s. In contrast, in the simulation with $z_0=46$ $Mm$
we find a time evolution where the boundary shell is extended above $z_0$ from $t=500$ s at about $0.2$ Mm/s.
In summary, the simulations with a uniform boundary shell and with $z_0=130$ $Mm$ both show a final volume expansion of about 2 by the end of the simulation, whereas the simulation with $z_0=46$ $Mm$ shows a maximum expansion of $2.7$. 
Interestingly, this process also leads to a more efficient boundary shell with respect to the phase-mixing, as our measure of the efficiency of the boundary shell is about 35\% higher when we consider a non-uniform loop.
The time evolution pattern of the efficiency is very similar for all simulations (Fig.\ref{tiploops_expa}).
Additionally, these boundary shells also extend in the $z$-direction while expanding in the $x-y$ plane, and therefore their consistency is affected too.
Fig.\ref{tiploops_cons} compares the consistency evolution for the three simulations, relative to their respective initial consistency. We find that the consistency of the boundary shell of a uniform loop quickly drops as the simulation starts.
The simulation with $z_0=130$ $Mm$ undergoes a smaller initial decrease, followed by an increase to a value close to 1 for about $500$ $s$, before decreasing again to smaller values and finally ramping up again when the loops settles to a new equilibrium near the end of the simulation, after the boundary driver stops. Although fluctuating, some consistency is clearly present for a significant fraction of the time evolution. Finally, the simulation with $z_0=46$ $Mm$ shows a nearly constant boundary shell consistency throughout the simulation. Although it evolves, the extension in the $z$-direction remains relatively constant. Near the end of the simulation, because the original boundary shell has extended in the $z$-direction and because of the newly reached equilibrium, the consistency measure is twice as large as at $t=0$.

In conclusion, the evolution of the boundary shell is significantly affected by the initial structure of the loop. We find that the presence of an interior region as a central uniform density enhancement is not required to form loop-like structures. A shallow initial boundary shell can be sufficient to develop a larger, steeper, and more consistent boundary shell around the loop.
In fact, we find that an initially small density enhancement (here represented by the simulation with $z_0=46$ $Mm$) relatively develops a more favourable boundary shell compared to a fully formed uniform loop.

\section{Heating the corona}
\label{haetingthecorona}

In this section, we analyse how the heating resulting from the dissipation of transverse waves depends on a number of parameters.
In particular, we investigate the role of the initial loop structure, the extent of the footpoint driver region, the reflection of the waves at the upper footpoint and the dissipation coefficients.
In Sect.\ref{MHDsimulation}, we have shown that the dissipation 
of MHD waves contributes to at least about 1\% of the energy budget needed to maintain the thermal structure of the corona by comparing the ohmic heating in the simulation with the estimated radiative losses in the boundary shell of the loop.
This finding is in agreement with our previous theoretical studies \citep{PaganoDeMoortel2017,Pagano2018,Pagano2019}.
However, the estimates are sensitive to the specific modelling setup and there are several physical reasons why the heating contribution could be enhanced.
In this section, we examine when the heating contribution from the dissipation of MHD waves can be more effective using the simulation described in Sect.\ref{MHDsimulation} as a reference simulation.

\subsection{Density structure}
\label{haetingthecorona_densitystructure}
\begin{figure}
\centering

\includegraphics[scale=0.24]{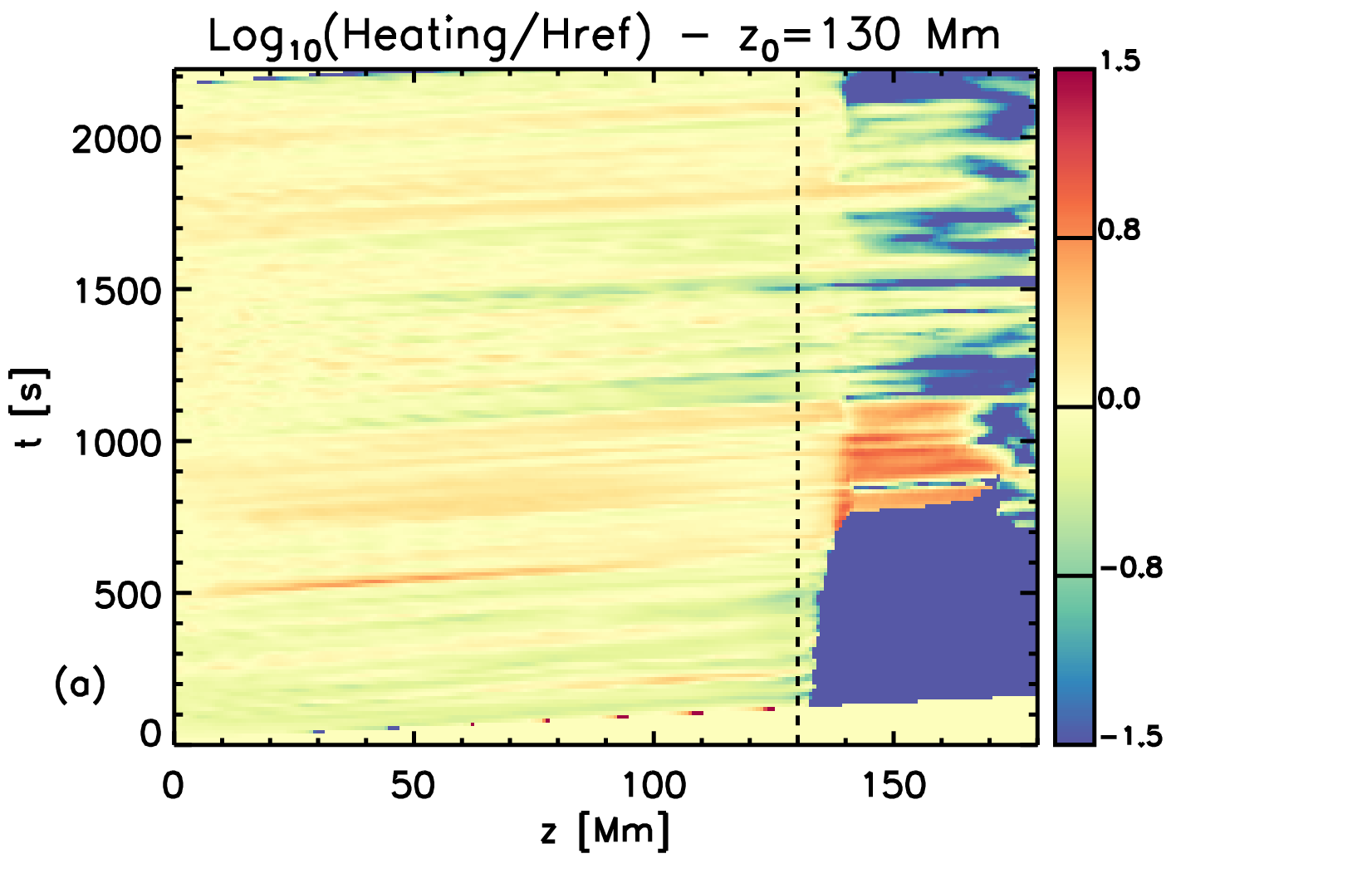}
\includegraphics[scale=0.24]{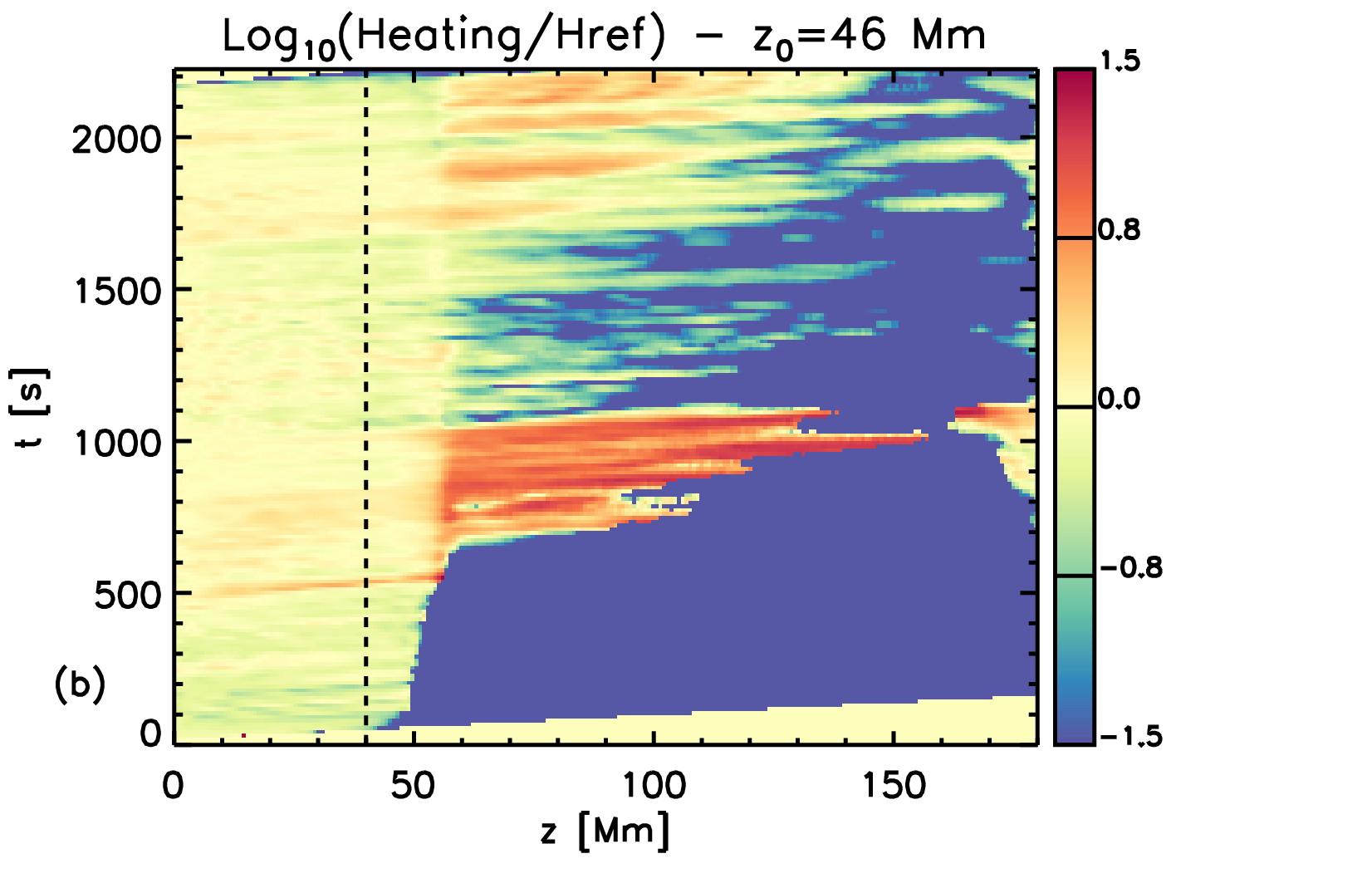}

\caption{Maps of the logarithm of the average heating in the boundary shell in the simulation with (a) $z_0=130$ $Mm$ or (b) $z_0=46$ $Mm$ normalised to the reference simulation and plotted as a function of position along the loop and time.}
\label{heatingtiploop}
\end{figure}

The total heating input is not significantly affected by the initial loop structure, however its distribution is.
Fig.\ref{heatingtiploop} shows the logarithm of the average heating in the boundary shell as a function of the coordinate along the loop and time, normalised by the same heating quantity in our reference simulation, for the two simulations with $z_0=130$ $Mm$ and $z_0=46$ $Mm$.
For the most part, we find that the heating is not significantly different with respect to the reference simulation (i.e.~with a uniform boundary shell); noticeable differences are only found in the portion of the loop above $z_0$
where the heating is substantially smaller during the time when the uniturbulence is not yet effective ($t<60$ $s$). Near the end of the simulation, it settles to similar values. At the same time, the density is lower in the non-uniform density structures and, hence, radiative losses are smaller in comparison to the reference simulation. Hence, the heating profile can now contribute to up to 8\% of the radiative losses in these non-uniform configurations.

\subsection{Role of the footpoint motions}
\label{haetingthecorona_footpoints}
\begin{figure}
\centering
\includegraphics[scale=0.24]{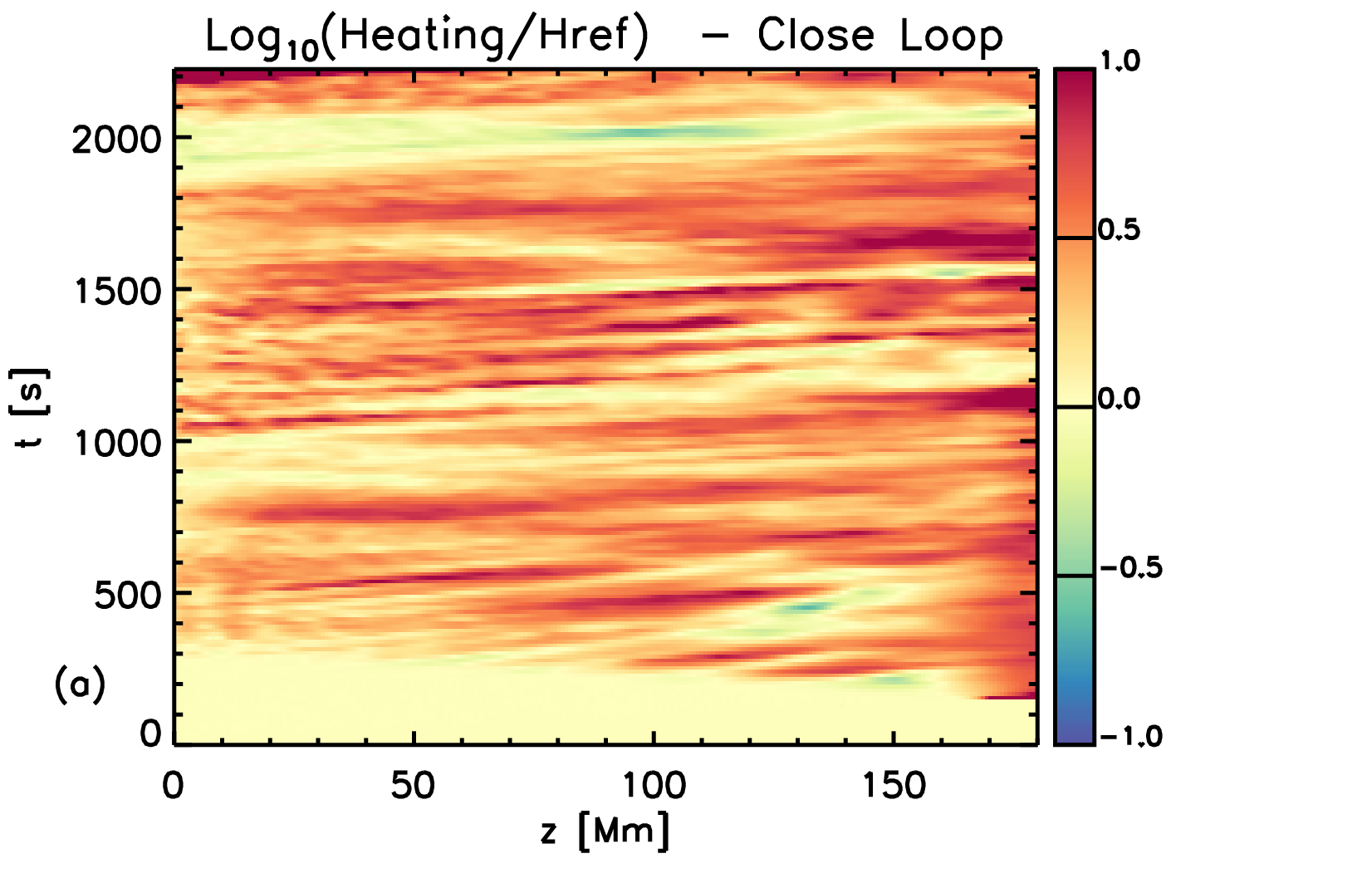}
\includegraphics[scale=0.24]{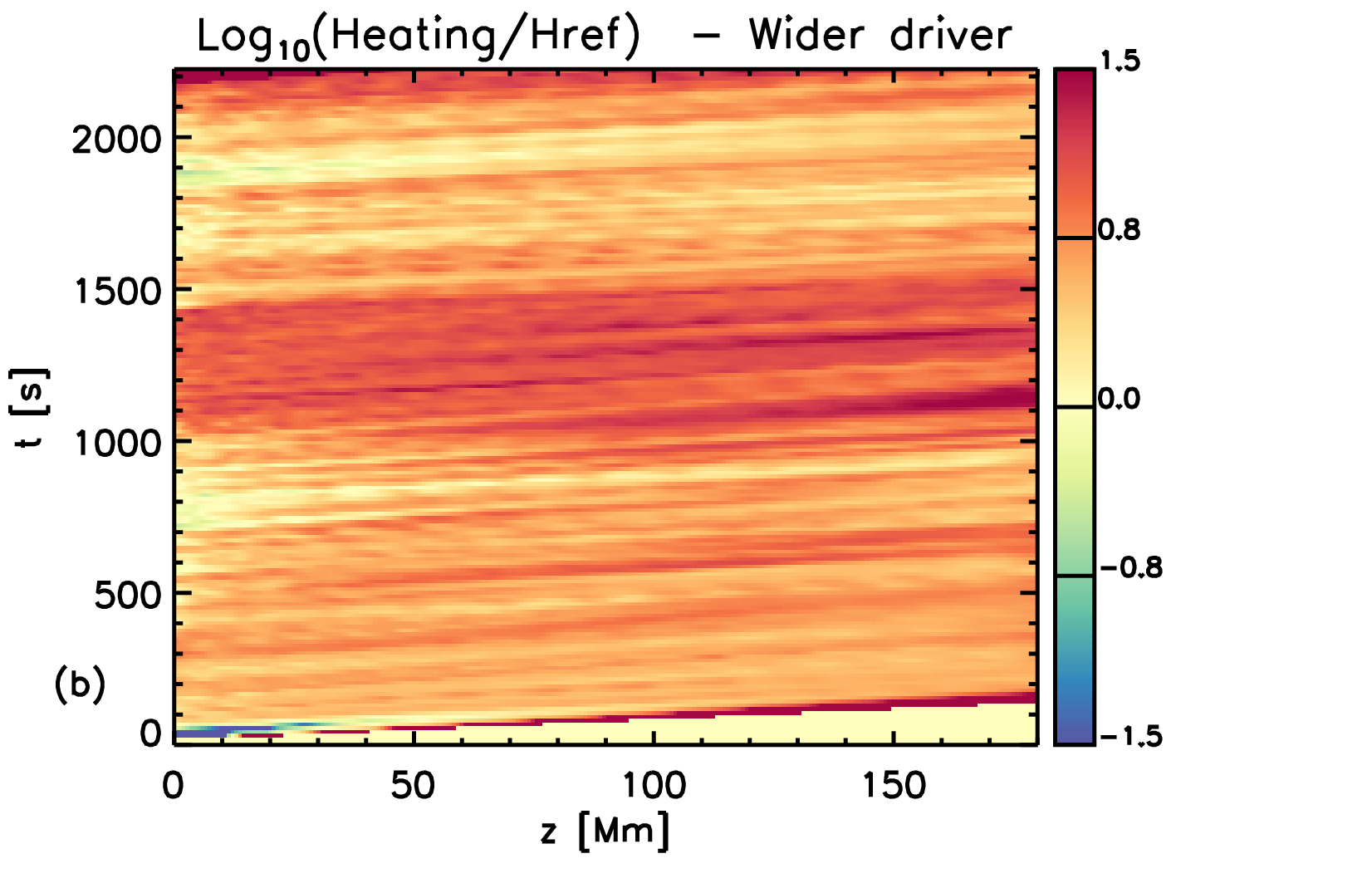}
\caption{Maps of the logarithm of the average heating in the boundary shell in the simulation with (a) a closed loop or (b) a wider driver normalised to the reference simulation and plotted as a function of position along the loop and time.}
\label{haetingub}
\end{figure}

We have so far used a driver which only affects the central part of the lower boundary (where the loop structure is initially situated) and outflow boundary conditions at the upper boundary.
This scenario represent a situation where only one footpoint of a loop is subject to transverse oscillations and any wave energy not dissipated along the loop flows out at the opposite footpoint (or the structure is an open field coronal structure). Fig.\ref{haetingub} shows the heating profiles normalised to the reference simulation (as in Fig.\ref{heatingtiploop}) for two simulations where we change these assumptions, i.e. where the outer boundary is reflective or where the driver acts on the entire lower boundary.

In the simulation with reflective upper boundary conditions, the transverse waves cannot leave the domain from that boundary and the propagating waves now interact with counter-propagating ones. This can make the phase mixing more efficient and further accelerate the expansion of the boundary shell.
Indeed, we find that the volume of the boundary shell in this simulation expands by a factor of 3, which is larger than previously found. In particular, the boundary shell expansion now also happens closer to the upper boundary. From an energetic point of view, we find that the thermal energy included in the domain in this simulation is always larger than the thermal energy enclosed in the respective ideal simulation, showing that the excess in thermal energy is due to the heating from the dissipation of waves.
Moreover, the heating pattern differs from the one found in the simulation with open boundary condition, as stronger heating signatures are now also present at the upper footpoint (see Fig.\ref{haetingub}a). In this simulation, the time integrated heating is enhanced by a factor of 2.2.

\begin{figure*}
\centering
\includegraphics[scale=0.32]{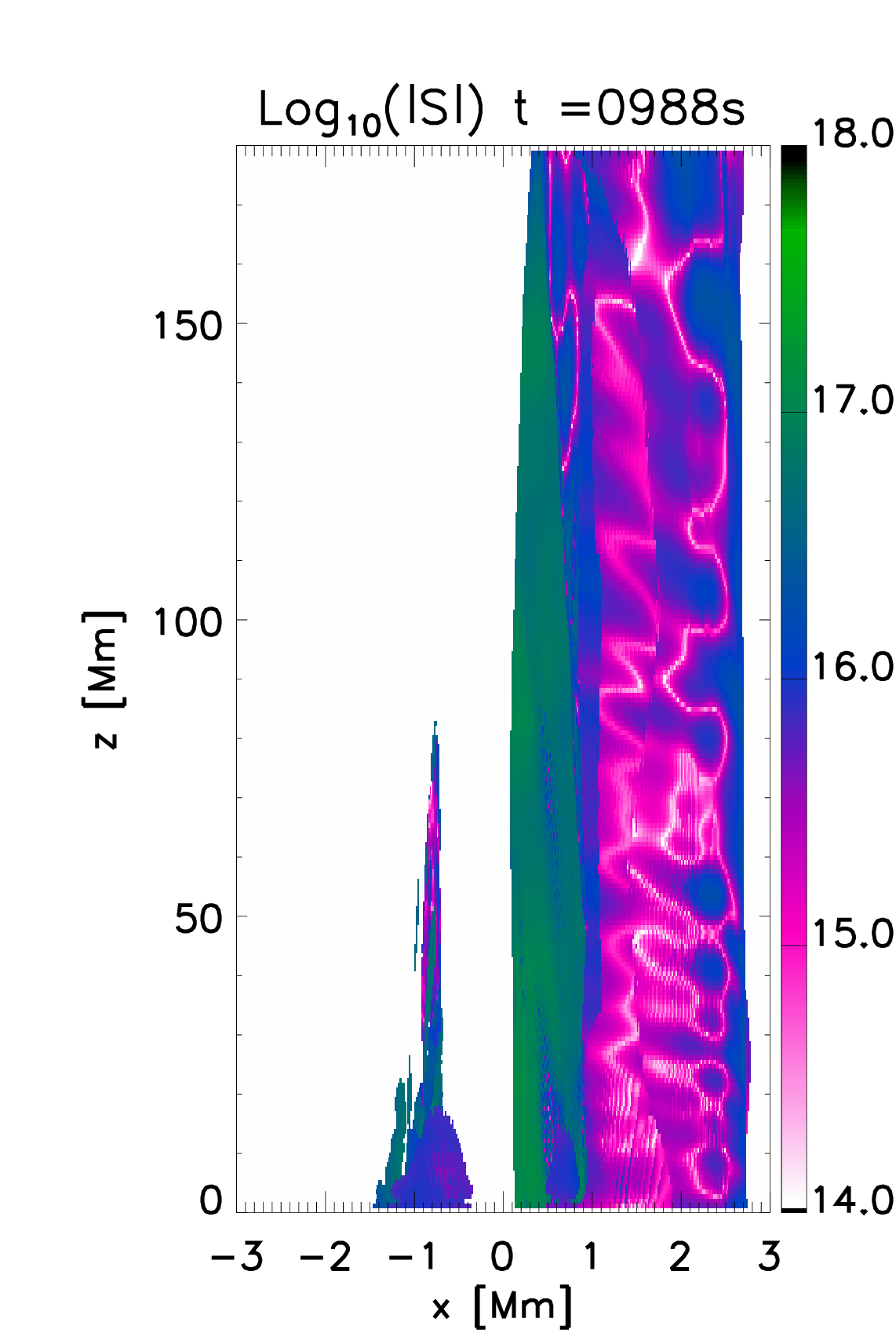}
\includegraphics[scale=0.32]{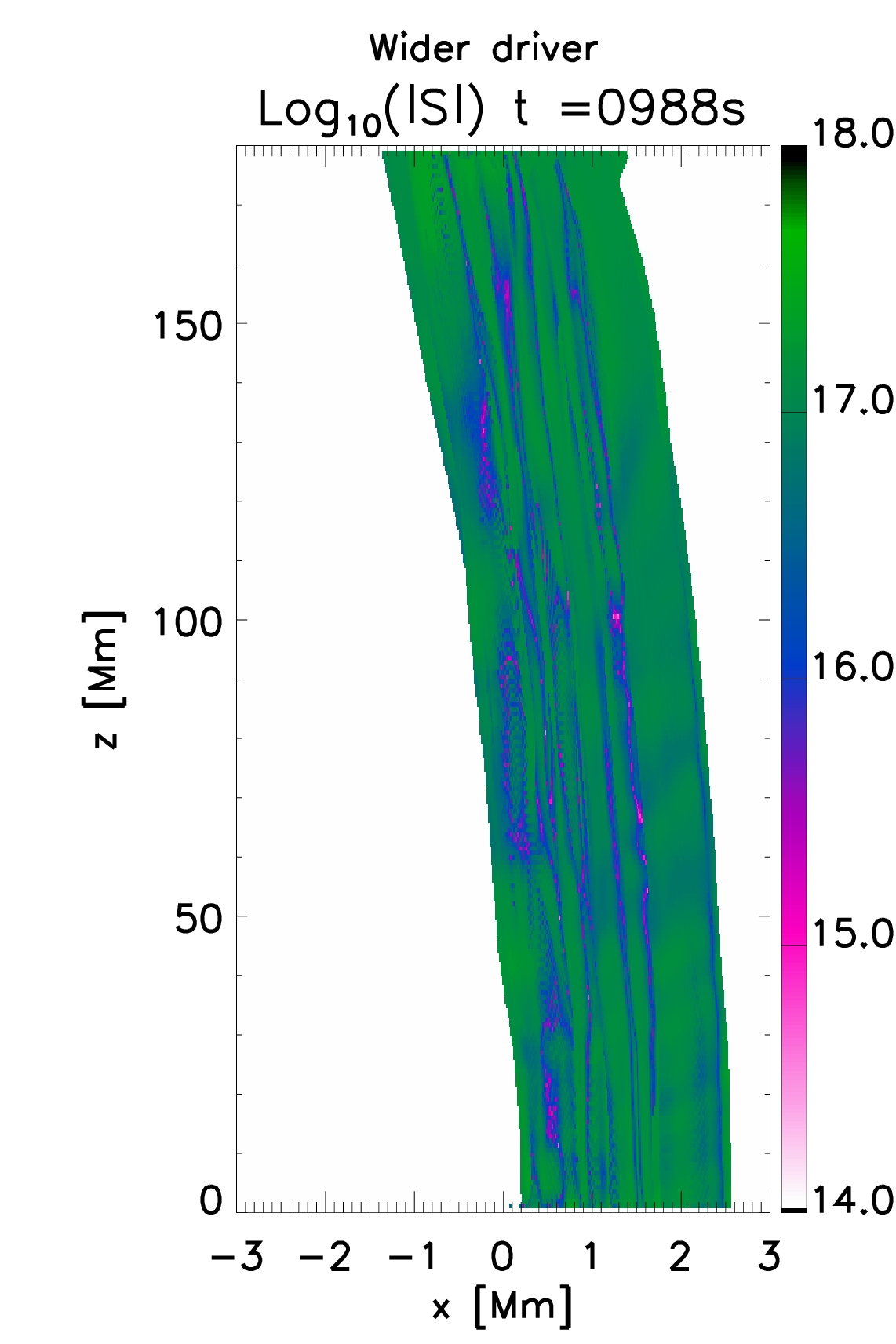}
\caption{Isosurface of the gradient of the Alfv\'en speed for the reference simulation (left hand side) and for the wider driver simulation (right hand side) at $t=988$ $s$, coloured by the modulus of the component of the Poynting flux across the isosurface.}
\label{poynting}
\end{figure*}
At the same time, when we consider a different extension of the boundary driver, the heating output differs significantly. Fig.\ref{haetingub}(b) shows the heating deposition for
the simulation where the boundary driver is applied to the entire lower boundary. The heating is now always larger than in the reference simulation, by a factor of at least 5 and with peaks of up to a factor of 50. For reference, in this configuration, about 8 times more energy is injected into the domain by extending the driver to the full lower boundary, in particular including the lower-density loop exterior. There are two key differences with respect to the reference simulations. Firstly, the loop interior and exterior move in phase near the lower boundary
and there is no damping of the oscillations of the fluxtube due to its interaction with the background. This allows more wave energy to be converted into thermal energy via phase mixing and thus leads to higher Ohmic heating.
Additionally, as already shown in \citet{Pascoe2010}, because of the mode coupling between the kink and Alfv\'en modes, wave energy flows as Poynting flux from the exterior to the boundary shell, where it can then be dissipated through phase mixing. Fig.\ref{poynting} compares the Poynting flux component across the external surface of the boundary shell for our reference simulation and for the simulation where the driver is applied to the entire lower boundary at the same time $t=988$ $s$ (i.e.~when the boundary shell has not yet been fragmented). We find that the Poynting flux is up to two orders of magnitude larger in the latter simulation.

\subsection{Higher velocity driver}
\label{haetingthecorona_veldriver}

\begin{figure}
\centering
\includegraphics[scale=0.24]{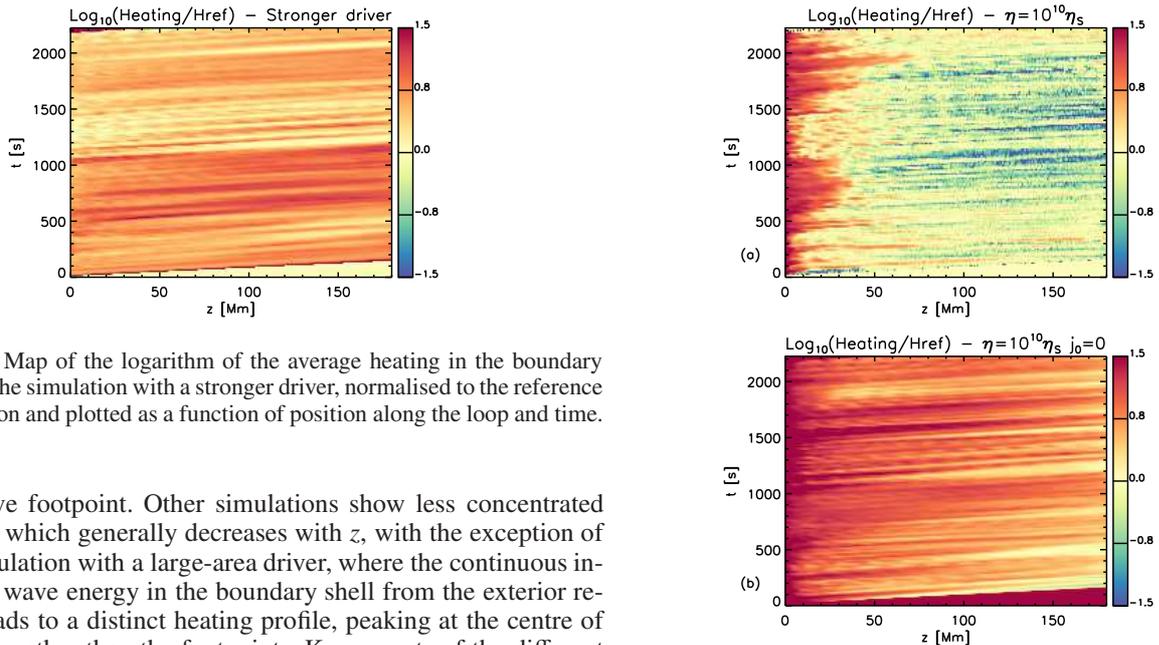}
\caption{Map of the logarithm of the average heating in the boundary shell in the simulation with a stronger driver, normalised to the reference simulation and plotted as a function of position along the loop and time.}
\label{haetingua}
\end{figure}

In our reference simulations, we adjust the amplitude of the driver
in order to have transverse velocities of the order of $v_x\sim v_y \sim 15$ $km/s$ beyond the damping layer, i.e. past $z=0$ Mm.
However, because of the time variation of our driver, the speeds can drop to only a few $km/s$. As larger transverse velocities are regularly observed in the corona, we also run a simulation where the speed of the driver is increased by a factor of 2.5 with respect to the simulation described in Sec.\ref{MHDsimulation}.
Fig.\ref{haetingua} shows the normalised heating for this simulation, which is uniformly higher than the corresponding heating in the reference simulation by a factor of at least 2, with peaks of up to 15.
In this simulation, the time integrated heating is enhanced by a factor of about 4.5, compared to an increase of a factor of 6.25 in the kinetic energy. Assuming we can extrapolate this scaling further, then the enhancement of the heating by a factor of 100 which we require, would imply that we need to increase the amplitude of the driven waves by a factor of 11.
However, non-linear effects could certainly play a role with larger amplitudes and such a dramatic increase might not be necessary in order to reach a similar heating enhancement. Either way, substantially larger amplitudes are needed to supply the entire heating requirement via transverse wave dissipation.

\subsection{Higher dissipation coefficient}
\label{haetingthecorona_eta}
Higher values of the resistivity coefficient or lower values for the critical current at which the resistivity is triggered might lead to more effective heating. We therefore run two further simulations: one where $\eta_0=10^{10}\eta_S$ (i.e.~enhanced by two orders of magnitude compared to the simulations so far) and one where this higher value of the resistivity coefficient is used uniformly, without a critical current trigger.

\begin{figure}
\centering
\includegraphics[scale=0.24]{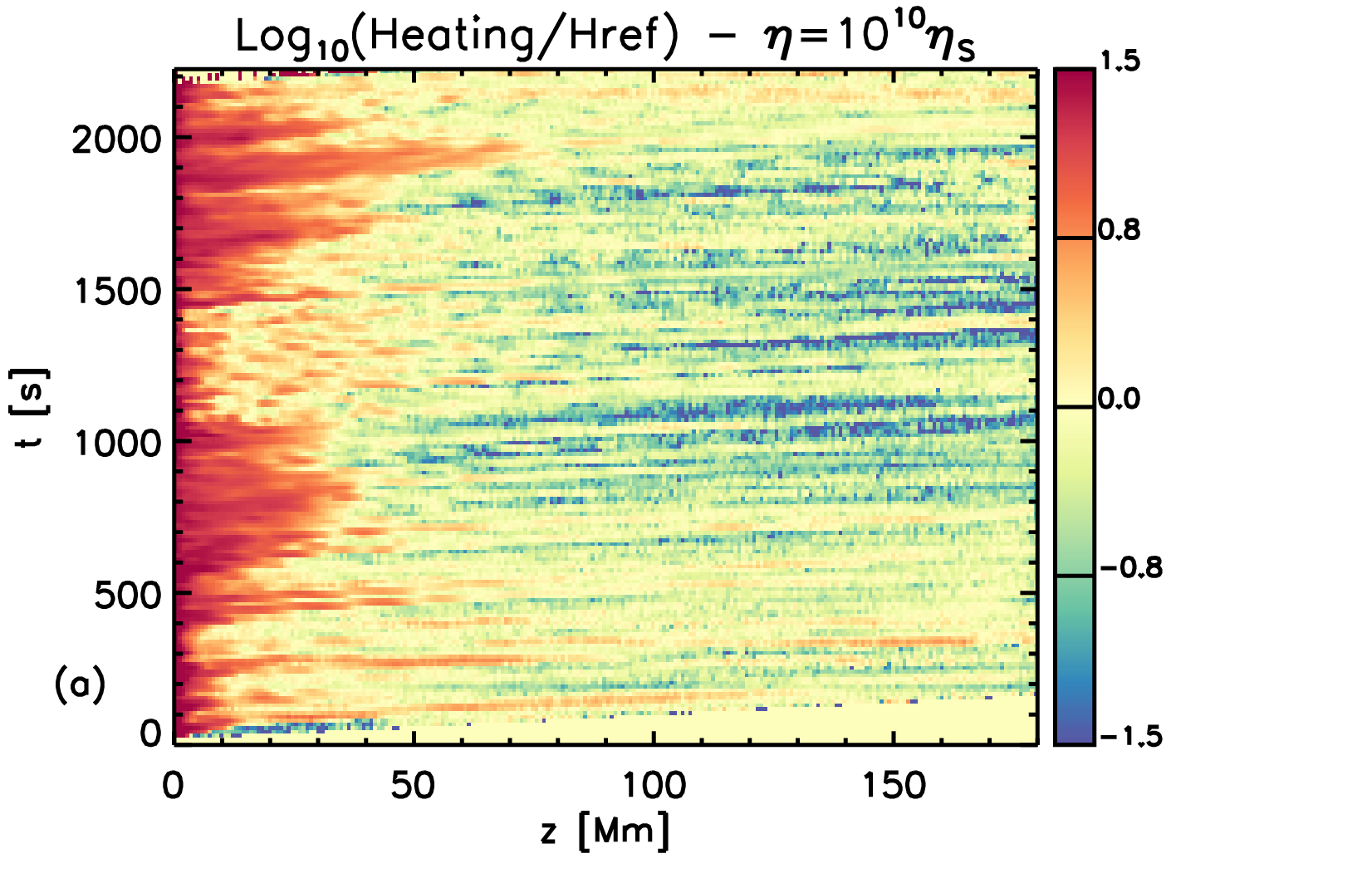}
\includegraphics[scale=0.24]{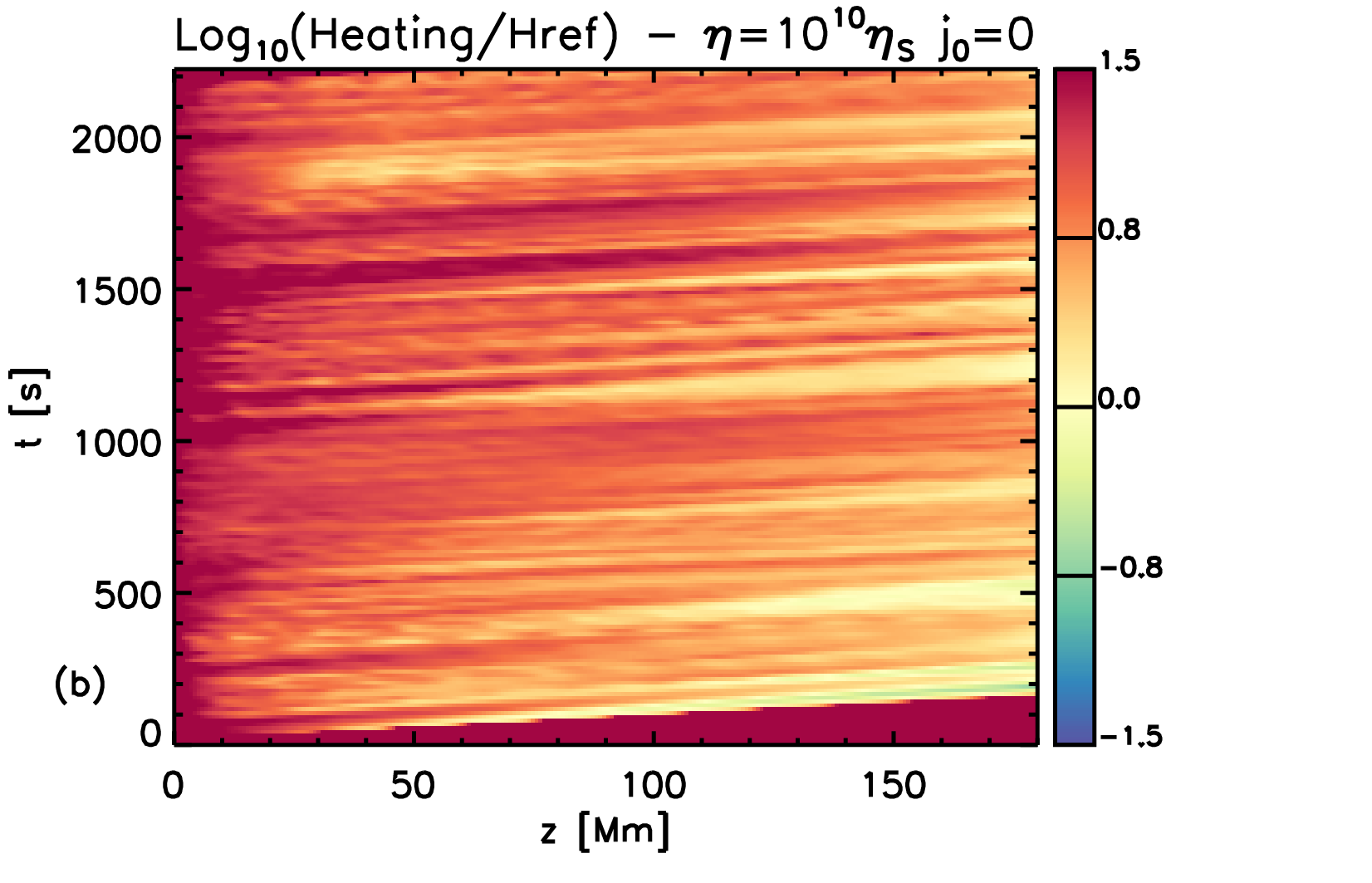}
\caption{Maps of the logarithm of the average heating in the boundary shell in the simulation with (a) $\eta=10^{10}\eta_S$ and (b) $\eta=10^{10}\eta_S$ and $j_0=0$, normalised to the reference simulation and plotted as a function of position along the loop and time.}
\label{heatingeta}
\end{figure}

Fig.\ref{heatingeta} shows the normalised heating for these two new simulations. The simulation where we simply enhance the resistivity coefficient (Fig.\ref{heatingeta}a) shows a larger overall heating compared to the reference simulation, particularly near the footpoint, whereas
the simulation without the critical current threshold (Fig.\ref{heatingeta}b) shows stronger heating all along the whole loop. For both simulations, the heating increases by more than one order of magnitude, where this is confined to the region near the footpoint when a critical current to trigger the resistivity is used.

\begin{figure}
\centering
\includegraphics[scale=0.24]{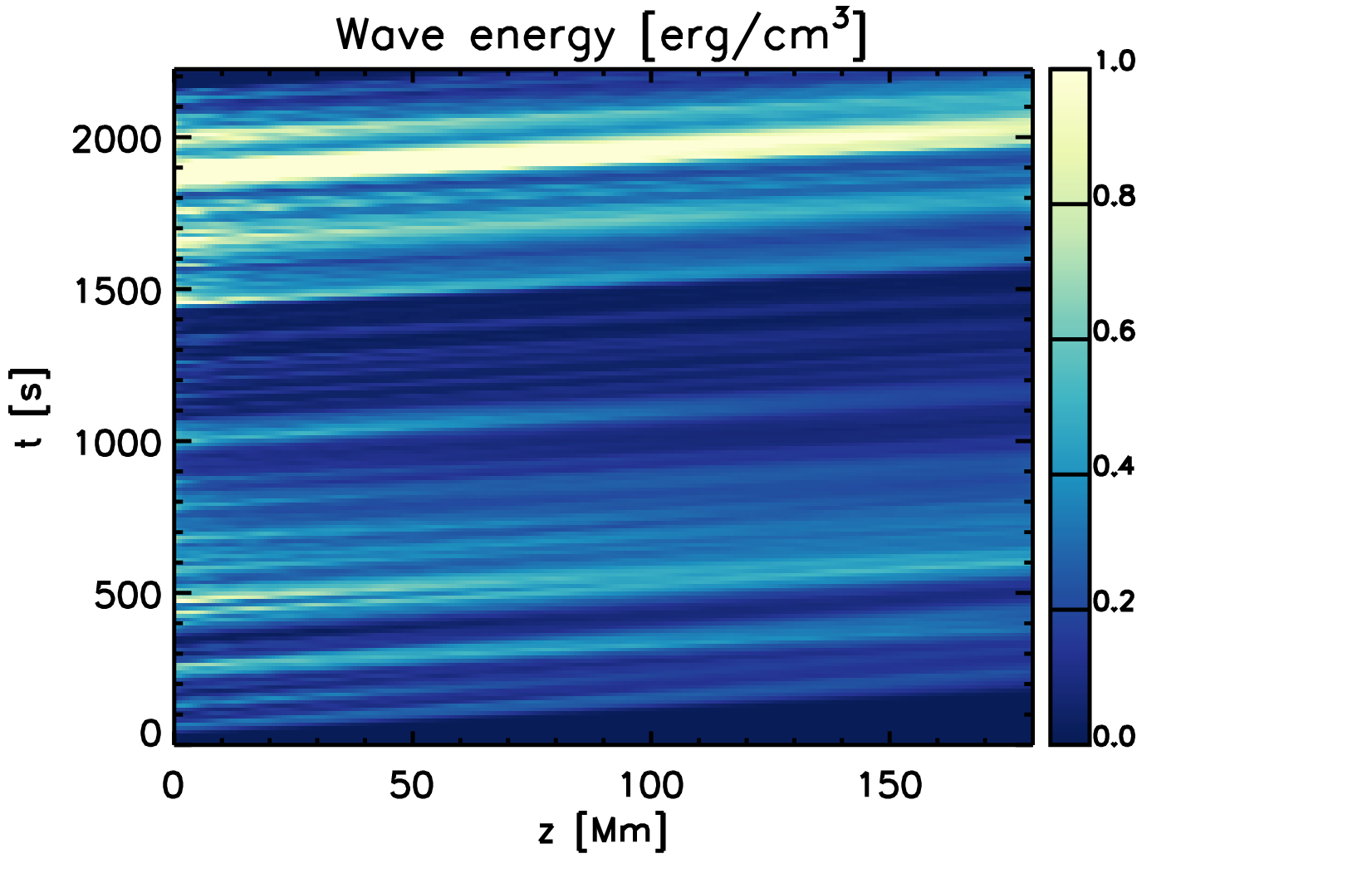}
\includegraphics[scale=0.24]{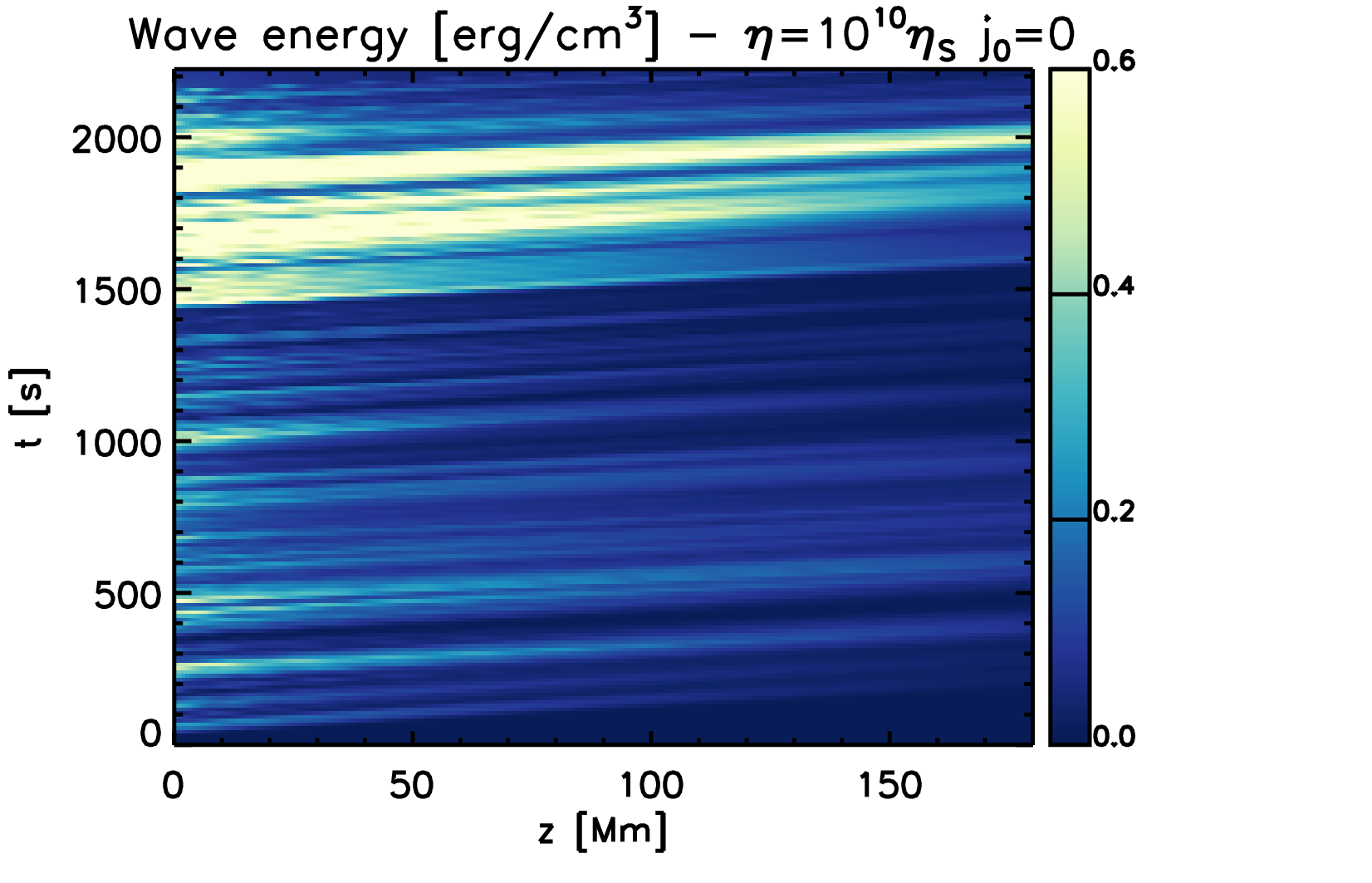}
\caption{Maps of the average wave energy in the boundary shell as a function of time and position along the loop for (left) the reference simulation and (right) the simulation with $\eta_0=10^{10}\eta_S$ and $j_0=0$.}
\label{waveener1000}
\end{figure}

This higher dissipation and heating happens at the expense of the wave energy. However, the occurrence of heating leads to the generation of additional transverse velocities from local transverse pressure gradients and therefore, some of the plasma kinetic energy cannot be ascribed to the propagation of MHD waves. However, we can use the property of equipartition of kinetic and magnetic energies in MHD waves to identify the energy associated with transverse waves in the simulation defined as:
\begin{equation}
\begin{split}
E_W=\frac{1}{2}\left[\frac{1}{2}\rho\left(v_x^2+v_y^2\right)+\frac{1}{8\pi}\rho\left(B_x^2+B_y^2\right)\right]\\
-\frac{1}{2}\sqrt{\left[\frac{1}{2}\rho\left(v_x^2+v_y^2\right)-\frac{1}{8\pi}\rho\left(B_x^2+B_y^2\right)\right]^2}
\end{split}
\end{equation}
Figure \ref{waveener1000} shows the average wave energy in the boundary shell along the loop as a function of time for the reference simulation and for the simulation with $\eta_0=10^{10}\eta_S$ and $j_0=0$.

We find that while the wave energy decreases only marginally along the loop in the reference simulation, it does so significantly in the other simulation, implying that the wave amplitude is damped substantially along the loop by the effect of the resistivity.

\subsection{Heating Summary}
\label{haetingthecorona_summary}

We can use our analysis of a number of different physical configurations to shed light on the conditions under which MHD waves can contribute significantly to the heating of the solar corona.
Generally, we find that the dissipation of MHD waves via phase mixing can certainly contribute to about 1\% of the energy that is radiated by the solar corona, but in certain configurations, the heating can be significantly increased.

\begin{figure*}
\centering
\includegraphics[scale=0.28]{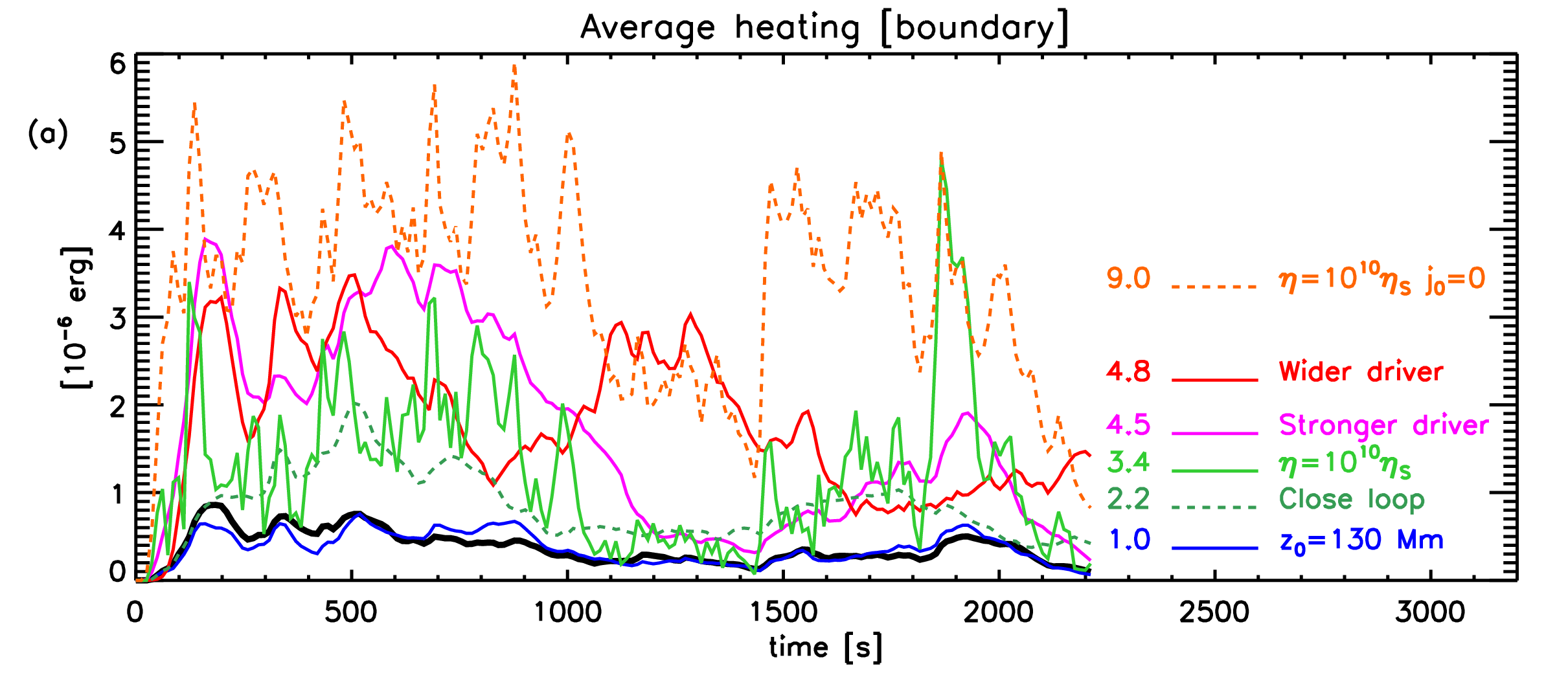}
\includegraphics[scale=0.28]{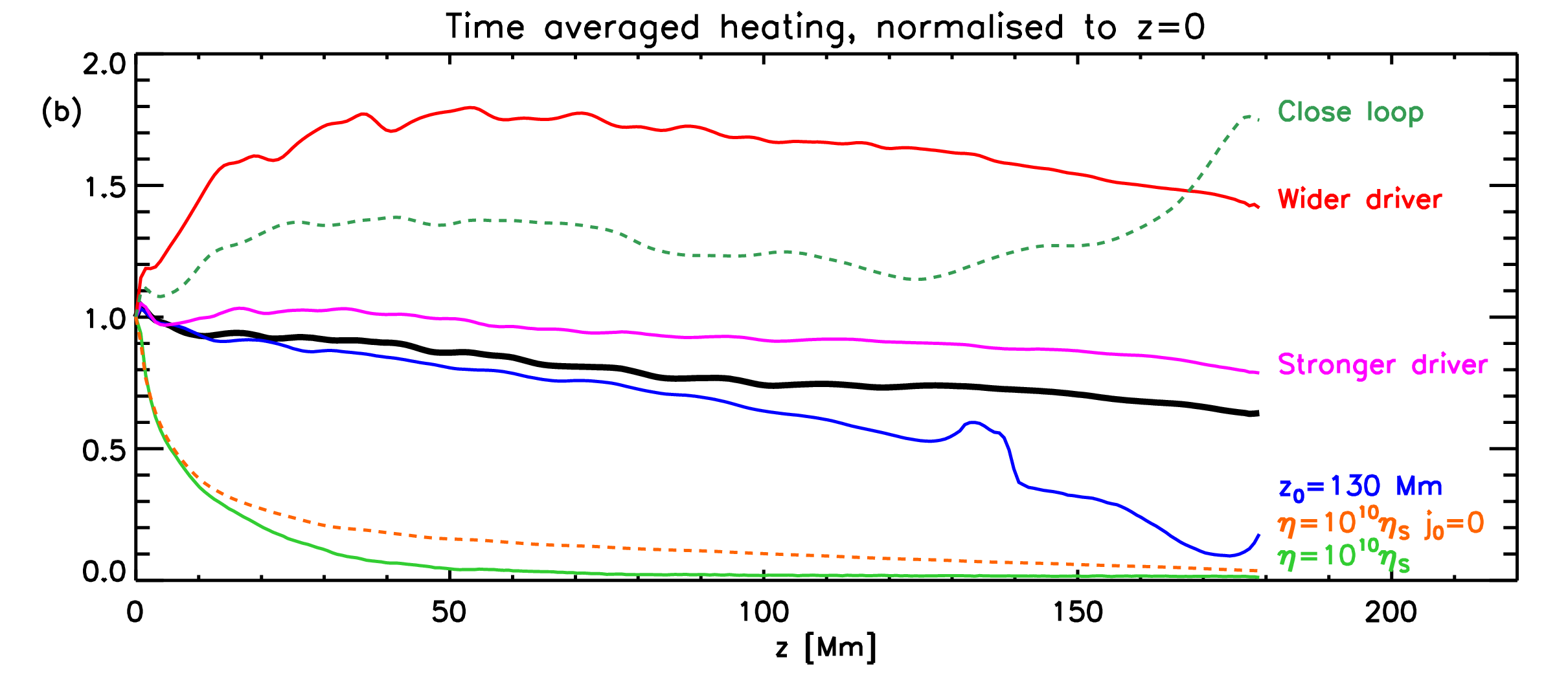}
\caption{(a) Average heating in the boundary shell as a function of time for a number of simulations listed in the colour legend with the average heating enhancement reported.
(b) Average heating in the boundary shell as a function of position along the loop for the same simulations normalised to the average heating at $z=0$.}
\label{heatings}
\end{figure*}

To show the enhancement in heating, in Fig.\ref{heatings}(a) we 
plot the average heating in the boundary shell as function of time
for some of the key simulations in this study. We report in the figure how much the heating is enhanced on average with respect to the reference simulation. We find that the heating is on average enhanced by a factor of at least $4.5$ when the extent of the footpoint driver is larger than the size of the loop interior or when a higher resistivity coefficient without a critical trigger current is used or for a higher amplitude driver.
One could argue that if several of these enhancements were to occur simultaneously, the wave heating may become sufficient to counteract the radiative losses.

These different configurations also have implications for the nature of the wave generation and propagation in the solar corona. In particular, different configurations lead to different heating locations. Fig.\ref{heatings}(b) shows the heating averaged in time as function of the coordinate along the loop, normalised to its value at $z=0$ $Mm$. When we have higher resistivity, the heating is highly localised at the footpoint. Similarly, for the simulation with a closed loop, most of the heating takes place near the reflective footpoint. Other simulations show less concentrated heating which generally decreases with $z$, with the exception of the simulation with a large-area driver, where the continuous inflow of wave energy in the boundary shell from the exterior region leads to a distinct heating profile, peaking at the centre of the loop rather than the footpoints. Key aspects of the different heating profiles for the different simulations are summarised in Table\ref{waveheatingtable}.

\begin{table*}[!h]
\centering
\begin{tabular}{|c | c | c | c|}
\hline                        
 \bfseries Configuration & \bfseries Heating & \bfseries Heating & \bfseries Wave\\
 & \bfseries Enhancement &\bfseries Location &\bfseries Dynamics \\
 \hline                        
  Wide driver & 4.8 & Centre & Large driver cells \\      
  \hline                        
  Closed loop & 2.2 & Footpoints &  Reflective footpoints \\      
  \hline                        
  Higher amplitude driver & 4.5 & Distributed & V x 2.5  \\      
  \hline                        
  Higher resistivity & 6.2 & Footpoints & Strong damping  \\      
 \hline                        
\end{tabular}
\caption{Summary of the heating enhancement in the simulations with respect to the reference simulation and the observable consequences for the wave dynamics.}             
\label{waveheatingtable}      
\end{table*}

\section{Relevance of observed power spectrum}
\label{spectrum}

Finally, we investigate how using different power spectra for
the boundary driver can affect the heating output. In particular, we use the different power spectra in Fig.\ref{spectrumtimeseries}(a) to investigate the importance of the key features associated with the power spectrum observed by \citet{2019NatAs...3..223M}, namely the excess in power around the 4 minute oscillations and the particular distribution of the power.

\begin{figure}
\centering
\includegraphics[scale=0.25]{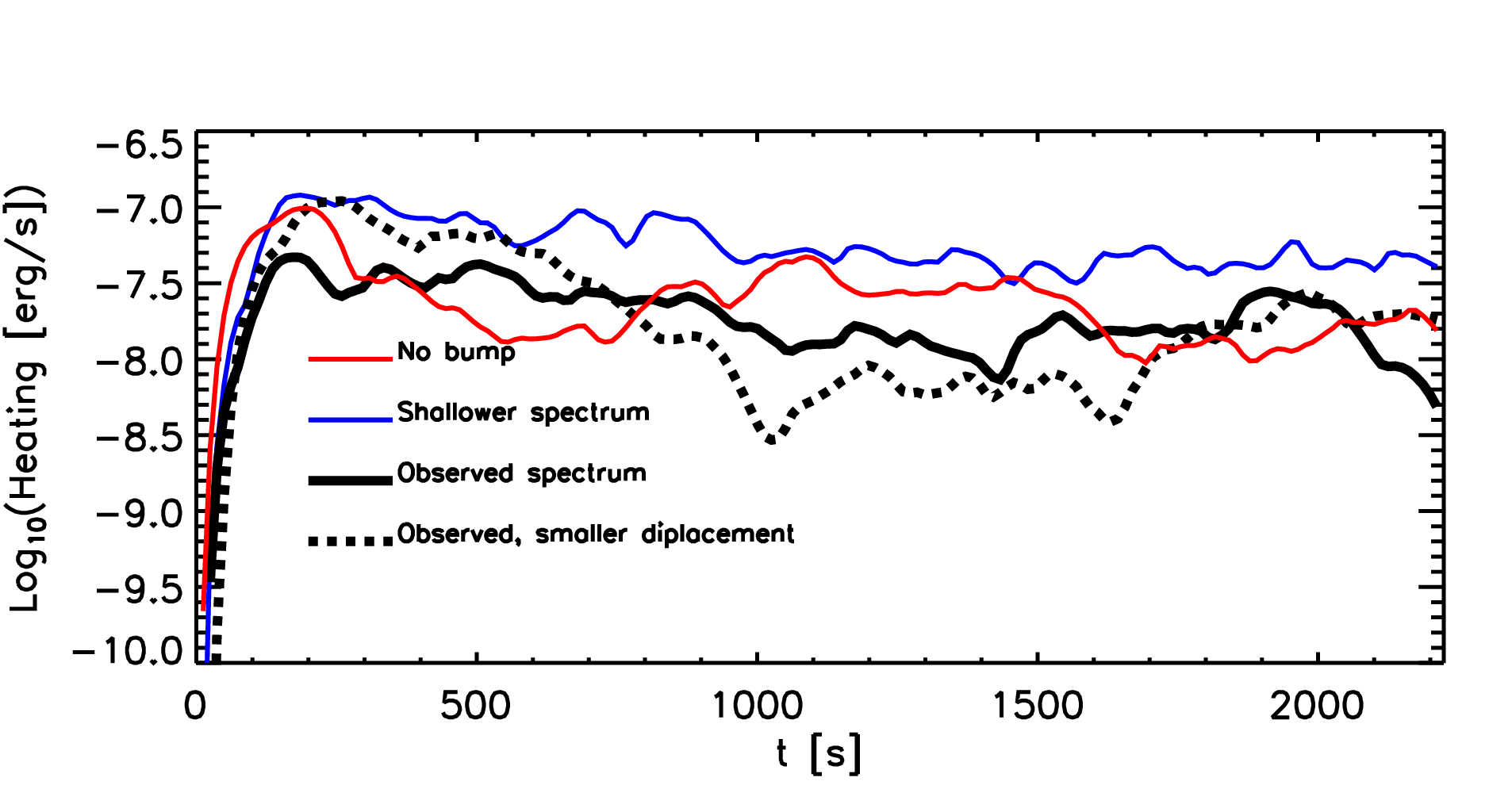}
\includegraphics[scale=0.25]{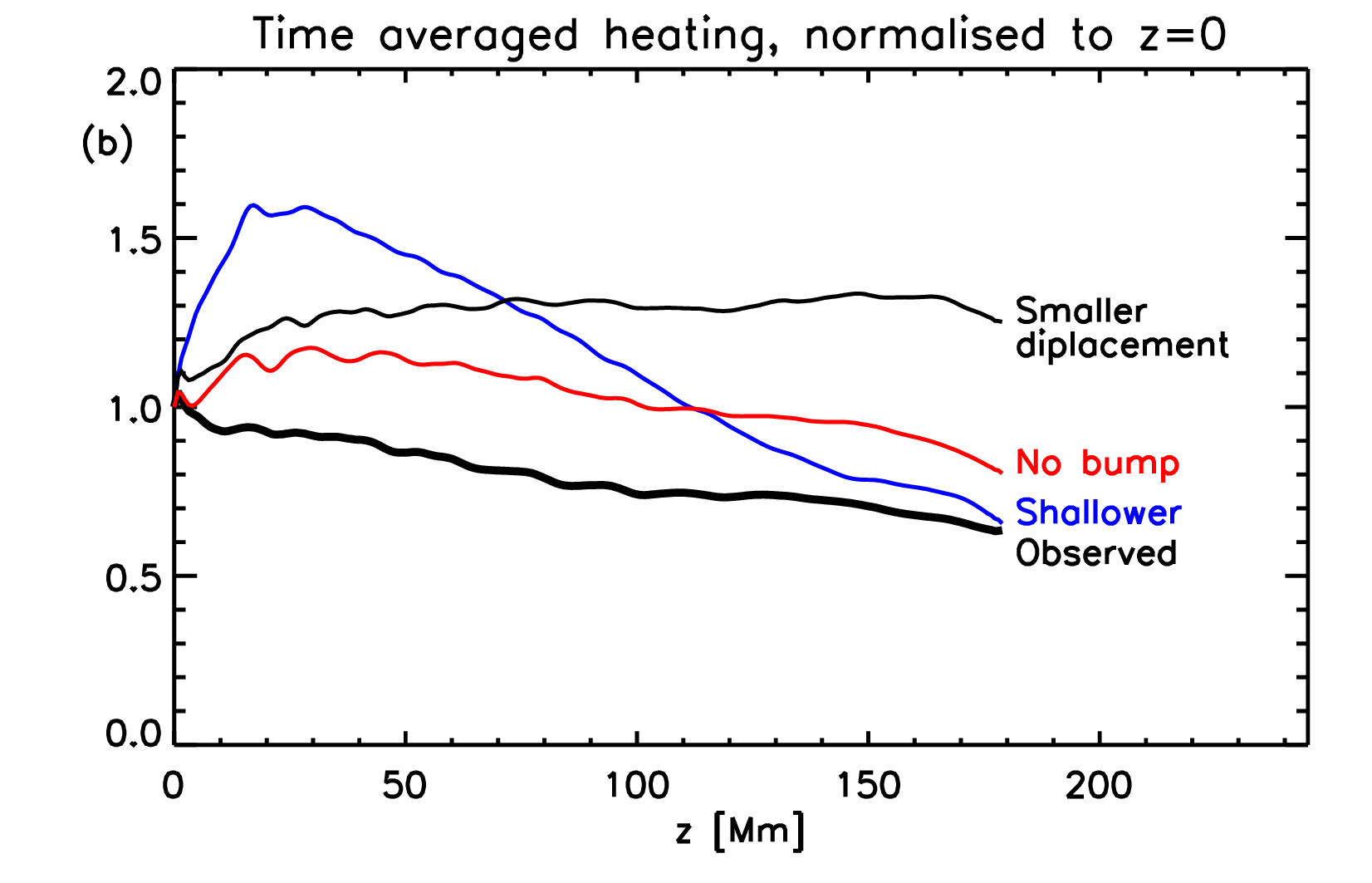}
\caption{(a) Average heating in the boundary shell as function of time 
for the simulations with four different time series, as explained in Fig.\ref{spectrumtimeseries}b.
(b) Average heating in the boundary shell as a function of the position along the loop for the four different time series, as explained in Fig.\ref{spectrumtimeseries}b}
\label{hespectra}
\end{figure}

Fig.\ref{hespectra}(a) shows the average heating in the boundary shell for the four simulations based on the power spectra shown in Fig.\ref{spectrumtimeseries}(a). The simulations all use a uniform boundary shell. Two of these simulations (black lines) are driven by timeseries which are based on the same (observed) power spectrum, the simulation without the excess power at 4 minutes is represented by the red lines and the blue lines correspond to the simulation for which the power was redistributed with a more shallow gradient (i.e.~the power associated with the long period oscillations is reduced and the power at the short period oscillations in increased). As the driver is always applied to the centre of the domain, the amount of energy (i.e.~Poynting flux) that is injected into the system and, hence, available for heating, crucially depends on the amount of energy in the driver, on the power spectrum, and the position of the waveguide. In
order to isolate the effect of the spectrum, we here compare simulations for which the maximum displacement during the time series is comparable. 

We find that the heating profiles for the simulations using the observed power spectrum and the simulation without the 4 minute excess power are quite similar. In contrast, the simulation with a shallower spectrum consistently leads to higher heating. The heating increase in this simulation is consistent with the additional power at the higher frequencies with respect to the observed power spectrum, confirming that high-frequency oscillations dissipate more efficiently.

Fig.\ref{hespectra}(b) shows the time averaged heating as a function of the coordinate $z$ along the loop for these simulations. For three of the simulations (red and black lines), we find mostly similar heating profiles, with the heating largely uniform along the loop.  The difference between the two black lines (both based on the observed power spectrum) can be explained by the different footpoint displacements (see Fig.\ref{spectrumtimeseries}(c)). For the loop with the larger displacement (thick black line), phase mixing is most efficient low down in the loop, as the larger footpoint displacements reduce the field-aligned consistency of the boundary shell at larger $z$. However, the heating profile is qualitatively different for the simulation based on a shallower spectrum. In this simulation, the combination of the enhanced power at higher frequencies and the more efficient dissipation of the high-frequency oscillations leads to a distinct heating profile, concentrated near the driven loop footpoint.

\section{Discussion and Conclusions}
\label{conclusions}

In this paper, we have expanded the investigation of how wave energy is tranformed into plasma heating when transverse MHD waves travel along coronal loops. In particular, we have based this investigation on an observed power spectrum of transverse waves as well as analysing how other parameters affect this process.
We have focused on (1) the initial loop structure where we consider
uniform loop structures composed of an interior region and a boundary shell or loop structures where the loop interior vanishes and the boundary shell is present only for a portion of the loop length, (2) the behaviour of the footpoints of the loop, (3) the energy available for dissipation, and (4) the power spectrum itself.
The aim of this work is to show how the wave heating contribution is affected by these parameter and how the loop structure (mostly the boundary shell) evolves in these circumstances.

The evolution of the boundary shell is a crucial aspect for coronal heating models. Small spatial-scale structures where electric currents can develop and be dissipated are a universal requirement of coronal heating models, regardless of the mechanism(s) responsible for the ultimate heat deposition. Examples include wave heating models which need thin boundary shells where the waves can quickly undergo phase-mixing or nanoflares models which need small structures where the magnetic field is sufficiently tangled  for magnetic reconnection to release the free energy. The boundary shells, where the Alfv\'en speed changes rapidly, are precisely those regions where we find the necessary small scales.
In this paper, we have shown not only that waves can produce such boundary shells, but also that the observed spectrum in the solar corona produces perturbations that contributed to the formation of boundary shells. We have analysed three different properties of the boundary shells:
its spatial extent (both in terms of volume and in terms of surface at different loop cross section), its efficiency (measured by the variation of the Alfv\'en speed across the boundary), and its field-aligned consistency (measured by the distance along the loop over which an Alfv\'en speed gradient is present).

We found that the extent and efficiency of the boundary shell increase in time when transverse waves propagate along the loop, where this process accelerates after the development of uniturbulence. On the other hand, the field-aligned consistency of the boundary shell is reduced by the displacement of the loop from its initial position and increases only when the loop has time to relax to a new equilibrium. It is not clear though, whether such a new equilibrium is possible in a scenario where waves are present at all times. At the same time, we find that an initially non-uniform loop where no loop interior is present and where the density enhancement is concentrated near the lower boundary can increase the spatial extent, efficiency, and consistency of its boundary shell. In simple terms, this happens because such an initial loop structure is more able to expand in 3 dimensions and to enhance its Alfv\'en speed gradient with respect to a uniform loop. This result is particularly important, as it shows that transitory density enhancement (such as e.g.~spicules) that are subject to the continuous propagation of waves, can have a boundary shell expanding around and above where either wave heating or other heating mechanisms can be triggered.
\begin{figure*}
\centering
\includegraphics[scale=0.40]{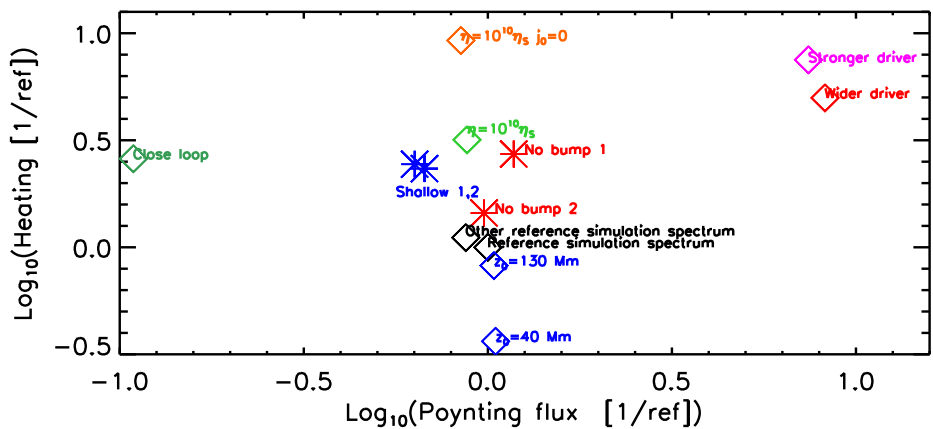}
\caption{Scattered plot of the total heating as a function of the total Poytning flux injected in the boundary shell. Both quantities are shown with respect to the reference simulation. Colours and symbols are clarified in the plot.}
\label{poyntingheating}
\end{figure*}

Fig.\ref{poyntingheating} summarises the heating output in our simulations (normalised to the reference simulation) versus an estimate of the Poynting flux injected directly in the boundary shell of the loop from the lower boundary. This scatter plot shows that, depending on the density structure and the various parameters we have considered here, the heating can be enhanced by more than one order of magnitude. The key result here is that to achieve stronger heating, injecting more energy in the loop structure is not necessarily required (as the stronger driver or wider driver simulations show), but that, for example, more effective dissipation can achieve the same result. Of particular interest is the simulation of the closed loop, which shows a very small Poynting flux with respect to other simulations, but significant heating. This occurs because at the lower boundary, where the Poynting flux is calculated, energy flux, from the waves reflected at the other boundary, is also leaving the domain.

Overall, the heating from the dissipation of transverse MHD waves does not seem to be sufficient to compensate for the energy radiated by the plasma.
We find that the wave heating can contribute only about 1\% of the radiative losses.
Factors that can lead to a more substantial contribution are (1) a non-uniform loop structure with a lower overall density, and thus less radiative losses, (2) a driver region that is larger than the loop cross section, in order to facilitate the generation of waves, a larger energy flow into the boundary shell and thus increase the available energy than can be converted into heating, (3) efficient reflection of the waves at the loop footpoints, (4) larger wave amplitudes or (5) a more effective dissipation process. Although these factors could increase the heating deposition from the transverse waves, their feasibility and effectiveness needs further investigation. For example, reflection and transmission at the transition region depends on the wavelength of the waves compared to the extent and density jump of the transition region but so does the efficiency of the wave dissipation through phase mixing; on the one hand, the dissipation of high-frequency waves is more efficient but at the same time, the energy of these waves is also more easily lost from the corona through transmission to the lower atmosphere \citep[see e.g.][]{Hollweg1984a,Hollweg1984b,Berghmans1995, DePontieu2001, VanDamme2020}. Hence, it is not a-priori obvious whether higher-frequency waves would indeed lead to increased heating.
Finally, in this work we are not focusing on the detailed development of the phase-mixing and the consequent energy conversion. A possible and interesting continuation of this work would be to devise a computationally feasible simulation to investigate in detail how much our approach using the anomalous resistivity departs from a detailed description of the phase-mixing process.

A more efficient wave energy dissipation mechanism would imply significant damping of the wave amplitude along the loop. Can such a decay be measured?
Some indications, coming from the analysis of propagating kink waves, suggest that the damping of waves can be measured in the corona and therefore provide fundamental constraints for these theoretical studies \citep[e.g.][]{2019ApJ...876..106T, Tiwariinprep}.
Finally, to have driver cells larger than the loop would also increase the Poynting flux and thus the energy available. However, this can be an elusive measurement, as the size of a single magnetic thread in coronal loops is still not clear \citep[see e.g.][]{Brooks2016a, Aschwanden2017,Williams2020} and it is difficult to measure the flow speeds in the tenuous background corona.
Moreover, it remains to be investigated how this scenario relates to the wave heating described in \citet{2011ApJ...736....3V, 2017ApJ...849...46V}, where a horizontal motion is applied over a spatial scale smaller than the cross section of the loop. In that granular driven scenario, driving happens below the corona on spatial scales smaller than magnetic bright points. 
However, a magnetic bright point will likely be the footpoint of a number of coronal loops, as they are at least $\sim100$ km in size in the photosphere \citep{2015ApJ...807..175A}, and coronal loops are about $\sim300-400$ km in width \citep{Aschwanden2017}, whereas the magnetic field expansion between the photosphere and the corona would cover a larger area. Hence, once the field expansion is taken into account, the coherence of the waves that reach the corona will be larger, and our wider driver scenario may not be in contradiction to the scenario modelled by \citet{2011ApJ...736....3V}. Additionally, if transverse waves are driven by p-modes, whose fronts are thought to be larger than the small-scale dynamics of the turbulent photospheric flows \citep{2007A&A...469.1155B,2010ApJ...722..131F}, the coherence would be even larger and more likely to occur.
Higher amplitude waves can also lead to more effective heating, but observations seem to measure velocities on average of the order of $\sim20$\,km/s \citep{Weberg2018}, possibly up to $\sim50$\,km/s \citep{2012ApJ...761..138M}, which would not be sufficient in the scenario we present here. However, line-of-sight integration might mean that higher velocities could be present, but so far, measurements of non-thermal line widths have not given conclusive evidence of the amplitudes of coronal waves \cite[e.g.][]{2008ApJ...686.1362D, McIntosh2012,Brooks2016b}.

Furthermore, our time-distance plots of the heating along the loops show that although there is a slight concentration near the footpoints, over time, the heating is mostly distributed along the loop. If we focus on a specific loop location, our simulations show impulsive heating patterns. Whereas the impulsive nature of the heating appears to agree, the location of the heating, on the other hand, seems in contradiction to at least some observational studies where the heating location has been found to be concentrated away from the loop footpoint \citep{2000ApJ...535..423R,2000ApJ...535..412R,2007ApJ...657.1127L,2015ApJ...810L..16S}. Overall, identifying whether any of the mechanisms suggested here to increase the efficiency of wave heating are viable in the solar corona will require further detailed modelling as well as higher cadence and higher resolution observations.

Our study also provides some predictions on the location of the heating along the loop for different loop and driver configurations. These results and the observable consequences are summarised in Fig.\ref{heatings}b and Table\ref{waveheatingtable}. \citet{1999SoPh..189...95G} provided a similar summary for heating by braiding, differentiating different heating profiles depending on the time interval between two braiding events. Similarly, \citet{2000ApJ...535..412R} highlighted the need to constrain the heating location with models to infer the heating mechanism and parameters. An example of this can be found in \citet{2000ApJ...539.1002P}, who tried to infer the heating mechanism from the (observed) temperature profile and concluded that heating is not localised at the footpoint.
More recently, \citet{2007A&A...471..311V} used a comparison between models and observations and suggested that heating localised near the footpoints was more likely to be caused by a resistive wave heating mechanism (rather than viscous). \citet{2017A&A...604A.130K} confirmed this results, noting though that the development of Kelvin-Helmholtz instabilities can displace the heating location closer to the apex of the loop.

Finally, we point out that a key limitation of this work is that we do not adopt a realistic loop structure in terms of gravitational stratification and a proper transition region at the foot point. In particular gravitational stratification could imply lower densities near the apex and hence smaller radiative losses, but at the same time might affect the efficiency of the wave heating, as the boundary shell gradients become less steep. However, if the heating is concentrated at the foot points, it might be able to counteract the smaller radiative losses from a stratified loop.

Finally, a key result from this paper is that transverse MHD waves can expand and sustain a boundary shell around or within loops. This is an important aspect of the debate on the role of waves as a heating mechanism, as transitory density enhancement are common features in the corona, as a consequence of, for example, spicules, plasma evaporation, coronal rain or transverse displacements. As footpoint motions continuously drive transverse waves, this scenario considerably facilitates the formation of small-scale structures allowing the conversion of wave energy into thermal energy. Even if the mechanism is not direct dissipation of MHD waves by phase mixing, the transverse waves clearly play a key role in the generation of small-scale structures \citep[see also e.g.][]{2020A&A...636A..40H}.

\begin{acknowledgements}
The authors would like to thank Tom Van Doorsselaere for the helpful discussions on uniturbulence.
This work has received support from the UK Science and Technology Facilities Council (Consolidated Grant ST/K000950/1), the European Union Horizon 2020 research and innovation programme (grant agreement No. 647214) and the Research Council of Norway through its Centres of Excellence scheme, project number 262622.
R. J. Morton is grateful for support from the UKRI Future Leader Fellowship (RiPSAW - MR/T019891/1) and STFC (ST/T000384/1).
This work used the DiRAC@Durham facility managed by the Institute for Computational Cosmology on behalf of the STFC DiRAC HPC Facility (www.dirac.ac.uk). The DiRAC@Durham equipment was funded by BEIS capital funding via STFC capital grants ST/P002293/1 and ST/R002371/1, Durham University and STFC operations grant ST/R000832/1. The DiRAC component of CSD3 was funded by BEIS capital funding via STFC capital grants ST/P002307/1 and ST/R002452/1 and STFC operations grant ST/R00689X/1. DiRAC is part of the National e- Infrastructure.
We acknowledge the use of the open source (gitorious.org/amrvac) MPI-AMRVAC software, relying on coding efforts from C. Xia, O. Porth, R. Keppens.

\end{acknowledgements}

\bibliographystyle{aa}
\bibliography{ref}

\end{document}